\documentclass[aps,noshowpacs,twocolumn,superscriptaddress,nofootinbib]{revtex4-1}

\usepackage{graphicx}
\usepackage{xr}
\usepackage{rotate}
\usepackage{epsfig}
\usepackage[pdftex]{hyperref}
\usepackage{amssymb,amsfonts,amsmath}
\usepackage{multirow}
\usepackage{xcolor,colortbl}
\usepackage[mathscr]{eucal}
\usepackage{paralist}
\usepackage{pdflscape}
\usepackage{textcomp}
\usepackage{dcolumn}
\usepackage[normalem]{ulem}

\usepackage[utf8]{inputenc}
\usepackage{graphicx,epsfig}
\usepackage{times}
\usepackage{dcolumn,bm,fleqn,epic,eepic,float}
\usepackage{booktabs,tabularx,ltablex,longtable}
\usepackage[procnames]{listings}
\usepackage{url}
\usepackage{capt-of}

\usepackage{multibib}

\makeatletter
\AtBeginDocument{\let\LS@rot\@undefined}
\makeatother

\newcommand*{\set}[1]{\ensuremath{\left \lbrace #1 \right \rbrace}}

\newcommand{\avg}[1]{\langle #1 \rangle}
\newcommand{\norm}[1]{\bigl\lVert #1 \bigr\rVert}
\newcommand*{\Prob}[1]{\boldsymbol{\mathbf{#1}}}
\newcommand{\pcondd}[3]{#1 ( #2 \,\Vert\, #3)}
\newcommand{\pcond}[3]{#1 ( #2 \,\vert\, #3)}

\renewcommand{\eqref}[1]{Eq.~\textbf{\ref{#1}}}

\newcommand{\etal}{\emph{et al.} }
\newcommand{\ie}{\emph{i.e.} }
\newcommand{\eg}{\emph{e.g.} }

\newcommand{\arxiv}{\texttt{arXiv} }

\makeatletter

    \def\CT@@do@color{%
      \global\let\CT@do@color\relax
            \@tempdima\wd\z@
            \advance\@tempdima\@tempdimb
            \advance\@tempdima\@tempdimc
    \advance\@tempdimb\tabcolsep
    \advance\@tempdimc\tabcolsep
    \advance\@tempdima2\tabcolsep
            \kern-\@tempdimb
            \leaders\vrule
                    \hskip\@tempdima\@plus  1fill
            \kern-\@tempdimc
            \hskip-\wd\z@ \@plus -1fill }
    \makeatother

\graphicspath{{figures/}}

\newcommand{\polilog}[1]{\text{Li}_{#1}(\text{e}^{-\lambda})}

\newcommand{\ulbf}[1]{\underline{\textbf{#1}}}
\newcommand*{\abr}[1]{\ensuremath{\left( #1 \right)}}

\definecolor{keywords}{RGB}{255,0,90}
\definecolor{comments}{RGB}{0,0,113}
\definecolor{codered}{RGB}{160,0,0}
\definecolor{codegreen}{RGB}{0,150,0}

\newcommand*{\FormatDigit}[1]{\textcolor{codegreen}{#1}}

\newcolumntype{d}[1]{D{.}{.}{#1}}

\begin{document}

\title{Entropic selection of concepts unveils hidden topics in documents corpora}

\author{Andrea Martini}
\email{andrea.martini@epfl.ch}
\affiliation{Laboratory for Statistical Biophysics, \'Ecole Polytechnique F\'ed\'erale de Lausanne (EPFL), CH-1015 Lausanne, Switzerland}

\author{Alessio Cardillo}
\email{alessio.cardillo15@gmail.com}
\affiliation{Laboratory for Statistical Biophysics, \'Ecole Polytechnique F\'ed\'erale de Lausanne (EPFL), CH-1015 Lausanne, Switzerland}
\affiliation{GOTHAM Lab -- Institute for Biocomputation and Physics of Complex Systems (BIFI), University of Zaragoza, E-50018 Zaragoza, Spain}

\author{Paolo De Los Rios}
\email{paolo.delosrios@epfl.ch}
\affiliation{Laboratory for Statistical Biophysics, \'Ecole Polytechnique F\'ed\'erale de Lausanne (EPFL), CH-1015 Lausanne, Switzerland}

\begin{abstract}
The organization and evolution of science has recently become itself an object of scientific quantitative investigation, thanks to the wealth of information that can be extracted from scientific documents, such as citations between papers and co-authorship between researchers. However, only few studies have focused on the concepts that characterize full documents and that can be extracted and analyzed, revealing the deeper organization of scientific knowledge. Unfortunately, several concepts can be so common across documents that they hinder the emergence of the underlying topical structure of the document corpus, because they give rise to a large amount of spurious and trivial relations among documents. To identify and remove common concepts, we introduce a method to gauge their relevance according to an objective information-theoretic measure related to the statistics of their occurrence across the document corpus. After progressively removing concepts that, according to this metric, can be considered as generic, we find that the topic organization displays a correspondingly more refined structure. 
\end{abstract}

\maketitle

%
%

The recent advent of ``\emph{big data}'' is having a transformative on many disciplines \cite{lynch-nature-2008,lazer-science-2009,evans-science-2011}. \emph{Science of science}, \ie the scientific study of scholar activities, makes no exception by leveraging the availability of large amount of information to provide a new and quantitative view of the dynamical organization of the scientific community and its activities. The availability of detailed metadata (\ie data about the data) associated to publication records constitutes an authentic treasure trove. Information like date, title, abstract, affiliations, keywords, and bibliographies have been used, for example, to study the patterns of citations between research articles \cite{de_solla_price-science-1965, radicchi-pnas-2008, kuhn-prx-2014}, the structure of scientific collaborations \cite{newman-pnas-2004,milojevic-pnas-2014}, their stratification and geographical distribution \cite{jones-science-2008, grauwin-scientometrics-2011,gargiulo-scirep-2014,kumar_pan-scirep-2012}, and to identify the best contributions and most successful actors \cite{chen-jinfom-2007,wang-science-2013,petersen-pnas-2014,ke-pnas-2015}.
Unfortunately, the increasing volume of data that makes the science of science possible is associated with a fast growing number of publications, which is turning, in recent times, into a serious issue for scientists \cite{van_noorden-nature-2014,bornmann-j_ass_inf_sci_tec-2015,ginsparg-nature-2011}. It is indeed clear that, in order to stay up to date with the advances within a given discipline, reading all the newly published documents would require an excessive amount of time, possibly leading to reading choices focusing only on those documents that can be considered of \emph{relevance}, and possibly missing some important work that does not seem relevant after a first, superficial perusal. 

To assist researchers in such selection process, several tools have been developed throughout the years \cite{gibney-nature-2014}. Most of them make use of the meta-information attached to the documents (title, abstract, keywords, references and so on) to recommend selected contents. One crucial aspect is the topical classification of documents through semantic analysis, which has captured the interest of the scientific community \cite{griffiths-pnas-2004,mane-pnas-2004,liu-proc_sigir-2004,boyack-pone-2011,steyvers-proc_sigkdd-2004,blei-proc_icml-2006,thuc-proc_pikm-2008,small-res_pol-2014,lancichinetti-prx-2015,silva-jinfo-2016,jensen-jinfor-2016,gerlach-arxiv-2017}, and constitutes one of the core missions/concerns of information retrieval \cite{jurafsky-book-2000,manning-book-2008,evans-science-2011,leskovec-book-2014}. However, the amount of information available in a title, or an abstract, may not be enough to identify the main topic of a given document. The semantic analysis of full documents by extracting its relevant \emph{concepts} might provide more complete and reliable information \cite{frantzi-ijdl-2000}.

One way to map the topical structure of a collection of documents is to consider them as the nodes of a network \cite{boccaletti-scirep-2006,newman-book-2010,latora-book-2017}, while the weight of the edges captures the similarity between documents with respect to their characterizing concepts \cite{boyack-pone-2011} (and references therein). However, the presence of ``common concepts'' appearing in almost every document results in a network which is very dense, akin to an almost complete graph (see Sec.~\ref{s_sssec:phys_2013_similarity_nets} and \ref{s_sssec:climate_network_similarity} of Supplementary Information). An alternative approach to find topics, which is considered the state-of-the-art in information retrivial, is using the so-called \emph{Latent Dirichlet Allocation} (LDA) \cite{blei-jmalea-2003,lancichinetti-prx-2015}, which is nonetheless equally affected by the presence of common concepts.

One of the most recent attempts to simultaneously analyze large corpora of documents  by automatically extracting concepts and by having  experts tag common, non-informative concepts is the  ScienceWISE platform\footnote{\url{http://sciencewise.info}} (SW). 
Nevertheless, the manual curation of common concepts requires the allocation of a considerable portion of time by the users -- assuming their willingness to cooperate. Also, the massive amount of documents, often from domains that are only weakly related with each other (as, \eg, subdomains in physics), demands the presence of a large number of experts with vastly different competences. Furthermore, what can be considered common for an expert in a context may not be so for others, leading to ambiguities. Hence, the definition of common concepts \emph{tout-court} without any objective approach may lead to biases and errors. Given these premises, an automatic filtering method able to discriminate common concepts based on objective, measurable observables would be highly desirable.

In the present manuscript, we propose an approach toward the solution of these problems. More specifically, we design a method that -- given a set of documents -- identifies generic concepts according to their statistical features. This, in turn, allows the reshaping of the relations among words/documents, fostering the emergence of the corpus' underlying topic structure. After introducing the method, we apply it on a collection of physics articles as well as on a collection of web texts on climate change. By performing LDA-based topic modeling on the filtered systems, we identify specific topics in a way that goes beyond a broad area classification, like \arxiv categories. Our findings highlight the fact that being common is an attribute of a concept that strongly depends on the context of the collection under study, and that is non-trivially associated to its frequency within the collection itself.

%
%

\section*{LDA topic modeling and concept filtering}
\label{sec:lda_topic-concept_filtering}

Here, we study the classification of manuscripts into topics induced by the concepts appearing within their whole text and extracted using the SW platform (see Materials and Methods for details). We consider two distinct datasets: scientific manuscripts submitted to the \arxiv\footnote{\url{http://arxiv.org}} e-print archive in the Physics section, and web texts on climate change. 

Given a corpus of documents, each document is parsed and its concepts are automatically extracted and weighted according to their relevance using, for example, their frequency across the document corpus (\textit{document frequency}, $df$) and the number of times they appear in each document (\textit{term frequency}, $tf$) (see Methods for details). The set of concepts pertaining to document 
$\alpha$ is denoted by $\mathcal{C}_{\alpha}$.
Concepts, weighted by their individual $df$ and $tf$, are then related to each other by their co-occurrences within documents, revealing the \textit{topical} organization of the corpus, namely groups of concepts and groups of documents associated with specific subjects. Topics can be obtained using, for example, the LDA algorithm \cite{blei-jmalea-2003} which is considered the state of the art in topic modeling (see Methods). If we count the number of distinct topics $N_T$ (Tabs.~\ref{tab:stats_topics_arxiv} and \ref{stab:stats_topics_climate}) identified by LDA, we observe that it is not very large, suggesting that, on average, each article is similar to a significant fraction of the others. Such paucity of topics is due to the presence of the so-called ``\emph{common concepts}'' (hereafter CC), which enhance the similarity between documents, and consequently reduces the ability to identify specific topics. Therefore, the widespread presence of CC is responsible for the lack of a fine grained classification of documents, \ie specificity. Finally, it is worth mentioning that in the case of similarity networks between documents, the presence of CC is responsible for the proliferation of spurious similarities among documents (see Sec.~\ref{s_sssec:phys_2013_similarity_nets} and \ref{s_sssec:climate_network_similarity} of Supplementary Information).

The SW platform has a built-in list of CC that has been prepared with the collaboration of users who are expert in Physics. In particular, these users can either tag as common some of the concepts that are already present on the platform, or suggest/recommend new ones. Obviously, updating the CC list requires the active cooperation of users. This task could become quite taxing, given the amount of documents and concepts to validate, and the rate at which they are deposited. More importantly, the tag of CC relies solely on the verdict of experts and does not take into account the topic composition of the corpus under scrutiny. As an example, the concept \emph{graphene} could be considered as common within a corpus composed mainly of articles about Material Science. Instead, it should possibly be treated as a specific one in a corpus focused on Biophysics. Thus, simply removing concepts that are manually declared as common might be inappropriate. The aforementioned naive example highlights the weaknesses of the current approach and, thus, calls for an alternative solution to CC tagging. 

Can we design a method to automatically select relevant concepts which also accounts for the composition of the collection? Given a corpus with $N_a$ documents, and being $\mathcal{C} = \bigcup_{\alpha=1}^{N_a} \mathcal{C}_{\alpha}$ the set of all its concepts, a \emph{relevant concept} should neither be too rare, in order for its properties to be statistically well characterized, nor too frequent, to be able to \emph{discriminate} between different documents. Furthermore, a concept is relevant for a document if it is mentioned several times in it. These properties are quantifiable using two well-know indicators of information retrieval \cite{jurafsky-book-2000,manning-book-2008,leskovec-book-2014}. The discriminative power of a concept $c \in \mathcal{C}$ appearing in $N_c$ documents is its \emph{document frequency} $df_c = \tfrac{N_c}{N_a}$. Its relevance for a given document $\alpha$, instead, is measured as the \emph{term frequency} $tf_c(\alpha)$, which is the number of times $c$ appears in $\alpha$. The average term frequency of $c$ is $\avg{tf_{c}} = \tfrac{1}{N_c} \sum_{\alpha=1}^{N_c} {tf_c(\alpha)}$.

In a two-dimensional representation of concepts, based on their $df$ and $\avg{tf}$ (Figs.~\ref{sfig:tessellation-base} and \ref{sfig:bidim_tess_arxiv}) it might be tempting to impose thresholds on both axis to define a region where relevant concepts are most likely to be found. Yet, imposing thresholds on $df$ and $\avg{tf}$ is problematic for several reasons.

Concepts that have been manually tagged as common in the Physics dataset do not fall in any particular location in the $df/\avg{tf}$ plane (black diamonds in Fig.~\ref{sfig:bidim_tess_arxiv}). At best, they tend to follow a law that is not a simple combination of $df$ and/or $\avg{tf}$. Furthermore, as seen in \cite{ferrer_i_cancho-pnas-2003,font_clos-njp-2013,visser-njp-2013,gerlach-njp-2014,yan-pone-2015} the probability that a word appears inside a text or a corpus $n$ times tends to follow the Zipf's law \cite{zipf-book-1949} or, in general, to be broad, as we show in Fig.~\ref{sfig:distro_df_avg_tf_arxiv}. Hence, imposing a characteristic scale on scale-free quantities is not only subjective, but likely right away incorrect. These limitations call for alternative ways to filter concepts based on their \textit{microscopic} behaviour.

\begin{figure*}
\centering
\includegraphics[width=0.7\textwidth]{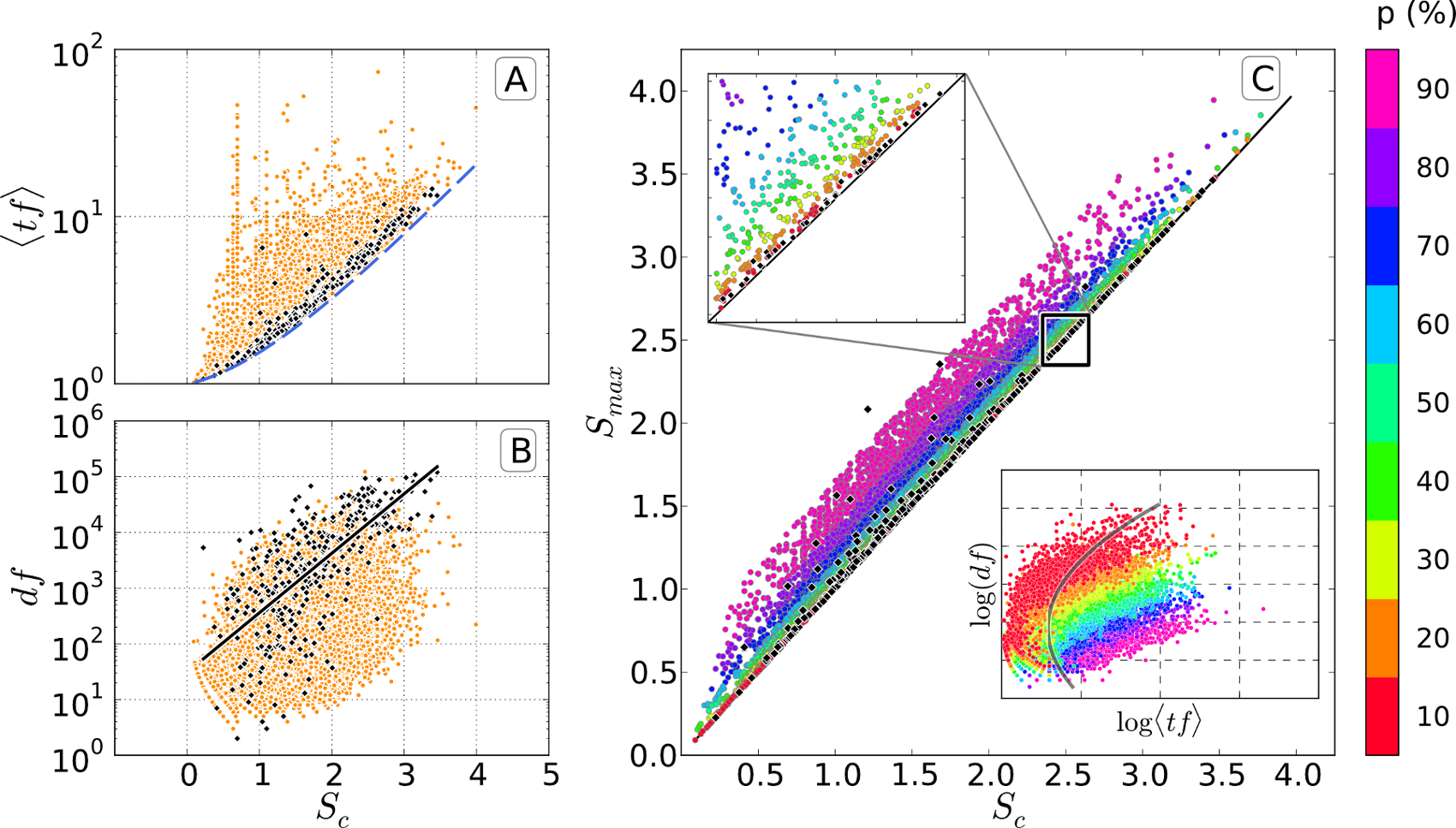}
\caption{Relations between entropy and other features for the concepts of the Physics dataset. SW common concepts (CC) are represented as black diamonds. (A) Relation between the entropy, $S_c$, and the average value of the term-frequency, $\avg{tf}$. The dashed line corresponds to the maximum entropy computed fixing only $\avg{tf}$. (B) Relation between the $df$ and the entropy $S_c$. The solid line is the linear least-squares regression between $\log{df}$ and $S_c$ for CC (Pearson correlation coefficient $r = 0.701$). (C) Organization of the concepts in the $(S_c, S_{max})$ plane. The color of the points encodes the percentile $p$ of the residual entropy distribution $P(S_d)$ to which concepts belong (concepts with $p > 90\%$ are omitted). The solid line is the $S_c = S_{max}$ curve. The Pearson correlation coefficient between $S_{max}$ and $S_c$ for CC is $r = 0.983$. The lower inset is the projection of the percentile information on the $(\log(df), \log\avg{tf})$ plane.}
\label{fig:bidim_entropies_arxiv-conditioned_entropy_vs_avg_tf_and_df}
\end{figure*}

\begin{figure}
\centering
\includegraphics[width=0.5\textwidth]{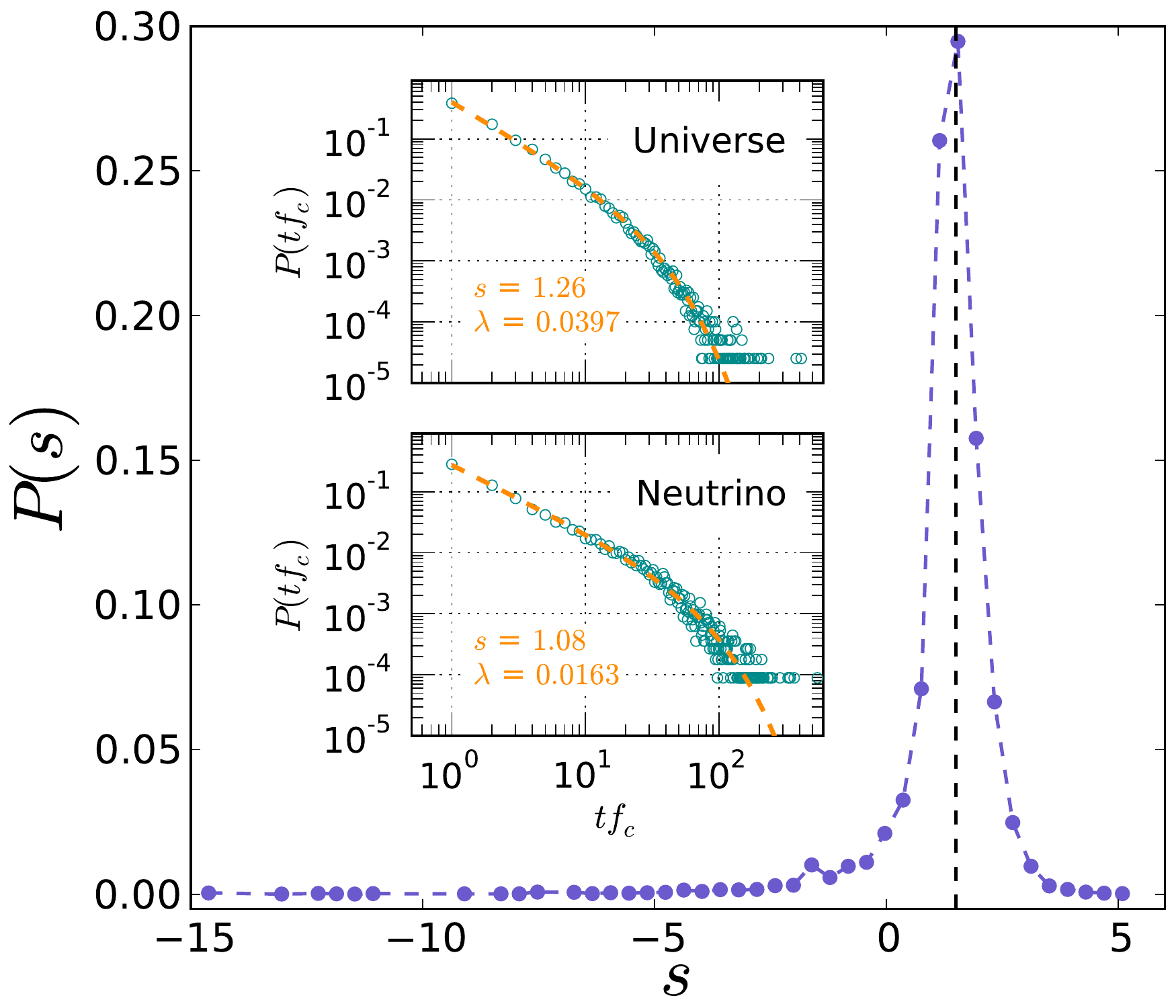}
\caption{Distribution of the power-law exponent, $s$, of the maximum entropy $tf$ distribution for the Physics dataset. Data for $s \leq -15$ are not shown. The dashed vertical line denotes $s=3/2$. The insets display the $tf$ distributions for concepts ``Universe'' (top) and ``Neutrino'' (bottom), together with the corresponding values of $s$ and $\lambda$. The first concept is a CC, while the second one is an example of concepts identified as generic by the entropic filtering. The distribution of the other parameter, $\lambda$, is reported in Fig.~\ref{sfig:histogram_lambda_phys}.}
\label{fig:histo_alpha_inset_fit_selected_concepts}
\end{figure}

Using the  notion of Shannon \emph{information entropy} \cite{shannon-bell-1948}, we can associate the importance of a concept to its \emph{entropy}, $S_c$,  \cite{berger-complin-1996,hotho-jcllt-2005,baek-njp-2011}, defined as  
\begin{equation}
\label{eq:entropy}
S_c = - \sum_{t=1}^{\max(tf_{c})} q_{c}(t) \, \ln q_{c}(t) \,,
\end{equation}
where $q_c(t) = \tfrac{N_c(t)}{N_c}$ is the probability of finding a document where concept $c$ has $tf=t$. It is worth mentioning that if the length of the documents in a corpus is not roughly constant, the same approach could still be used considering the $tf$ \emph{density}, $rtf = \tfrac{tf}{L}$, where $L$ is the length of the document. Interestingly, concepts hand-marked as common tend to have $S_c$ higher than others with the same value of $\avg{tf}$ (Fig.~\ref{fig:bidim_entropies_arxiv-conditioned_entropy_vs_avg_tf_and_df}, panel A), and tend to accumulate toward an ideal hull of the distribution of the concepts in the $(S_c,\avg{tf})$ plane. The same behavior, even more pronounced, is observed in the case of $\avg{\ln (tf)}$ (Fig.~\ref{sfig:ent_cond_vs_log_avg_tf}). This observation suggests that there could be some underlying mechanism that pushes the entropy of common concepts toward its maximum possible value. We have thus checked if a similar behavior could be reproduced with a Maximum Entropy Principle (MEP) approach \cite{baek-njp-2011,yan-pone-2015} and, we have computed for each concept its \emph{maximum entropy} $S_{max}(c)$ constrained by the values of $\avg{tf_c}$ and $\avg{\ln{(tf_c)}}$ (see Materials and Methods). Indeed, direct inspection for several concepts, and in particular for the ones manually tagged as common, reveals that $q_c(t)$ is well described by a power-law with cutoff, $q_c(t) \propto t^{-s} e^{-\lambda t}$ (see insets in Fig.~\ref{fig:histo_alpha_inset_fit_selected_concepts}), which is precisely the functional form expected from a maximum entropy principle with the above mentioned constraints.

%
%
%

Remarkably, the MEP approach reveals that $s=3/2$ is the most frequent value of the power-law exponent (Fig.~\ref{fig:histo_alpha_inset_fit_selected_concepts}), which is the exponent typical of critical branching processes \cite{harris-book-1963}. Although further investigations in this direction go beyond the scope of the present work, it is suggestive to picture the appearance of papers in the \arxiv as a process where older manuscripts ``inspire'' (``generate'', in the branching process language) new papers containing a similar number of concepts, each appearing a similar number of times.

%
%
%

Using entropies as a new set of coordinates, we arrange concepts on the $(S_c, S_{max})$ plane as reported in Figs.~\ref{fig:bidim_entropies_arxiv-conditioned_entropy_vs_avg_tf_and_df} and \ref{sfig:entropies_rescaled_tf_climate}. As expected, the majority of common concepts lays close to the $S_c = S_{max}$ line. We exploit this feature to design a new criterion to discriminate concepts. For each concept, we define its \emph{residual entropy}, $S_d(c)$, as the difference between its maximum and measured entropies $S_d(c) = S_{max}(c) - S_{c}(c)$. This quantity is equivalent to the Kullback-Leibler divergence between the observed probability distribution, $q_c(t)$, and the maximum entropy one, $p_c(t)$, reported in \eqref{eq:pmf} \cite{kullback-ann_mat-1951} (see Sec.~\ref{s_ssec:equiv-kl-diff_ent} of SI for details). We then assign concepts to different percentiles $p$ of the probability distribution of $S_d$. The color of the dots in Fig.~\ref{fig:bidim_entropies_arxiv-conditioned_entropy_vs_avg_tf_and_df}, panel C, accounts for the value of $p$ and the lower inset is the projection of the percentile information on the $(df, \avg{tf})$ space. Finally, we can use $p$ as a sort of ``distance'' from the maximum entropy curve $S_c = S_{max}$, considering as \emph{significant} those concepts having $p \geq \tilde{p}$, thus using them to find topics through LDA and, in turn, classify documents. 

A closer inspection of Fig.~\ref{fig:bidim_entropies_arxiv-conditioned_entropy_vs_avg_tf_and_df}, panel C, reveals that the manual annotation of common concepts is inadequate. On the one hand, there are concepts that were marked as common by SW experts but that are located far from the $S_c = S_{max}$ line. Examples are `M87 jet', `mechanical advantage', `FitzGerald-Lorentz contraction', `Boyle's law', `special linear group', and `double pendulum', which could be thought as generic only within a very selected range of topics, but become quite specific in a corpus spanning a wider range of subjects as in our case (see Fig.~\ref{sfig:donut_categories_physics}). On the other hand, there are many concepts close to the diagonal but not marked as common such as `dimensions', `statistics', `Hamiltonian', `degree of freedom', `intensity', `counting', and `luminosity' that have slipped through the attention of experts without being tagged as generic albeit being obviously so. 

Expectedly, the use of entropy to quantify the relevance of words within single texts is not new in natural language processing  \cite{herrera-epjb-2008, yan-pone-2015, carpena-pre-2016, altmann-jstat-2017}. However, apart from focusing on single documents, these studies seek to understand the role played by the position of words within texts. More importantly, none of them use entropies to assess the relevance of words for discriminating the content of documents within a collection, which instead constitutes the cornerstone of our approach. Our ranking method differs also from another well know approach of natural language processing, namely the ranking based on the \emph{Inverse Document Frequency} $IDF$ \cite{jones-jdoc-1972,robertson-jdoc-2004}, albeit the two quantities are not completely unrelated (see Sec.~\ref{s_sssec:phys_diff_ent_idf_2collections} and \ref{s_sssec:climate_diff_ent_idf} of SI). Finally, as shown in Tabs.~\ref{stab:concepts_ranked_in_papers_various_measures} -- \ref{stab:concepts_ranked_in_papers_various_measures_optimal_filtering}, our method can also be used to gauge the relevance of a concept within a given document.

%
%

\section*{Results}

The entropy-based objective criterion allows us to discard concepts before using them to extract the organization of documents into topics using LDA. The consequences of concept filtering on the topic mapping are displayed in Tabs.~\ref{tab:stats_topics_arxiv} and \ref{stab:stats_topics_climate}. In the case of the Physics dataset, both the total number of concepts $N_{con} = \lvert \mathcal{C} \rvert$ and the number of documents having at least one concept, $N_a$, decrease with $p$, albeit the latter remains pretty constant up to $p = 20\%$. The number of topics found by LDA, $T$, increases considerably, contrarily to the case of \emph{meaningful topics} (see Methods), $T^*$, which displays a rise and fall with a maximum at $p = 50\%$. The proliferation of topics is, to some extent, expected and mimics the existence of ``\emph{cultural holes}'' among distinct branches of Physics and, more in general, science itself \cite{vilhena-soc_sci-2014}. However, the monotonic decrease of the average number of documents, $\langle N_a \rangle_{T^{*}}$, and concepts, $\langle N_{con} \rangle_{T^{*}}$, per meaningful topic  denotes that initially topics become more specific, but afterwards they resemble the byproduct of spurious relations among concepts. Finally, the fraction of documents assigned to a meaningful topic, $F = \tfrac{T^* \langle N_a \rangle_{T^{*}}}{N_a(p=0)}$ quantifies the overall price we have to ``pay'' to retrieve a more refined topic mapping.

Except for $F$, a similar trend -- although not monotonic -- can be observed also for the climate dataset (see Tab.~\ref{stab:stats_topics_climate}). Therefore, a moderate reduction of the pool of concepts implies that topics become more specific. As a direct consequence, the overlap between the content of a document and its topic increases, as could be inferred from Figs.~\ref{sfig:topicmapping_statistics_phys_2009-2012}--\ref{sfig:ccdfs_topics_given_docs_phys2013} and Figs.~\ref{sfig:topicmapping_statistics_climate} and \ref{sfig:ccdfs_topics_given_docs_climate}.
%
%
%
%
\begin{table*}
\centering
%
\newcolumntype{d}[1]{D{.}{.}{#1} }
\begin{tabular*}{\hsize}{@{\extracolsep{\fill}}cccccccd{2}}
$p \, (\%)$ & $N_{con}$ & $N_a$ & \multicolumn{1}{c}{$T$} & \multicolumn{1}{c}{$T^*$} & \multicolumn{1}{c}{$\langle N_a \rangle_{T^{*}}$} & \multicolumn{1}{c}{$\langle N_{con} \rangle_{T^{*}}$} & \multicolumn{1}{c}{$F$} \\ \hline
{\cellcolor{gray!20}} 0 & {\cellcolor{gray!20}} 15040 & {\cellcolor{gray!20}} 189759 & {\cellcolor{gray!20}} 10 & {\cellcolor{gray!20}} 10 & {\cellcolor{gray!20}} 18976  & {\cellcolor{gray!20}} 6185 & {\cellcolor{gray!20}} 1.00\\
10 & 11807 & 187165 & 39 & 15 & 11705 & 714 & 0.93 \\
20 & 10496 & 183530 & 57 & 24 & 7036 & 386 & 0.89 \\
30 & 9184 & 174813 & 121 & 24 & 5657 & 267 & 0.72 \\
40 & 7872 & 157951 & 202 & 27 & 3308 & 139 & 0.47 \\
50 & 6560 & 130472 & 289 & 29 & 1885 & 77 & 0.29 \\
60 & 5248 & 96936 & 520 & 28 & 1214 & 50 & 0.18 \\
70 & 3936 & 60597 & 1255 & 17 & 753 & 32 & 0.07 \\
80 & 2624 & 32397 & 3103 & 4 & 565 & 17 & 0.01 \\
90 & 1312 & 9865 & 5747 & 0 & 0 & 0 & 0.00 \\
\hline
\end{tabular*}
\caption{Characteristics of the topic modeling on Physics dataset. The row $p=0\%$ corresponds to the original corpus/dataset, while $p>0\%$ to those filtered using the maximum entropy. In the columns we report: the percentage of filtered concepts $p$, the number of concepts $N_{con}$, of documents having at least one concept $N_a$, of topics found by LDA, $T$, and number of ``meaningful'' ones $T^*$. For the latter, we report also the average number of documents $\langle N_a \rangle_{T^{*}}$, and concepts $\langle N_{con} \rangle_{T^{*}}$ per topic. Finally, we report the fraction of documents assigned to a meaningful topic, $F$ (we define as meaningful those topics for which $\pi(t) > 0.01 $).}
\label{tab:stats_topics_arxiv}
\end{table*}
%

%
%

\subsection*{Organization of documents into topics}

\begin{figure*}
\centering
\includegraphics[width=\textwidth]{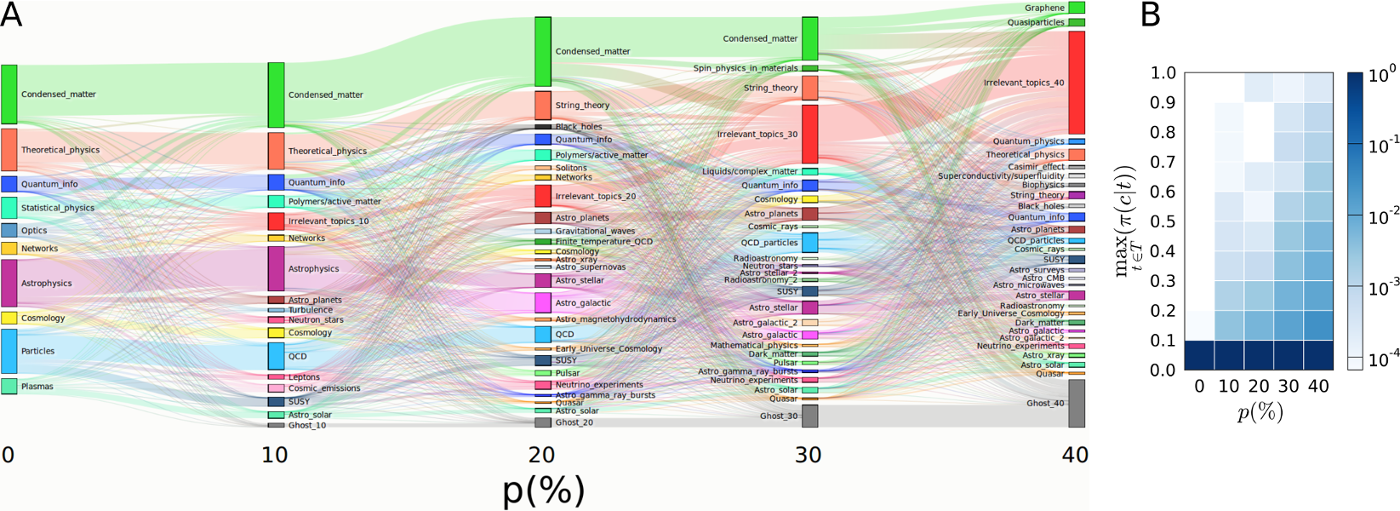}
\caption{Organization of the Physics corpus into topics. (A) Static Sankey diagram where each topic is represented as a colored box whose height is proportional to the number of papers it contains. A title is assigned to each box according to the ten most used concepts, \ie those appearing in more papers. The thickness of the bands between boxes indicates the number of shared articles. Each column denotes a different intensity of filtering $p$. Interactive version available at \cite{sankey-interactive}. (B) Heatmap for the maximum probability that a concept $c$ appears in a topic $t$, $\max_{t \in T}(\pcond{\pi}{c}{t})$, for different intensity of filtering $p$.}
\label{fig:sankey_phys_2009_2012}
\end{figure*}

Knowing the topic structure of a corpus is of utmost importance in platforms like Amazon, aNobii or Reddit\footnote{\url{https://www.amazon.com/} $\quad$ \url{http://www.anobii.com/} $\quad$ \url{https://www.reddit.com/}} where recommendations rely on the successful classification of documents according to their contents. In the case of ScienceWISE and Physics documents, inferring the topic structure has -- at least -- two possible implications. On the one hand, it could be used to recommend contents to users or to cross-validate the PhySH system recently adopted by the American Physical Society to classify manuscripts \cite{physh-website}. On the other hand, it could be used to portray the fine graining process of specialization undergoing in Physics and in all Science in general. To this aim, we study the evolution of the topic mapping of the corpus when we progressively reduce the pool of concepts using our entropic filtering. The result of the analysis is reported in the Sankey/alluvial diagram\footnote{The interactive version of these diagrams displaying additional information is available at \cite{sankey-interactive}} of Figs.~\ref{fig:sankey_phys_2009_2012}A, \ref{sfig:sankey_TM_phys2013}, and \ref{sfig:sankey_TM_climate} \cite{rosvall-pone-2010}. The topic structure of our collections has been obtained using an improved version of LDA named \emph{TopicMapping} introduced in \cite{lancichinetti-prx-2015}.

%
%

In such diagram, each box represents a topic and its height is proportional to the number of documents associated to it. The name of each box refers to its main subject. Each column identifies a different filter intensity $p$. The evolution of the topics reveals some intriguing and interesting features. When $p=0$, the detected topics clearly correspond to the major areas/sectors in physics. No finer grouping is possible at this level because of the presence of common concepts, that act as a ``glue'' within large subjects. As $p$ increases, there is a progressive fragmentation of topics passing from broad areas of Physics -- not exactly overlapping with the \arxiv classification as shown by \cite{palchykov-epj_ds-2016} -- to increasingly specific subjects at $p=20\%$. An example is the fragmentation of Astrophysics ($p=0\%$) into Stellar, Planetary, Galactic, and X-rays Astrophysics which progressively unfolds up to $p=40\%$. Since according to LDA each concept $c$ is associated to a topic $t$ with a certain probability $\pcond{\pi}{c}{t}$ (see Methods), the progressive specialization of topics for increasing values of $p$ is reflected in an increasing number of concepts having high probability of being assigned to few topics, thus reducing ambiguity (Fig.~\ref{fig:sankey_phys_2009_2012}B).

Although the pruning of common concepts allows to detect smaller and more specific topics, pushing it to overly large values of $p$ deteriorates the results. This is due to a superposition of different effects. On the one hand, when the pool of concepts shrinks too much the statistical significance of an increasing fraction of topics found by LDA is below an acceptable level (see Methods and Figs.~\ref{sfig:topicmapping_statistics_phys_2009-2012}A, \ref{sfig:topicmapping_statistics_phys2013}A and \ref{sfig:topicmapping_statistics_climate}A of SI). Documents belonging to such irrelevant topics are gathered into the ``Irrelevant\_topics'' box. On the other hand, a growing fraction of papers gets stripped of all its concepts and vanishes from the collection (``ghost'' papers). Therefore, heuristically, filtering should not exceed a level $p_{opt}$ such that the fraction of documents assigned to a meaningful topic, $F$, does not fall below a certain threshold (other complementary criteria are shown in Sec.~\ref{s_ssec:optimal_filtering} of SI). Assuming a threshold value of 0.75, for the Physics dataset $20 \% \leq p_{opt} \leq 30 \%$ while for climate we have $p_{opt} \sim 5 \%$ as reported in Tabs.~\ref{tab:stats_topics_arxiv} and \ref{stab:stats_topics_climate}.

As mentioned before, we have also repeated the same analysis presented hiterto for Physics on a set of web documents about climate change. In this case, we had to slightly modify the approach by considering the probability distribution of the rescaled $tf$, namely the $tf$ of concept $c$ in paper $\alpha$ ($tf_{c}(\alpha)$) divided by the length of the paper ($L(\alpha)$), \ie $rtf_c(\alpha) = \tfrac{tf_c(\alpha)}{L(\alpha)}$, because of the presence in the set of groups of documents of vastly different size. Correspondingly, the entropy had to be redefined as an integral (see Sec.~\ref{s_sssec:maxent_rescaled_tf} of SI for methodological details, Secs.~\ref{s_sssec:entropic_filter_climate} and \ref{s_sssec:topicmapping_climate} for the results). Furthermore, the climate dataset was parsed for \textit{keywords}, without an ontology structure. As a consequence the similarity between documents is less precise as reflected by the extremely high fraction of documents falling in the ``Irrelevant\_topics'' category in the original dataset. This notwithstanding, filtering based on the entropy of keywords can still be used to generate a more suitable pool of concepts to feed the \emph{TopicMapping} algorithm. At the same time, the identification of concepts of different degrees of generality might be used to generate an ontology for this document corpus.

In general, the phenomenology of filtering can be grouped into two classes:
\begin{inparaenum}[i)]
 \item preservation with specialization (\eg Condensed\_matter/Quantum\_info) \ie when the topic of a community remains unaltered but the concepts used to characterize it are more specific;
 \item splitting with specialization (\eg Astrophysics $\rightarrow$ Stellar + Galactic + Planets) \ie when the removal of generic concepts ends up in the fragmentation of the original topic into more specific sub-topics.
\end{inparaenum} 
Finally, we want to stress that the observed phenomenology does not change even when topics are extracted performing community detection on the networks of similarities between articles as shown in Secs.~\ref{s_sssec:phys_2013_similarity_nets} and \ref{s_sssec:climate_network_similarity} (Figs.~\ref{sfig:sankey_net_phys2013} and \ref{sfig:sankey_climate_rtf}) of the SI, confirming that the emergence of hidden topics is not an artifact of the method used to retrieve topics but rather a characteristic of the concept filtering \emph{per se}. Moreover, rankings of concepts using either $IDF$ or residual entropy $S_d$ are different especially for values of $p$ higher than $10\%$, as shown in Figs.~\ref{sfig:inters_percentiles_idf_vs_maxent_tf} and \ref{sfig:comparison_comm_physics}.

\section*{Discussion}

The access to the semantic content of whole documents grants an unprecedented opportunity for their classification and can improve the search of contents within huge collections. However, such opportunity comes at a hefty price: the similarity relations among documents based on their concepts are cluttered due to the presence of common concepts, hindering the retrieval of the topic structure/landscape. In the present manuscript we have presented a method based on maximum entropy to filter the pool of concepts by automatically selecting the relevant ones and improve the topic modeling of big document corpora. According to the method, common concepts are those whose entropy is closer to their maximum one. The definition of \emph{common} stemming from our method is less subjective than the one used by the SW platform since it does not rely on user validation and, more importantly, depends on the content of the documents under scrutiny. We presented the benefits of selective concept pruning on two different corpora: scientific preprints on Physics and web documents on climate change (Figs.~\ref{fig:sankey_phys_2009_2012}--\ref{sfig:sankey_TM_phys2013} and \ref{sfig:sankey_TM_climate}). Finally, the entropic filtering proposed here could be applied in a recursive way on sub-corpora of documents or can be used to study the evolution in time of the generality of a concept (like Graphene or Python). Last but not least, the method could be used also to improve already existing ontologies.


\section*{Materials and Methods}

\subsection*{Data}

We consider two collections of documents. One containing scientific manuscripts from {\small \arxiv\footnote{\url{https://arxiv.org/}}}, a repository of electronic preprints of scientific articles, and another made of web articles on climate change extracted using the underlying machinery of the ScienceWISE platform. In the case of scientific manuscripts, we selected documents submitted from year 2009 to 2012 under the physics categories either as primary or secondary subjects resulting in a corpus of 189,759 articles (Tab.~\ref{tab:stats_topics_arxiv}). The composition of the corpus in terms of {\small \arxiv} categories is reported in Tab.~\ref{stab:primary_categories}. We have considered also a smaller corpus of 52,979 manuscripts submitted in 2013 within the same categories. However, the results corresponding to this collection are displayed only in the SI. The climate change corpus has been built selecting web documents written in English with at least 500 words, whose URLs are mentioned by -- at least -- 20 distinct tweets (see Sec.~\ref{s_sssec:climate_source} of SI). Texts are parsed and keywords are extracted using KPEX algorithm \cite{constantin-thesis-2014}. After that, keywords are matched with concepts available in a crowdsourced ontology accessible on the platform. The ontology has been built by initially collecting scientific concepts from online encyclopedias and subsequently refined with manual inspection by experts. The second step is missing for climate web documents since no ontology is available in SW. Overall, the climate collection has 18,770 articles. The Physics dataset possesses 15,040 concepts, from which we discarded those appearing always with the same $tf$ ending up with 13,124 concepts, 348 of which have been marked as ``common'' by SW. For the climate dataset, instead, we have 152,871 keywords. By deleting those having a $\max(rtf) - \min(rtf) \leq 0.005$, only 9222 keywords are left.

\subsubsection*{Relevance of concepts}

Given the set of concepts $\mathcal{C}= \bigcup_{\alpha=1}^{N_a} \mathcal{C}_{\alpha}$ used in a corpus having $N_a$ documents, the relevance of a concept $c$ in a document $\alpha$ is given by its {\small \emph{boosted term frequency}}, $tf_c$, \ie the number of times $c$ appears in $\alpha$ modulated according to the location (title, abstract, body) where it appears. The relevance of $c$ to discriminate documents in the corpus corresponds, instead, to its {\small \emph{Inverse Document Frequency}}. The product of these two estimators is nothing else than the so-called {\small \textsc{TF-IDF}} and is commonly used in information retrieval to quantify the relevance of a concept in a document \cite{jones-jdoc-1972,robertson-jdoc-2004}. Hence, we have:
\begin{equation}
\label{eq:tfidf}
\text{\textsc{TF-IDF}}_c (\alpha) = tf_{c}(\alpha) \cdot IDF_c = tf_{c}(\alpha) \cdot \ln \left( \dfrac{N_a}{N_c} \right)\,.
\end{equation}
where $IDF_c$, is the {\small \emph{Inverse Document Frequency}} and penalizes concepts used frequently, and $N_c$ is the number of papers containing concept $c$.

\subsection*{Maximum entropy principle}

To gauge how informative a concept can be, we calculate (using \eqref{eq:entropy}) its entropy $S_c$ based on the term-frequencies $tf_c$. We have observed that concepts labeled as common in the SW platform tend to have a higher entropy with respect to other concepts having the same $\avg{tf}$ (Fig.~\ref{sfig:ent_cond_vs_log_avg_tf}). To corroborate such regularity, we have applied the maximum entropy principle to the distribution of the term-frequencies of a concept, $tf_c$, to determine the associated probability mass function that satisfies certain constraints. As shown in Supplementary Information, Sec.~\ref{s_ssec:maxent_models}, the selection of the empirical values of the first moment and log-moment, $\langle tf_c \rangle$ and $\langle \ln \left({tf_c}\right) \rangle$ \cite{visser-njp-2013}, as constraints implies a probability mass function of the following form: 
\begin{equation}
 p_{c}(t) = \dfrac{\dfrac{\text{e}^{-\lambda t}}{t^s}} {\text{Li}_{s}(\text{e}^{-\lambda})} \, ,
 \label{eq:pmf}
\end{equation}
where $\text{Li}_{s}(\text{e}^{-\lambda})$ is the {\small \emph{polylogarithm}} of order $s$ and argument $\text{e}^{-\lambda}$, defined as:
\begin{equation}
 \text{Li}_{s}(\text{e}^{-\lambda}) = \sum_{t=1}^{\infty}{\dfrac{\text{e}^{-\lambda t}}{t^s}} \, .
 \label{eq:polylog}
\end{equation}
The parameters $\lambda$ and $s$ are determined, for each concept $c$, imposing the constraints to \eqref{eq:pmf} and solving numerically the system of equations:
\begin{equation}
\begin{array}{rcl} 
\dfrac{\text{Li}_{s-1}(\text{e}^{-\lambda})} {\text{Li}_{s}(\text{e}^{-\lambda})} & = & \langle tf_c \rangle \, ,\\ 
\\
- \dfrac{ \partial_s{\text{Li}_{s}(\text{e}^{-\lambda})}} {\text{Li}_{s}(\text{e}^{-\lambda})}  & = & \langle \ln \left({tf_c}\right) \rangle \, .
\end{array}
\label{eq:system}
\end{equation}
As a consequence, the maximum entropy $S_{max}$ is:
\begin{equation}
S_{max} = \ln{ \left[ \text{Li}_{s}(\text{e}^{-\lambda}) \right] } + \lambda \langle tf_c \rangle + s  \langle \ln { \left( {tf_c} \right) } \rangle \, .
\label{eq:max_ent}
\end{equation}

\subsection*{Latent Dirichlet Allocation \& topic mapping}

Over the years, several methods to retrieve the organization of groups of documents into distinct topics have been proposed in information retrieval \cite{jurafsky-book-2000,manning-book-2008}. The state-of-the-art is the Latent Dirichlet Allocation (LDA) method \cite{blei-jmalea-2003}, which is an evolution of Probabilistic Latent Semantic Analysis (PLSA), also known as Probabilistic Latent Semantic Indexing (PLSI) \cite{hofmann-puai-1999,hofmann-proc_sigir-1999}. We adopt an improved version of LDA named TopicMapping (TM) introduced by Lancichinetti \emph{et al.} in \cite{lancichinetti-prx-2015}. In a nutshell, the main idea behind LDA is that topics are nothing else than groups of related words and, consequently, documents are associated to mixtures of topics. The co-occurrence of words in documents is responsible for the emergence of topics.

Given a corpus of $N_a$ documents, its topics $T$ are subsets of the set of concepts $\mathcal{C}$. The probability that a concept $c$ belongs to a topic $t$ is $\pcond{\pi}{c}{t}$. Conversely, the topic mixture of an article $\alpha$ is described by the probability that a topic $t$ appears in $\alpha$, $\pcond{\pi}{t}{\alpha}$. According to LDA, both the probability distributions over the topics appearing in a document $\pcond{\Prob{\Pi}}{T}{\alpha} = \set{\pcond{\pi}{t}{\alpha}}_{t=1}^{T} $, and the one over the concepts belonging to a topic $t$, $\pcond{\Prob{\Pi}}{C}{t} = \set{\pcond{\pi}{c}{t}}_{c=1}^{C}$, are drawn from a Dirichlet distribution having the total number of topics $T$, and of concepts $C$ as parameters. In our case both $T$ and the assignment of concepts to topics are computed from the TM algorithm.

According to TM, the relations between concepts in a corpus can be mapped as a network where concepts are the nodes and edges accounts for their co-occurrence \cite{cong-phys_lif_rev-2014}. Considering two concepts $u$ and $v$, the weight of the edge connecting them, $z_{uv}$, is equal to the dot product similarity of their term-frequency vectors $tf_{u}$ and $tf_{v}$, $z_{uv} = \sum_{\alpha=1}^{N_{a}} tf_{u}(\alpha) \, tf_{v}(\alpha) \in [0,\infty)$. To reduce the impact of noisy interactions corresponding to concepts appearing too frequently, only edges with weight significantly higher than their randomized counterpart are retained for a $p$-value of 5 \%. The null model used to produce such randomization assumes that concepts are randomly distributed across documents while preserving the sum of the $tf$s across the whole corpus. Clusters/communities of concepts are then identified as topic prototypes using the Infomap algorithm applied on the pruned network \cite{rosvall-pnas-2008}. Then, a local optimization of the PLSA likelihood \cite{hofmann-puai-1999} is adopted to relax the exclusive assignment of a concept to a single topic, and to narrow the number of topics within documents. Finally, a further refinement is applied on the PLSA results by optimizing the LDA likelihood, thereby obtaining the final probabilities defined above.

After obtaining the topics from LDA, documents are composed by multiple topics (\eg interdisciplinary articles). Nevertheless, we can assign each document $\alpha$ to its \emph{dominant} topic $\tilde{t}$ that maximizes the probability $\pcond{\pi}{t}{\alpha}$, as we can see from Figs.~\ref{sfig:ccdfs_topics_given_docs_phys_2009-2012}, \ref{sfig:ccdfs_topics_given_docs_phys2013}, and \ref{sfig:ccdfs_topics_given_docs_climate}. Moreover, TM associates to each topic $t$ identified by LDA a measure of its statistical significance, $\pi(t) = \sum_{c} \pcond{\pi}{t}{c} \, \pi(c)$, which is given by the sum of the probability that word $c$ belongs to topic $t$ multiplied by the probability that the word $c$ appears in the whole corpus. We consider as \emph{meaningful} those topics having $\pi(t) \geq 0.01$ (see Figs.~\ref{sfig:topicmapping_statistics_phys_2009-2012}, \ref{sfig:topicmapping_statistics_phys2013}, and \ref{sfig:topicmapping_statistics_climate} of SI).

\section*{acknowledgments}
All the authors acknowledge the financial support of SNSF through the project CRSII2\_147609. AC acknowledges the support of \emph{Ministerio de Economia y Competitividad} (MINECO) through grant RYC-2012-01043. The authors also thank Alex Constantin and Sta\v{s}a Milojevic for many helpful discussions, and the developer team of ScienceWISE for its help.
%

%
%
%

%
%

\clearpage
\widetext
\begin{center}
\textbf{\large Supplementary Materials for the manuscript entitled:\\Entropic selection of concepts unveils hidden topics in documents corpora}
\end{center}
\setcounter{equation}{0}
\setcounter{figure}{0}
\setcounter{table}{0}
\setcounter{page}{1}
\setcounter{section}{0}
\makeatletter
\renewcommand{\theequation}{S\arabic{equation}}
\renewcommand{\thefigure}{S\arabic{figure}}
\renewcommand{\thetable}{S\roman{table}}
\renewcommand{\thesection}{S\Roman{section}}


\setlength{\abovedisplayskip}{10pt}
\setlength{\belowdisplayskip}{10pt}

\title{Supplementary Materials for the manuscript entitled:\\Entropic selection of concepts unveils hidden topics in documents corpora}

\maketitle

%
%
%
%

\section{Theory}
\label{s_sec:theory}

In this section we provide the theoretical details behind our maximum-entropy based filtering method. We begin introducing the two-dimensional tessellation filtering (Sec.~\ref{s_ssec:bid_tess}). Then, we prove the relation between full entropy, $S_f$, and conditional one, $S_c$, and we motivate why we based the filtering methodology on the conditional entropy instead of the full one (Sec.~\ref{s_ssec:rel-full_ent-cond_ent}). After that, we provide the details of the maximum entropy models used in the main text (Sec.~\ref{s_ssec:maxent_models}) and we demonstrate the equivalence between the residual entropy, $S_d$, and the Kullback-Leibler divergence between the probability distributions of empirical observations and maximum entropy model (Sec.~\ref{s_ssec:equiv-kl-diff_ent}).  Finally, we present the comparison between the concept lists ranked according to residual entropy $S_d$ and $IDF$ (Sec.~\ref{s_ssec:comparison_sc_idf}), between communities after filtering these concept lists (Sec.~\ref{s_ssec:comparison_communities}), and the correlation between ranked lists of concepts (Sec.~\ref{s_ssec:comparison_rankings_kendall}). Finally, in Sec.~\ref{s_ssec:similarity_network} we show how to generate a network of similarity between documents.

\subsection{Two-dimensional tessellation}
\label{s_ssec:bid_tess}

\begin{figure}[h]
\centering
\includegraphics[width=0.4\textwidth]{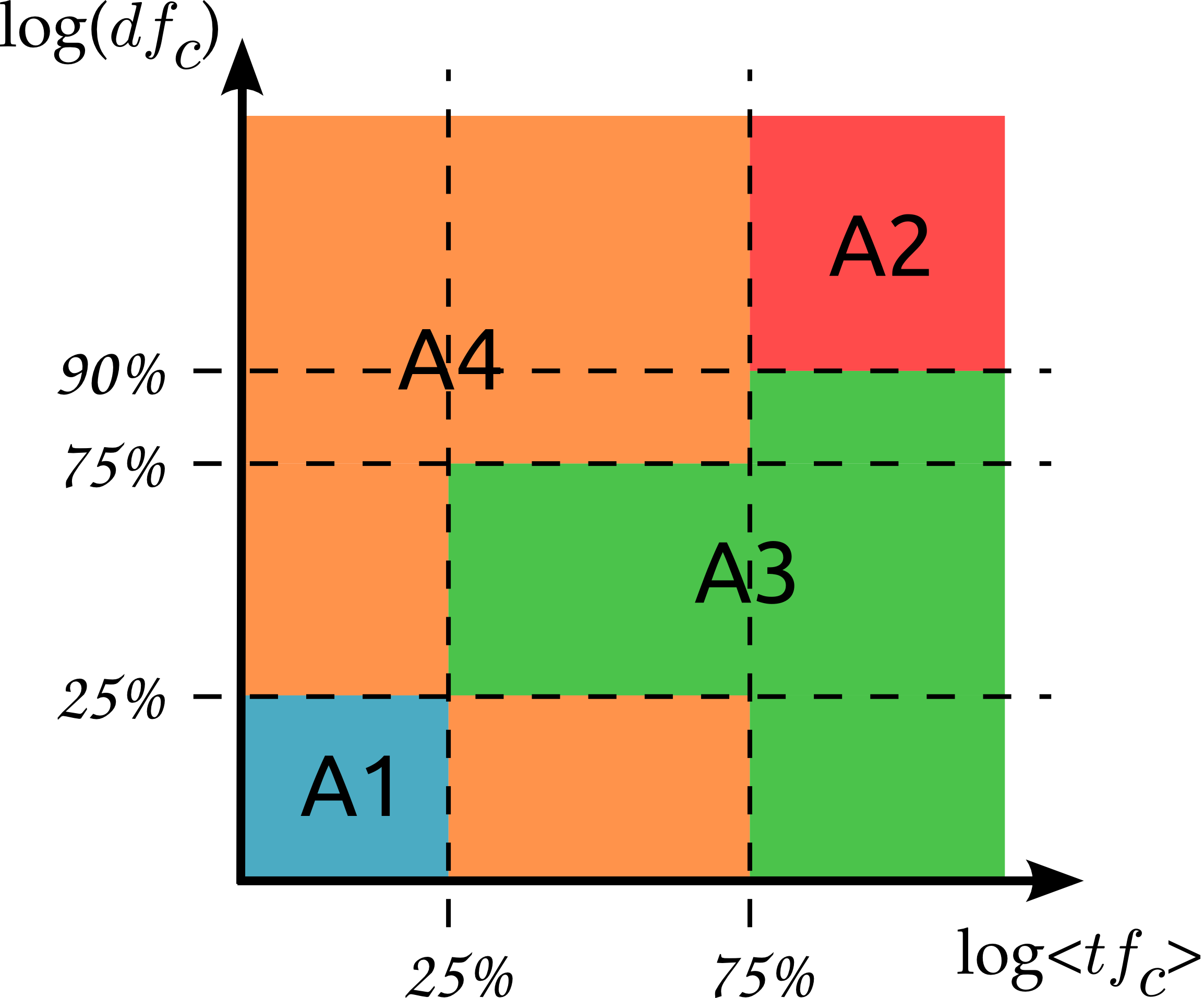}
\caption{Tessellation of the $\avg{tf_{c}} - df_c$ plane used to classify concepts. Each colored area defines one type of concept. Logarithmic scale is used to visualize neatly the results. The dashed lines are used to delimit the tiles.}
\label{sfig:tessellation-base}
\end{figure}

As we have commented in the main text, the document frequency $df$ and average term frequency $\avg{tf}$ (or its average density $\avg{rtf}$), have been used extensively in information retrieval to characterize the relevance of words/n-grams \cite{manning-book-2008}. More specifically, we can use such features as coordinates of a two-dimensional space, thus classifying a concept $c$ by the position it occupies in the $(\langle tf_{c} \rangle, df_c)$ plane. We tessellate the plane by imposing thresholds on the coordinates to delimit regions where \emph{ubiquitous}, \emph{rare} and \emph{relevant} concepts fall as shown in Fig.~\ref{sfig:tessellation-base}.

%
%
%

\noindent We define the following tiles on the $\avg{tf_{c}} - df_c$ plane:
\begin{description}
 \item[A1] The domain of \textbf{specific/rare} concepts characterized by having small values of both $df$ and $\avg{tf}$.
 \item[A2] The domain of \textbf{common/ubiquitous} concepts displaying high values of both $df$ and $\avg{tf}$.
 \item[A3] The domain of \textbf{relevant/informative} concepts corresponding to those having intermediate values of $df$ and $\avg{tf}$.
 \item[A4] All the remaining concepts appearing within documents not enough times (on average) to be considered relevant.
\end{description}
If we consider each trait separately, we can divide its space into three -- or more -- domains denoting low, medium and high values of such trait. Specifically, we consider two values for $\avg{tf_c}$ ($25\%$ and $75\%$), and three for $df_c$ ($25\%$,$75\%$, and $90\%$), instead. We use percentiles values since both quantities tend to have a broad distribution, hence raw values are not suitable to highlight their variability. Therefore, we can estimate the similarity between documents using only those concepts classified as \emph{relevant}. Nevertheless, despite its intuitive nature, the tessellation method is not a good filtering approach since it depends on too many parameters, whose numbers and values are arbitrary. Moreover, as shown later on in Fig.~\ref{sfig:bidim_tess_arxiv}, the tessellation is unable to reproduce all the features of those concepts tagged as generic by ScienceWISE. Such limitations drove us to abandon the two-dimensional filtering in favors of alternative approaches.

\subsection{Relation between full entropy and conditional entropy}
\label{s_ssec:rel-full_ent-cond_ent}

The probabilistic formulation of entropy used in the main text does not contemplate as an event the \emph{absence} of a concept $c$ in a document (\ie $tf = rtf = 0$). For this reason, in \eqref{eq:entropy} the sum starts from $tf = 1$ (in the continuous case, the integral has $rtf = \varepsilon > 0$ as lower bound). Such entropy, $S_c$, gets labeled as \emph{conditional} since it is computed under the condition that concept $c$ appears in the document. However, it is possible to define another probability distribution including the absence event which translates into another entropy, $S_f$, labeled as \emph{full}. To construct such distribution, we consider the total number of papers in the collection, $N_a$, while concept $c$ appears only in $N_c \leq N_a$. Then, we extend the $tf_c$ probability distribution by incorporating the absence event as a term that corresponds to the fraction of papers where the concept $c$ did not appear. Such term is exactly $1-df_c$, where $df_c = \tfrac{N_c}{N_a}$ is nothing else that the document frequency of concept $c$. In conclusion, the probability for the concept $c$ appearing $k \in [0, \infty]$ times is then $q_c(k) = \tfrac{N_c(k)}{N_a}$, where $q_c(0) = 1 - df_c$. Therefore, the full entropy associated to distribution $q_c(k)$ is:
\begin{align*}
 S_{f} & = - \sum_{k=0}^{\infty} q_c(k) \ln q_c(k) = \\
       & = - q_c(0) \ln q_c(0) - \sum_{k=1}^{\infty} q_c(k) \ln q_c(k) = - (1 - df_c) \ln(1 - df_c) - \sum_{k=1}^{\infty} \frac{N_c(k)}{N_a} \ln \left( \frac{N_c(k)}{N_a} \right) \,.
\end{align*}
Since $\tfrac{N_c(k)}{N_a} = \tfrac{N_c(k)}{N_c} \tfrac{N_c}{N_a} = \tfrac{N_c(k)}{N_c} \, df_c$, we have:
\begin{align}
\label{seq:full_entropy}
       \nonumber S_{f} &= - (1 - df_c) \ln(1 - df_c) - \sum_{k=1}^{\infty} \frac{N_c(k)}{N_c} \, df_c \ln \left( \frac{N_c(k)}{N_c} \, df_c \right) = \\
       & = - (1 - df_c) \ln(1 - df_c) - df_c \ln \left( df_c \right) \sum_{k=1}^{\infty} \frac{N_c(k)}{N_c} - df_c \sum_{k=1}^{\infty} \frac{N_c(k)}{N_c} \ln \left( \frac{N_c(k)}{N_c} \right) = \\
       \nonumber & = S_{bin} + df_c \; S_c \,. 
\end{align}
where we used the normalization condition over the $N_c(k)$, \ie $ \sum_{k=1} \frac{N_c(k)}{N_c} = 1$.  The full entropy $S_{f}$ is a linear combination of two entropies: the \emph{binary entropy}, $S_{bin}$, and the \emph{conditional entropy}, $S_c$, respectively. The first accounts for the probability of presence/absence of a concept in the collection. The second is the entropy computed in \eqref{eq:entropy} of the main text but modulated by the $df_c$.

\begin{figure*}[ht!]
\centering
\includegraphics[width=\columnwidth]{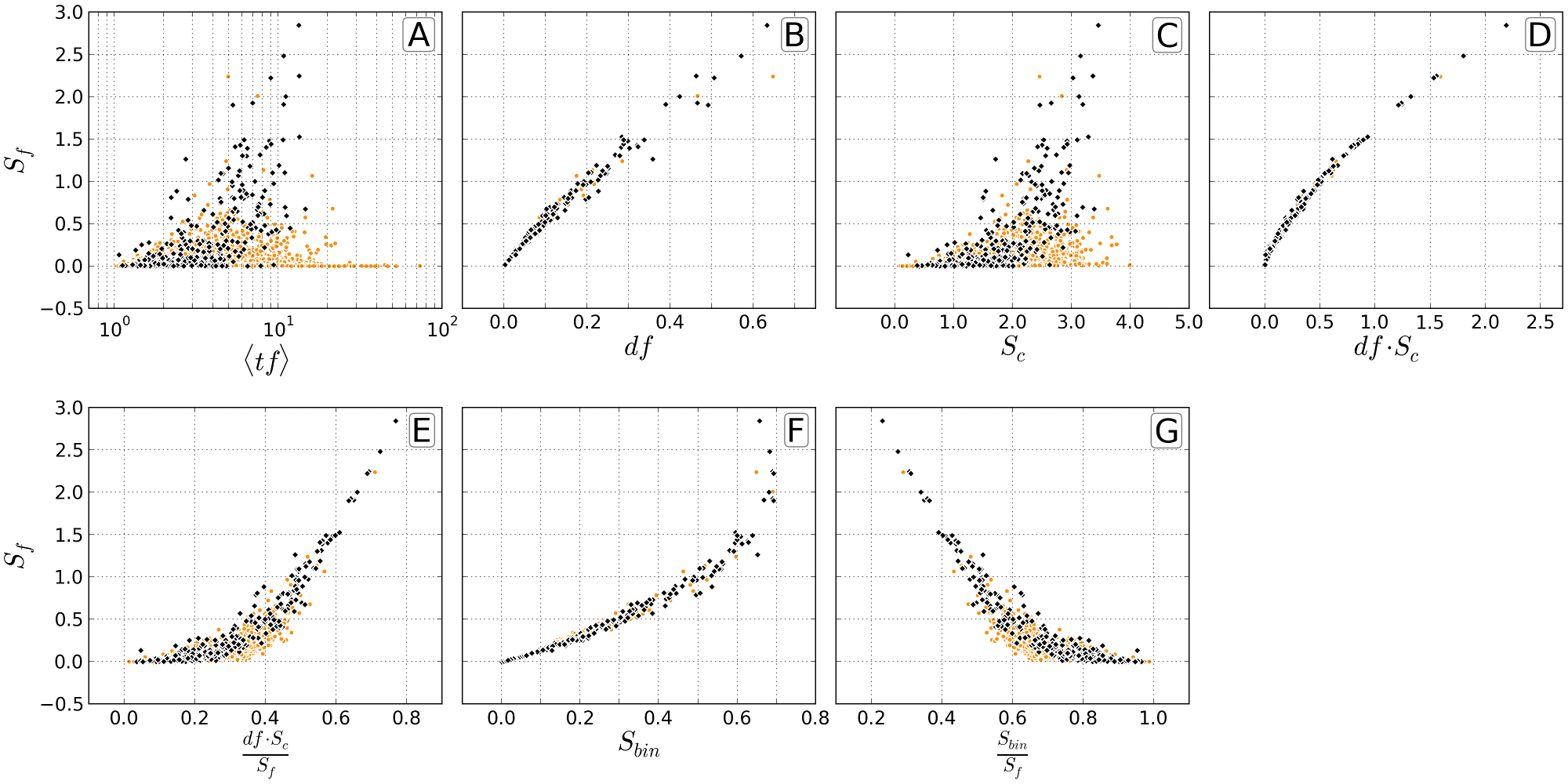}
\caption{Relation between the full entropy $S_f$ and other quantities. We consider: average term frequency $\avg{tf}$ (A), document frequency $df$ (B), conditional entropy $S_c$ (C), the contribution of conditional entropy to full one $df \cdot S_c$ (D), conditional entropy contribution over the full one $\tfrac{df \cdot S_c}{S_{f}}$ (E), binary entropy $S_{bin}$ (F) and binary entropy contribution over the full one $\frac{S_{bin}}{S_{f}}$ (G).} 
\label{sfig:full_entropy_vs_various_quantities}
\end{figure*}

It is natural to ask whether $S_f$ could be used to classify concepts as good as $S_c$, or not. To this aim, in Fig.~\ref{sfig:full_entropy_vs_various_quantities} we display the relation among $S_f$ and several quantities to assess if $S_f$ is a valid alternative to $S_c$ in discriminating generic concepts. 
More specifically, in panel (A) we display the relation between $\avg{tf}$ and $S_f$ -- in analogy with Fig.~\ref{fig:bidim_entropies_arxiv-conditioned_entropy_vs_avg_tf_and_df}(A). The comparison of the two figures strikingly highlights the ability of $S_c$ to grasp the tendency of generic concepts to display higher entropies for a given value of $\avg{tf}$. Following the parallelism with Fig.~\ref{fig:bidim_entropies_arxiv-conditioned_entropy_vs_avg_tf_and_df}, in panel (B) we inspect the relation between $S_f$ and $df$ confirming the inability of the former to discern generic concepts, alike to what displayed in Fig.~\ref{fig:bidim_entropies_arxiv-conditioned_entropy_vs_avg_tf_and_df}B. One may argue that the quantities used so far -- $\avg{tf}$ and $df$ -- are the most naive ones, and there might be others better suited to improve the performances of $S_f$. However, the adoption of quantities as: $S_c$ (panel C), $df \cdot S_c$ (panel D) and $\tfrac{df \cdot S_c}{S_{f}}$ (panel E) does not change situation: a clear separation between generic concepts and the others is missing. Finally, we perform the same analysis for the first term in \eqref{seq:full_entropy}, $S_{bin}$. The full entropy $S_f$ does not present a characteristic dependence for the generic concepts either on $S_{bin}$, panel (F), or on its fraction explained by $S_{bin}$, $\frac{S_{bin}}{S_{f}}$, panel (G).
In a nutshell, none of the relations displayed in Fig.~\ref{sfig:full_entropy_vs_various_quantities} seems to provide additional clues to design a classification criterion. Hence, the full entropy $S_f$ is unfit to distinguish generic concepts, since its discriminative power is weaker than the conditional one.

\newpage 
 
\subsection{Maximum entropy models}
\label{s_ssec:maxent_models}

The maximum entropy principle provides a framework to compare the amount of information carried by concepts, as encoded by the conditional entropy $S_c$. However, the raw value of $S_c$, which quantifies the actual information present in the data, is not enough to establish if a concept is generic or not. Indeed, we need to fairly assess if the observed entropy $S_c$ is small or not when confronted to an expected value. Such theoretical counterpart of the observed entropy is the maximum entropy and it is associated to a theoretical distribution where some features are fixed. The required features are quantities extracted from the data and the maximum entropy distribution is then the maximally random distribution that reproduce such features. In such a way, the established features uniquely determine the maximum entropy distribution. Operationally, in order to impose the constraints derived from the fixed features, we adopt the Lagrangian multipliers formalism, a tool that allow to easily establish the maximum entropy distribution that fulfills the constraints. In the rest of the section, we describe two different maximum entropy models with the respective features and we detail the calculations that lead to the associated distributions.

\subsubsection{Discrete \textsc{TF}}
\label{s_sssec:maxent_discrete_tf}

The first maximum entropy model is devised to characterize the observed term-frequency distribution of a concept, $tf_c$. The probability that concept $c$ appears $k$ times in the collection is simply the fraction of papers where it is present $k$ times, \ie $q_c(k) = \frac{N_c(k)}{N_c}$. Since the most typical feature of a distribution is its mean value, it is reasonable to consider the average term-frequency $\avg{tf_c}$ as a constraint of the maximum entropy distribution. Furthermore, in the literature there have been many evidences that the term-frequency distribution spans several orders of magnitude and has a fat tail profile (see references in the main text). However, to properly describe the observed behavior, we have to include another constraint which is the average logarithm of the term-frequency, $\avg{\ln(tf_c)}$. We denote the expected probability of concept $c$ occurring $k$ times as $p_c(k)$. The analytical form of $p_c$ is then calculated by maximizing its entropy $S_{max}$ under the constraints on the average term-frequency and the average logarithm of the term-frequency, which must be equal to $\avg{tf_c}$ and $\avg{\ln(tf_c)}$ respectively:

\begin{align}
\label{seq:maxent_tf_start}
  \tilde{S} & = - \sum_{k=1}^{\infty}{ p_{c}(k) \ln p_{c}(k) } + \lambda \left( \avg{tf_c} - \sum_{k=1}^{\infty}{ k \, p_{c}(k) } \right) + s \left( \avg{\ln(tf_c)} - \sum_{k=1}^{\infty}{ \ln(k) \, p_{c}(k) } \right) + \nu \left( 1 - \sum_{k=1}^{\infty}{ p_{c}(k) } \right) \, .
\end{align}
In the above equation, $\lambda$ is the Lagrange multiplier associated to the constraint $\avg{tf_c}$, $s$ is the one associated to $\avg{\ln(tf_c)}$ and $\nu$ is associated to the normalization condition of the probability mass function $p_c(k)$. The maximization of \eqref{seq:maxent_tf_start} with respect to $p_c(k)$ is performed as $\frac{\partial \tilde{S}}{\partial p_{c}(k)} = 0$, which gives:

\begin{equation}
\label{seq:maxent_tf_deriv}
  - \ln p_{c}(k) - 1 - \lambda k - s \ln(k) - \nu = 0 \, .
\end{equation}
Thus, the probability mass function $p_c$ is defined as:
\begin{equation}
\label{seq:maxent_tf_pmf_unnorm}
p_c(k) = \frac{\text{e}^{-(\nu + 1)} \text{e}^{-\lambda k}}{k^{s}} \, .
\end{equation}
This probability mass function corresponds to a power law with a cutoff. The power law $k^{-s}$ is responsible for the fat tail of the distribution, while the cutoff $\text{e}^{-\lambda k}$ is likely due to the finite size of the collection of articles under scrutiny. The maximization of \eqref{seq:maxent_tf_start} with respect to each Lagrangian multiplier allows to impose the respective constraint. In turn, such constraints determine the parameters that appear in \eqref{seq:maxent_tf_pmf_unnorm}. Thus, maximizing \eqref{seq:maxent_tf_start} with respect to $\nu$, $\frac{\partial \tilde{S}}{\partial \nu} = 0$, we recover the normalization condition:
\begin{align}
\label{seq:maxent_tf_normalization}
\nonumber \sum_{k=1}^{\infty}{ p_{c}(k) } & = \text{e}^{-(\nu + 1)} \sum_{k=1}^{\infty} { \frac{ \text{e}^{-\lambda k} }{ k^{s} } } = 1 \, , \\
\text{e}^{(\nu + 1)} & = \sum_{k=1}^{\infty} { \frac{ \text{e}^{-\lambda k} }{ k^{s} } }  = \text{Li}_{s}(\text{e}^{-\lambda}) \, . 
\end{align}
In the last equation, the summation is equal to the special function called polylogarithm of order $s$ and argument $\text{e}^{-\lambda}$. For any value of $s, \text{e}^{-\lambda} \in \mathbb{C}$, the validity of such expression is limited to the case when the modulus of the argument is smaller than one,  $\lvert \text{e}^{-\lambda} \rvert < 1$. However, in the present case we are interested only on real valued parameters. \eqref{seq:maxent_tf_normalization} allows to properly normalize the probability mass function in \eqref{seq:maxent_tf_pmf_unnorm} so that we obtain:
\begin{equation}
\label{seq:maxent_tf_pmf}
p_c(k;s,\lambda) = \frac{ \frac{\text{e}^{-\lambda k}}{ k^{s} } } {\text{Li}_{s}(\text{e}^{-\lambda})} \, .
\end{equation}
The maximization of \eqref{seq:maxent_tf_start} with respect to $\lambda$, $\frac{\partial \tilde{S}}{\partial \lambda} = 0$, allows to express the constraint $\avg{tf_c}$:
\begin{align}
\label{seq:maxent_tf_avg_tf}
\nonumber \sum_{k=1}^{\infty}{ k \, p_{c}(k) } = \frac{ \sum_{k=1}^{\infty} { \frac{ k \, \text{e}^{-\lambda k} } { k^{s} } } } { \text{Li}_{s}(\text{e}^{-\lambda}) } & = \frac{ \sum_{k=1}^{\infty} { \frac{ \text{e}^{-\lambda k} } { k^{s-1} } } } { \text{Li}_{s}(\text{e}^{-\lambda}) } = \avg{tf_c} \, , \\
\frac{ \text{Li}_{s-1}(\text{e}^{-\lambda}) } { \text{Li}_{s}(\text{e}^{-\lambda}) } & = \avg{tf_c} \, . \\
\end{align}
Note that in the last equation we used the definition of the polylogarithm and the normalization constant obtained in \eqref{seq:maxent_tf_normalization}.
Finally, the constraint on $\avg{\ln(tf_c)}$ is imposed by maximizing \eqref{seq:maxent_tf_start} with respect to $s$, $\frac{\partial \tilde{S}}{\partial s} = 0$:
\begin{align}
\label{seq:maxent_tf_avg_ln_tf}
\nonumber \sum_{k=1}^{\infty}{ \ln(k) \, p_{c}(k) } & = \frac{ \sum_{k=1}^{\infty} { \frac{ \ln(k) \, \text{e}^{-\lambda k} } { k^{s} } } } { \text{Li}_{s}(\text{e}^{-\lambda}) } = \avg{\ln(tf_c)} \, , \\
- \frac{ \partial_s{\text{Li}_{s}(\text{e}^{-\lambda})} } { \text{Li}_{s}(\text{e}^{-\lambda}) } & = \avg{\ln(tf_c)}
\, . \\
\end{align}
To derive the expression in the last line we used the identity:
\begin{equation*}
\sum_{k=1}^{\infty} { \frac{ \ln(k) \, \text{e}^{-\lambda k} } { k^{s} } } = - \frac{ \partial } { \partial s }  \sum_{k=1}^{\infty} { \frac{ \text{e}^{-\lambda k} } { k^{s} } } 
=  - \frac{ \partial } { \partial s } \text{Li}_{s}(\text{e}^{-\lambda}) \, .
\end{equation*}

From Eqs.~\textbf{\ref{seq:maxent_tf_avg_tf}} and \textbf{\ref{seq:maxent_tf_avg_ln_tf}} we see that both parameters $\lambda$ and $s$ are present in each of them. Since they are coupled in both equations they cannot be retrieved in an explicit form but we have to solve Eqs.~\textbf{\ref{seq:maxent_tf_avg_tf}} and \textbf{\ref{seq:maxent_tf_avg_ln_tf}} simultaneously through a numerical method. The details of the algorithmic implementation of the system, along with some snippets of code, are provided in Sec.~\ref{s_sssec:filter_implementation_tf}. The maximum entropy $S_{max}$ associated to the probability in \eqref{seq:maxent_tf_pmf} is then:

\begin{align}
\label{seq:maxent_tf}
\nonumber S_{max} & = - \sum_{k=1}^{\infty}{ p_{c}(k) \ln p_{c}(k) } \\
& = \ln{ \left[ \text{Li}_{s}(\text{e}^{-\lambda}) \right] } + \lambda \avg{tf_c} + s  \avg{ \ln { \left( {tf_c} \right) } } \, .
\end{align}

\subsubsection{Density of \textsc{TF}}
\label{s_sssec:maxent_rescaled_tf}

The second maximum entropy model is conceived to represent the rescaled version of the term-frequency distribution of a concept $c$, denoted as $rtf_c$. In particular, the density of the term-frequency accounts for the length $L(\alpha)$, in terms of words, of document $\alpha$ where concept $c$ is present, $rtf_c(\alpha) = \tfrac{tf_c(\alpha)}{L(\alpha)}$. The term-frequency density is better suited to describe the relevance of a concept within documents when their length is inhomogeneous. In the opposite case documents exhibit a typical length scale and the usage of raw term-frequency $tf_{c}$ is more appropriate since it does not alter the observed frequency. 

Being the term-frequency density a continuous variable we have to adopt a probability density function $p_c(x)$ to define its maximum entropy distribution. The two constraints that we set are the average and variance of the logarithm of the term-frequency density, $\avg{\ln(rtf_c)}$ and $\sigma^{2}({\ln(rtf_c)})$. We take the logarithm of the term-frequency density since it is more appropriated to describe a broad distribution of values: the average of the logarithm identifies the most likely value of the distribution while the variance characterize its variability scale. The analytical expression of the probability density function $p_c(x)$ is determined by maximizing its entropy $S_{max}$ under the constraints on the average and variance of the logarithm of the term-frequency density that must equate $\avg{\ln(rtf_c)}$ and $\sigma^{2}({\ln(rtf_c)})$ respectively:

\begin{align}
\label{seq:maxent_rtf_start}
  \nonumber \tilde{S} & = - \int_{0}^{\infty}{ p_{c}(x) \ln p_{c}(x) \; dx } \\ 
  \nonumber & + \gamma \left( \avg{\ln(rtf_c)} - \int_{0}^{\infty}{ \ln(x) \, p_{c}(x) \; dx } \right) \\
  \nonumber & + \eta \left[ \sigma^{2}({\ln(rtf_c)}) - \int_{0}^{\infty}{ \left( \ln(x) - \int_{0}^{\infty}{ \ln(x) \, p_{c}(x) \; dx } \right)^{2} \, p_{c}(x) \; dx } \right] \\
  & + \nu \left( 1 - \int_{0}^{\infty} { p_{c}(x) \; dx } \right) \, .
\end{align}
In \eqref{seq:maxent_rtf_start} we introduced the Lagrange multipliers $\gamma$, $\eta$ and $\nu$ that are correspondingly associated to the constraints $\avg{\ln(rtf_c)}$, $\sigma^{2}({\ln(rtf_c)})$ and the normalization condition of the probability density function $p_c(x)$. From the maximization of \eqref{seq:maxent_rtf_start} with respect to $p_c(x)$, $\frac{\partial \tilde{S}}{\partial p_{c}(x)} = 0$, we obtain:
\begin{equation}
\label{seq:maxent_rtf_deriv}
  - \ln p_{c}(x) - 1 - \gamma \ln(x) - \eta \left( \ln(x) - \mu \right)^{2} - \nu = 0 \, ,
\end{equation}
where we defined the constant $\mu = \int_{0}^{\infty}{ \ln(x) \, p_{c}(x) \; dx}$, which is the average value of the logarithm of $x$ according to the maximum entropy distribution $p_c(x)$.
As a consequence, the probability density function $p_c$ is defined as:
\begin{equation}
\label{seq:maxent_rtf_pdf_unnorm}
p_c(x) = \frac{\text{e}^{-(\nu + 1)} \text{e}^{-\eta \left( \ln(x) - \mu \right)^{2}}}{x^{\gamma}} \, .
\end{equation}
As we did in the previous case, Sec.~\ref{s_sssec:maxent_discrete_tf}, we must impose the normalization condition on the probability density function \eqref{seq:maxent_rtf_pdf_unnorm}, \ie the analogous of \eqref{seq:maxent_tf_normalization}, and we must calculate the parameters $\eta$ and $\gamma$, similarly to what we performed in Eqs.~\textbf{\ref{seq:maxent_tf_avg_tf}} and \textbf{\ref{seq:maxent_tf_avg_ln_tf}}. Since we have already detailed the process to calculate the parameters in Sec.~\ref{s_sssec:maxent_discrete_tf}, we do not report here the intermediate steps but directly the full expression of the probability density function:
\begin{equation}
\label{seq:maxent_rtf_pdf}
p_c(x;\mu,\sigma) = \frac{1}{ \sqrt{2 \pi} \, \sigma \, x} \exp \left[-\frac {(\ln x - \mu)^{2}} {2 \, \sigma^{2}} \right] \qquad \ \text{with} \; x>0 \,.
\end{equation}

Such probability density function corresponds to the lognormal function describing the distribution of a positive random variable $x$ whose logarithm $\ln(x)$ follows a normal distribution. Thanks to the constraints imposed on the average and the variance of the logarithm of $x$, the parameters $\mu$ and $\sigma^{2}$ that appear in \eqref{seq:maxent_rtf_pdf} can be directly calculated from the observed data:
\begin{equation}
\label{seq:maxent_rtf_parameters}
\mu = \int_{0}^{\infty}{ \ln(x) \, p_{c}(x) \; dx } \equiv \avg{\ln(rtf_c)} \, , \quad 
\sigma^{2} = \int_{0}^{\infty}{ \Bigl( \ln(x) - \mu \Bigr)^{2} p_{c}(x) \; dx } \equiv \sigma^{2}({\ln(rtf_c)}) \, .
\end{equation}
Note that $\mu$ is a constant determined from the actual data and is not a function of $\sigma^{2}$. In contrast, in Eqs.~\textbf{\ref{seq:maxent_tf_avg_tf}} and \textbf{\ref{seq:maxent_tf_avg_ln_tf}} the parameters $\lambda$ and $s$ were coupled. The maximum entropy $S_{max}$ associated to the probability in \eqref{seq:maxent_rtf_pdf} is then:
\begin{align}
\label{seq:maxent_rtf}
\nonumber S_{max} & = - \int_{0}^{\infty}{ p_{c}(x) \ln p_{c}(x) \; dx } \\
& = \ln{ \left( \sqrt{2 \pi} \, \sigma \right) } + \mu + \frac{1}{2} \, .
\end{align}

\subsection{Equivalence between Kullback-Leibler divergence and entropy difference, $S_d$}
\label{s_ssec:equiv-kl-diff_ent}

In main text, for each concept $c$, we defined its \emph{residual entropy}, $S_d(c)$, as the difference between the maximum entropy $S_{max}(c)$ and the conditional entropy $S_c(c)$ (unless explicitly indicated, we avoid to specify concept $c$ in the notation). Here we show that $S_d$, used to characterize the generality of the concepts, is exactly equivalent to the Kullback-Leibler divergence (KL), a widely used measure used to compute the difference between two probability distributions \cite{kullback-annmathstats-1951}. More specifically, we consider the KL between the maximum entropy probability distribution $p$ (see \eqref{eq:pmf}) and the empirical observed one $q$. For the sake of simplicity, we demonstrate here the case where $p$ and $q$ are probability mass functions describing discrete random variables, although the same reasoning can be used in the case of probability density functions. In particular, we recall that $q(k)$ denotes the observed probability that the $tf$ of a given concept is equal to $k$, which corresponds to $\tfrac{N(k)}{N_a}$, where $N(k)$ is the number of papers where the concept appears $k$ times and $N_a$ is the total number of papers. The Kullback-Leibler divergence from $p$ to $q$ is defined as:
\begin{equation}
\label{seq:kl_divergence}
\pcondd{D_{\mathrm{KL}}}{q}{p} = \sum_{k=1}^{\infty}{q(k) \ln \left( \frac{q(k)}{p(k)} \right) } = - \sum_{k=1}^{\infty}{ q(k) \ln p(k) } + \sum_{k=1}^{\infty}{ q(k) \ln q(k) }  \, .
\end{equation}
The last term in the \eqref{seq:kl_divergence} is nothing else, apart for the sign, than the conditional entropy $S_c$:
\begin{equation}
\label{seq:conditional_entropy}
S_c = - \sum_{k=1}^{\infty}{ q(k) \ln q(k) }  \, .
\end{equation}
The first term, instead, can be rewritten using the maximum entropy probability $p$ (see \eqref{eq:pmf}) as:
\begin{align}
\label{seq:cross_entropy}
\nonumber - \sum_{k=1}^{\infty}{ q(k) \ln p(k) } & = - \sum_{k=1}^{\infty}{ q(k) \ln \left( \cfrac{\cfrac{\text{e}^{-\lambda k}}{k^{s}}} {\text{Li}_{s}(\text{e}^{-\lambda})} \right) } =
\nonumber \sum_{k=1}^{\infty}{ q(k) \ln \left[ \text{Li}_{s}(\text{e}^{-\lambda}) \right] } - \sum_{k=1}^{\infty}{ q(k) \ln \left( \dfrac{\text{e}^{-\lambda k}}{k^s} \right) } \\
\nonumber & = \ln \left[ \text{Li}_{s}(\text{e}^{-\lambda}) \right] - \sum_{k=1}^{\infty}{ q(k) \ln \left( \dfrac{\text{e}^{-\lambda k}}{k^s} \right) } =
\nonumber \ln \left[ \text{Li}_{s}(\text{e}^{-\lambda}) \right] + \lambda \sum_{k=1}^{\infty}{ q(k) k } + s \sum_{k=1}^{\infty}{ q(k) \ln k } \\
& = \ln \left[ \text{Li}_{s}(\text{e}^{-\lambda}) \right] + \lambda \avg{k} + s \avg{\ln k}
\equiv S_{max}  \, .
\end{align}
Plugging the results of Eqs.~\textbf{\ref{seq:conditional_entropy}} and \textbf{\ref{seq:cross_entropy}} into \eqref{seq:kl_divergence} we get:
\begin{equation}
\label{seq:kl_divergence_final}
\pcondd{D_{\mathrm{KL}}}{q}{p} = - \sum_{k=1}^{\infty}{ q(k) \ln p(k) } + \sum_{k=1}^{\infty}{ q(k) \ln q(k) } = S_{max} - S_{c} \, .
\end{equation}
Hence, for a given concept $c$, the KL divergence of between $p$ and $q$ coincides with the residual entropy $S_{d} = S_{max} - S_{c}$.

\newpage
\subsection{Comparisons between sets}
\label{s_ssec:comparison_sets}

\subsubsection{Overalap between concepts' sets based upon residual entropy or $IDF$}
\label{s_ssec:comparison_sc_idf}

Despite being both defined on the collection under scrutiny, the $IDF$ and the residual entropy $S_d$ encode different information. One penalizes concepts present in many papers; the other is more intrinsically related to the distribution of the frequency of a concept. As a consequence, despite being different, we expect to observe some correlation between them. To compare the list of concepts obtained according to their residual entropy, $S_d$, or their inverse document frequency, $IDF$, we compute the overlap, $O_{n,m}$, between the set of concepts falling in the $n$-th slice of the percentile of $IDF$, ${\cal{A}}_n$, and the set of concepts falling in the $m$-th slice of the percentile of $S_d$, ${\cal{B}}_m$. By $k$-th \emph{slice} of the percentile $\tilde{P}$ of a probability distribution $p(x)$, we refer to the range of values of $x$ such that $x \in \bigl[ \tilde{P}_{k-10}, \tilde{P}_k \bigr], \; k \in \{ 10, 20, \ldots, 90 \}$. Thus, we have:
\begin{equation}
\label{seq:overlap}
O_{n,m} = \dfrac{\lvert {\cal{A}}_n \cap {\cal{B}}_m \rvert}{\lvert {\cal{A}}_n \rvert} \,,
\end{equation}
%
with $O_{n,m} \in [0,1]$. As usual, $O_{n,m} = 1$ denotes complete overlap between the two sets, while $O_{n,m} = 0$ denotes that the sets have no elements in common.

\subsubsection{Comparison between communities}
\label{s_ssec:comparison_communities}

One of the standard measures used to compute the overlap between two communities of a networked system is the \emph{Jaccard score} $J$ \cite{hric-pre-2014}. Given a graph $G$ with $N$ nodes, several methods can be used to retrieve its organization into modules \cite{fortunato-physrep-2016}. The Jaccard score between a pair of sets of nodes (\ie modules) $\cal{A}$ and $\cal{B}$, $J_{\cal{A},\cal{B}}$ is given by:
\begin{equation}
\label{seq:jaccard}
J_{\cal{A},\cal{B}} = \dfrac{\lvert {\cal{A}} \cap {\cal{B}} \rvert}{\lvert {\cal{A}} \cup {\cal{B}} \rvert} \quad J_{\cal{A},\cal{B}} \in [0,1] \,.
\end{equation}
A value of one denotes complete overlap between the two sets (\ie the sets are the same), while a value of zero denotes that the sets completely different.

\subsubsection{Correlation between sets of ranks -- the Kendall coefficient}
\label{s_ssec:comparison_rankings_kendall}

To delve into the similarity of the sets of concepts used to label topics, we consider the Kendall coefficient, $\tau_b$,  \cite{kendall-book-1948}. Given a set $\mathcal{A}$ with $N$ elements, $\mathcal{A} = \left\lbrace a_1, a_2, \ldots, a_N \right\rbrace$, we consider two ordered sequences, $\mathcal{X} = \left\lbrace x_1, x_2, \ldots, x_N \right\rbrace$ and $\mathcal{Y} = \left\lbrace y_1, y_2, \ldots, y_N \right\rbrace$, of its elements (\ie $x_i,y_j \in \mathcal{A} \; \forall i,j=1,\ldots,N$.) We indicate with $(x_i, x_j)$ and $(y_i,y_j)$ the \emph{ordered pairs} of elements of those sequences. The Kendall coefficient, $\tau_b$, measures the correlation (similarity) among the rankings $\mathcal{X}$ and $\mathcal{Y}$ \cite{knight-jasa-2012}.
\begin{equation}
\label{seq:kendall_tau_b}
\tau_{b}(\mathcal{X},\mathcal{Y}) = \dfrac{R - S}{\sqrt{ \abr{R + S + X_{0}} \abr{R + S + Y_{0}} } }  \quad \tau_b \in [-1,1]\,.
\end{equation}
Where $R$ ($S$) is the number of \emph{concordant} (\emph{discordant}) pairs, \ie pairs $(x_i, x_j), (y_l, y_m) \; i,j,l,m = 1,\ldots,N$ such that $x_i \gtrless x_j$ and $y_l \gtrless y_m$ (equivalently, $x_i \gtrless x_j$ and $y_l \lessgtr y_m$ or viceversa.) $X_0$ is the number of \emph{ties} in $\mathcal{X}$, \ie those pairs $(x_i,x_j)$ for which $x_i = x_j$; $Y_0$, instead, is the analogous of $X_0$ but for set $\mathcal{Y}$. A value of $\tau_b = 1$ denotes maximal correlation meaning that the two rankings are exactly the same. On the contrary, $\tau_b = -1$ denotes complete anticorrelation (the rankings are inverted.) Finally, $\tau_b = 0$ indicates that there is no correlation at all among the two rankings. In our case, we will compute $\tau_b$ to measure the similarity among the ranked lists of concepts belonging to the topics/communities of the similarity networks, and those of concepts belonging to the topics obtained using the TopicMapping (TM) algorithm. More specifically, given a community $C$ of the similarity network, we select the all concepts appearing in the documents belonging to $C$, and we rank them in decreasing order of their \emph{local document frequency}, $ldf(w)$, \ie the fraction of documents of $C$ in which concept $w$ appears. In the case of TM, the ranking of concepts within a given topic $T$ is obtained ordering them in decreasing order of their probability to belong to that topic, $\pcond{\pi}{w}{T}$. Since the concepts appearing in a given community, $C$, and those appearing in one of the TM topics, $T$, do not generally coincide; to compute $\tau_b$ we consider only those concepts appearing in both $C$ and $T$.
%

%
%
%
%

\subsection{Networks of similarity between documents}
\label{s_ssec:similarity_network}

A naive attempt to infer the organization of documents into topics is to study the community structure of the network of similarities among those documents \cite{cong-phys_lif_rev-2014}. Documents are the nodes and the weights of links between pairs of documents account for their similarity. Amidst the plethora of documents similarity measures available (see \cite{boyack-pone-2011} and reference therein), we chose cosine similarity based on their concept vectors $\vec{d}$. Hence, for each document $\alpha$, we denote its set of concepts as $\mathcal{C}_{\alpha}$. The concepts vector $\vec{d}_\alpha$ is then composed by the {\small \textsc{TF-IDF}} of its concepts \cite{manning-book-2008}. Given the set of concepts used in a corpus with $N_a$ documents, $\mathcal{C}= \bigcup_{\alpha=1}^{N_a} \mathcal{C}_{\alpha}$, the relevance of a concept $c$ in a document $\alpha$ is given by
\begin{equation}
\label{seq:tfidf}
d_{\alpha}(c) =
\begin{cases}
tf_{c}(\alpha) \cdot IDF_c& \qquad \text{if } c \in \mathcal{C}_{\alpha} \,, \\
0& \qquad \text{otherwise,}
\end{cases}
\end{equation}
where $tf_{c}(\alpha)$ is the {\small \emph{boosted term frequency}}, \ie the number of times $c$ appears in $\alpha$ modulated according to the location (title, abstract, body) where it appears. The other factor, $IDF_c$, is the {\small \emph{Inverse Document Frequency}} and accounts for the frequency with which a concept appears in the corpus. In particular, $IDF$ penalizes concepts used frequently since:
\begin{equation}
\label{seq:idf}
 IDF_c = \ln \left( \dfrac{N_a}{N_c} \right) \,,
\end{equation}
where $N_c$ is the number of papers that contain concept $c$. Thus, the pairwise similarity between documents $\alpha$ and $\beta$ is given by:
\begin{equation}
\label{seq:tfidf-weight}
w_{\alpha \beta} = \dfrac{\vec{d}_{\alpha} \cdot \vec{d}_{\beta}}{ \norm{\vec{d}_{\alpha}} \, \norm{\vec{d}_{\beta}} } \,,
\end{equation}
where $\cdot$ denotes the scalar product and $\norm{\ldots}$ is the Euclidean norm. $w_{\alpha \beta}$ falls in the interval $[0,1]$, where $w=0$ indicates documents sharing any concept at all (\ie are completely different), and whose vectors form an angle $\theta = 90$\textdegree. A value $w=1$ is found if the documents not only have the same set of concepts, but use them in the same way (\ie are identical) thus having vectors forming an angle $\theta = 0$\textdegree. We remark that using the $tf$ or its density, $rtf$, to compute weights is equivalent. To ease the computation, we decided to prune all connections whose weight was below 0.01, corresponding to vectors having an angle $\theta = 89.43$\textdegree. However, the widespread presence of common concepts (CC) is responsible for the loss of one of the major advantages of framing the system as a network, \textit{i.e.} sparsity. Several solutions to the \emph{network sparsification} problem have been proposed \cite{serrano-pnas-2009,radicchi-pre-2011,gemmetto-arxiv-2017}. All of them ensure the conservation of the statistical properties of the original network acting in an \emph{ex-post} way. A more suitable approach, instead, would be to act \emph{ex-ante}, directly on the process responsible for the generation of the weights. This translates into acting directly on the concepts to select only the \emph{relevant} ones before computing the weights.
\indent Despite the removal of lightweight connections, the construction and analysis of the similarity network can become quite taxating computationally. For this reason, we do not perform the topological analysis of the similarity networks for all the datasets available. However, when available, the properties of the networks are reported in the first row ($p = 0 \%$) of Tabs.~\ref{stab:net_topol_prop_phys2013} and \ref{stab:net_topol_prop_climate}.

%
%
%
%
\newpage
\section{Datasets}
\label{s_sec:datasets}

\begin{figure*}[h!]
\centering
\includegraphics[width=\columnwidth]{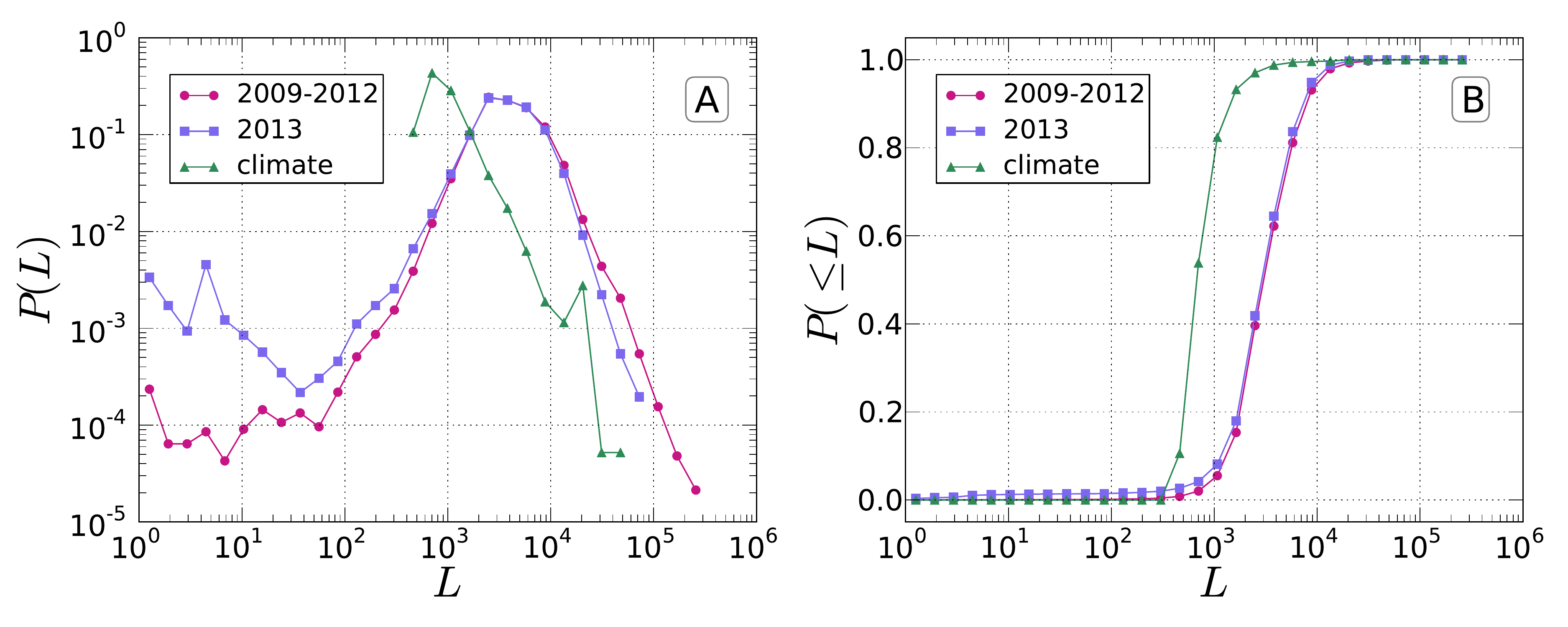}
\caption{(panel A) Probability distribution, $P(L)$, of the number of words per document, $L$, for all the collections considered. The Physics collections show a clear presence of a typical length of the documents around $L \simeq 3000$ words, while this is not the case for climate change ones. (panel B) Cumulative probability distribution functions, $P(\leq L)$, of the document length, $L$, for the same collections of panel A.}
\label{sfig:distro_num_words_both_data}
\end{figure*}

In this section we provide the details of the collections of documents used in our study, namely: \emph{Physics} (Sec.~\ref{s_ssec:physics}) and \emph{climate change} (Sec.~\ref{s_ssec:climate}), respectively. Here, we anticipate that the maximum entropy model based on term-frequency $tf_c$, described in Sec.~\ref{s_sssec:maxent_discrete_tf}, will be used for the \emph{Physics} collections, while the model based on term-frequency density $rtf_c$ will be adopted for the \emph{climate change} collection. We decided to work with two different maximum entropy models since the distribution of the document length exhibits different traits for the two collections, as shown in Fig.~\ref{sfig:distro_num_words_both_data}. More specifically, the document length for \emph{Physics} collections displays a typical value of $\simeq 3000$ words, while for the \emph{climate change} collection it does not present a characteristic scale and is quite inhomogeneous. In the following, for each collection of documents, we comment how the data have been collected and how the progressive pruning of concepts reveals their organization into topics.

%
%
%
%

\subsection{Physics corpus}
\label{s_ssec:physics}

\subsubsection{Source}
\label{s_sssec:phys_source}

\begin{figure*}[h!]
    \begin{minipage}[b]{0.30\columnwidth}
    \begin{tabular}{@{\vrule height 10.5pt depth4pt  width0pt}r|cd{2}|cd{2}|}
    \cline{2-5}
       & \multicolumn{2}{c|}{2009--2012} &  \multicolumn{2}{c|}{2013} \\ \hline
    Category & $N_a$ & \multicolumn{1}{c|}{$\%$} & $N_a$ & \multicolumn{1}{c|}{$\%$} \\ \hline
    astro-ph & 46922 & 24.73 & 12458 & 23.51 \\
    cond-mat & 45345 & 23.90 & 12679 & 23.93 \\
    hep & 37074 & 19.54 & 9661 & 18.24 \\
    physics & 21436 & 11.30 & 7407 & 13.98 \\
    quant-ph & 14018 & 7.39 & 4039 & 7.62 \\
    gr-qc & 8189 & 4.31 & 2273 & 4.29 \\
    nucl & 6955 & 3.66 & 1819 & 3.43 \\
    math-ph & 6776 & 3.57 & 1767 & 3.34 \\
    nlin & 3044 & 1.60 & 876 & 1.65 \\ \hline
    Total & 189759 & \multicolumn{1}{c|}{100} & 52979 & \multicolumn{1}{c|}{100} \\ \hline
    \end{tabular}
    \captionof{table}{Number of articles $N_a$ and percent size, $\%$, of each \texttt{arXiv} principal category in the Physics datasets.}
    \label{stab:primary_categories}
    \end{minipage}
    \hfill    
    \begin{minipage}[b]{0.69\columnwidth}
	\centering
        \includegraphics[width=0.49\textwidth]{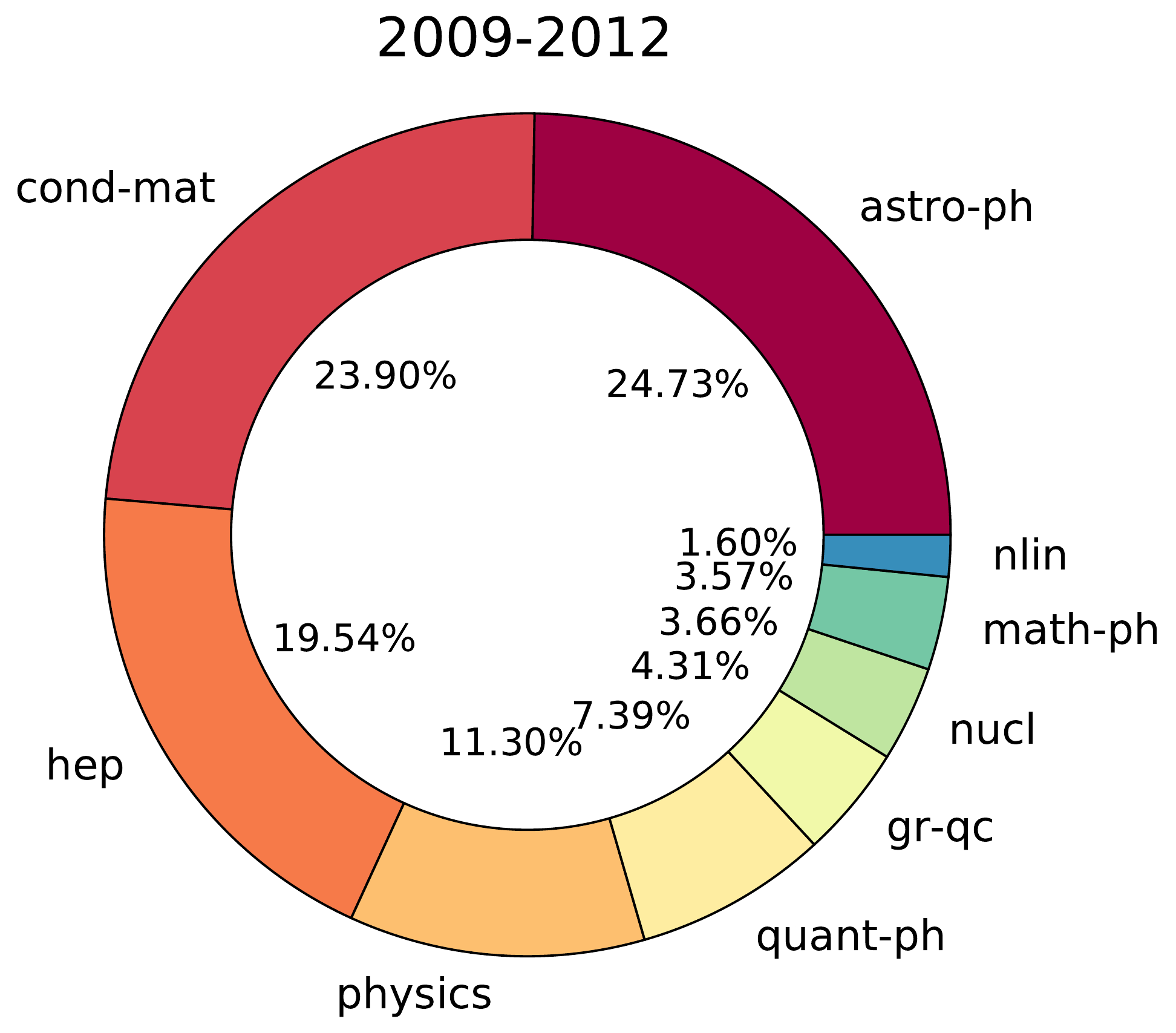}
	\includegraphics[width=0.49\textwidth]{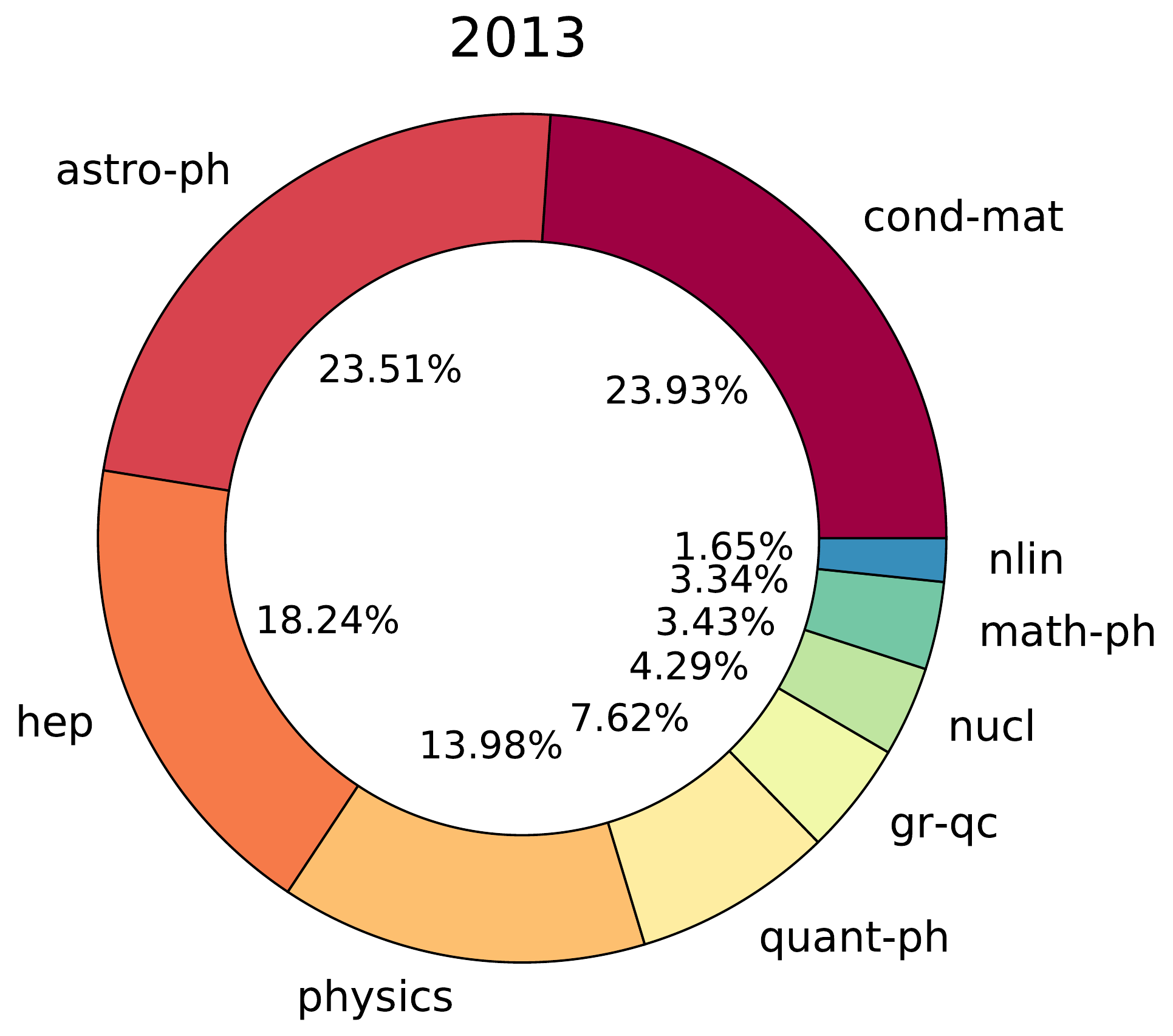}
	\captionof{figure}{Donut chart of the composition of the Physics datasets in terms of \texttt{arXiv} principal categories. (Left) 2009-2012, (Right) 2013.}
        \label{sfig:donut_categories_physics}
    \end{minipage}
\end{figure*}

The documents of the Physics datasets are scientific manuscripts submitted to the \texttt{arXiv}\cite{arxiv-web} archive, a repository of electronic preprints of scientific articles spanning several domains of science. We extracted two non-overlapping subsets of manuscripts submitted under the {\small \texttt{physics}} categories either as primary or secondary subjects. The first is made of $N_a = 189,758$ documents submitted between years 2009 and 2012. The second, instead, is made of $N_a = 52,979$ documents submitted during year 2013. In Tab.~\ref{stab:primary_categories} and Fig.~\ref{sfig:donut_categories_physics} we report the composition of the collections in terms of these main categories: physics (\textbf{physics}), condensed matter (\textbf{cond-mat}), astrophysics (\textbf{astro-ph}), quantum physics (\textbf{quant-ph}), mathematical physics (\textbf{math-ph}), nonlinear sciences (\textbf{nlin}), general relativity and quantum cosmology (\textbf{gr-qc}), nuclear physics (\textbf{nucl}), and high-energy physics (\textbf{hep}). The last two further divides into theory (\textbf{nucl-th}) and experiment (\textbf{nucl-ex}) for nuclear, and theory (\textbf{hep-th}), phenomenology (\textbf{hep-ph}), lattice (\textbf{hep-lat}), and experiment (\textbf{hep-ex}) for high-energy. The interested reader can check \cite{arxiv-web} for a more detailed description of each category. From the donut charts (Fig.~\ref{sfig:donut_categories_physics}) it is clear that the collections are highly inhomogeneous with cond-mat and astro-ph categories summing together almost half of the entire collection, while gr-qc, nucl, math-ph and nlin not even reaching the 14\% in both collections. 

%
%
%

\subsubsection{Two-dimensional tessellation}
\label{s_sssec:phys_tessellation}

\begin{figure}[ht!]
\centering
\includegraphics[width=\columnwidth]{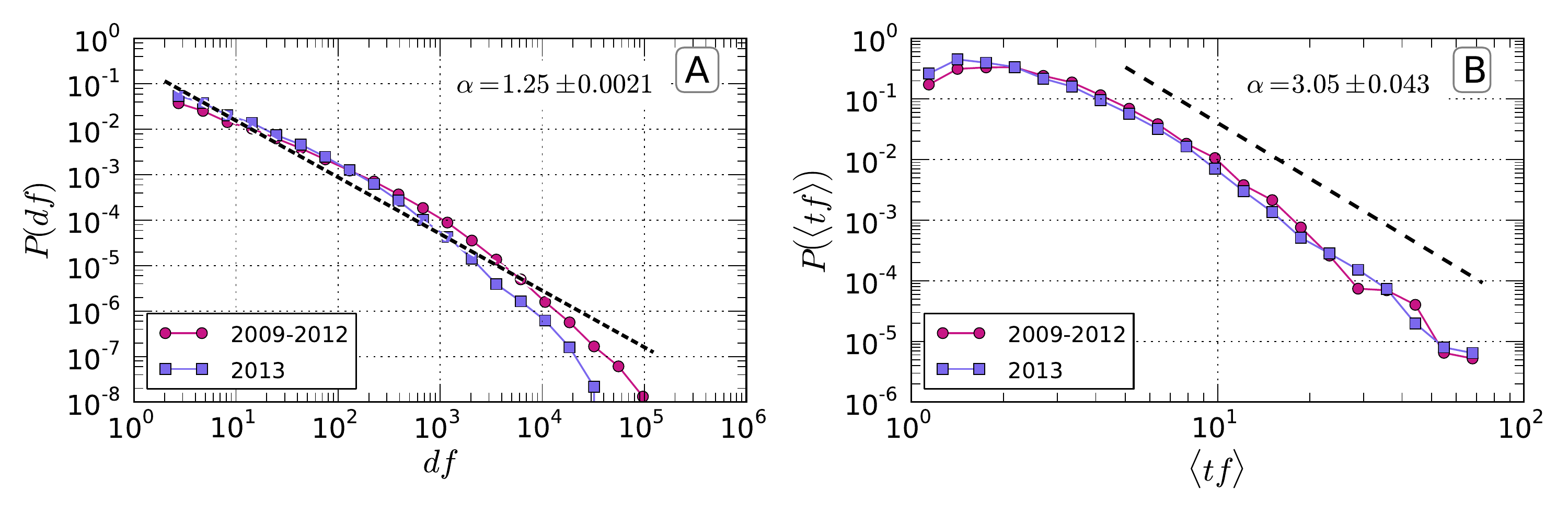}
\caption{Distribution of concept features for the Physic datasets of 2009-2012 (dots) and 2013 (squares). Panel A shows the distribution of the document frequency, $df$, while panel B refers to the distribution of the average term-frequency, $\avg{tf}$. In both panels, we reported -- as a dashed line -- the powerlaw fit of the distribution along with the values of its exponent $\alpha$ and its standard deviation. Powerlaw fits are displayed to highlight the trend of the distributions but are not meant to be the best fitting models. Nevertheless, both distributions have a broad shape, denoting high variability of their values.}
\label{sfig:distro_df_avg_tf_arxiv}
\end{figure}

\begin{figure}[h!]
\centering
\includegraphics[width=\columnwidth]{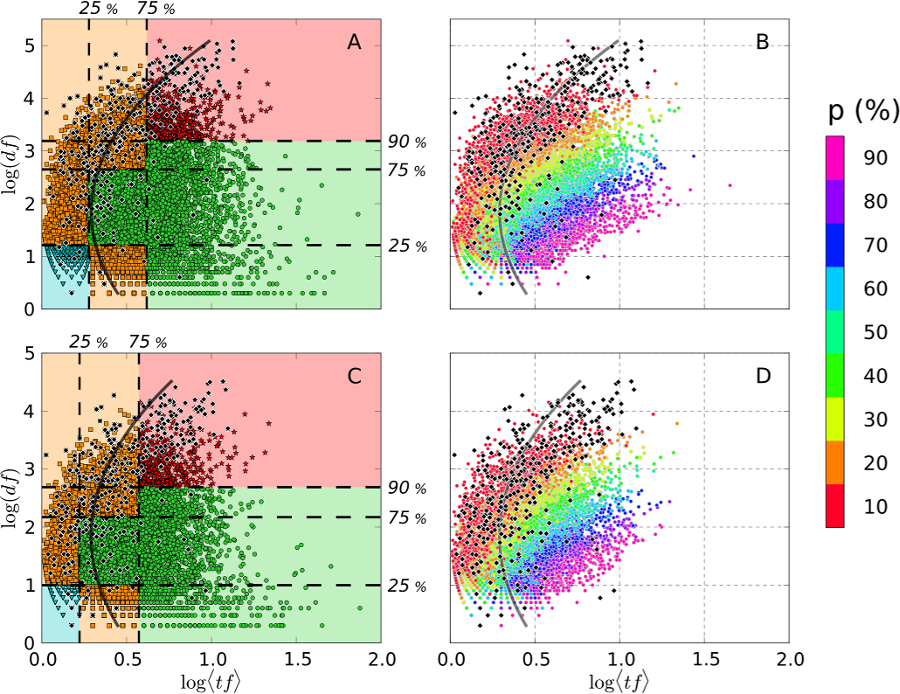}
\caption{Two-dimensional tessellation filtering for the Physics datasets of 2009--2012 (panels A and B), and 2013 (panels C and D). In panel A (C) the shape and color of the points account for the region they fall into (see Fig.~\ref{sfig:tessellation-base}). The classification/percentage of 2009--2012 (2013) concepts belonging to each class is the following. \textbf{Ubiquitous concepts} -- red stars -- are 4.4 \% (4.6 \%), \textbf{rare/specific concepts} -- cyan triangles -- are 10 \% (11 \%), \textbf{significant concepts} -- green circles -- are 46 \% (46 \%), while the \textbf{remaining concepts} -- orange squares -- are 39 \% (39 \%). Black diamonds represent concepts marked as generic in the ScienceWISE ontology, and the solid black line is a guideline of their average position. In panel B and D, instead, points are colored by the percentile $p$ of residual entropy they belong to, as reported in the color bar. Concepts in $p > 90\%$ are omitted. Panel B is also displayed in the bottom right corner of panel of Fig.~\ref{fig:bidim_entropies_arxiv-conditioned_entropy_vs_avg_tf_and_df}C.}
\label{sfig:bidim_tess_arxiv}
\end{figure}

Although the entropic filtering outperforms the two-dimensional tessellation one, we have decided to compute anyway the position of concepts on the $(\avg{tf}, df)$ plane to check if such representation highlights some interesting features. First of all, we remark that the distributions of such quantities are both heterogeneous, as shown in Fig.~\ref{sfig:distro_df_avg_tf_arxiv}. Thus, establishing a threshold on quantities that miss a characteristic scale is not properly justifiable. Then, in Fig.~\ref{sfig:bidim_tess_arxiv}, we report the position of concepts on the $(\avg{tf}, df)$ plane for the Physics datasets: panels A and B refer to 2009-2012, while panels C and D to 2013, instead. More specifically, in panels A and C, we colored concepts according to the class they belong to, as outlined in Sec.~\ref{s_ssec:bid_tess}. In both datasets, 46\% of the concepts (the green circles) could be considered \emph{significant} according to the tessellation scheme. On the other hand, in panels B and D we color each concept according to the value of its residual entropy percentile $p$. The colored dots show a pattern that grasps much better the location of CCs than the two-dimensional tessellation.
%
%
%
%
%
%
%

\begin{figure}[ht!]
\centering
\includegraphics[width=\columnwidth]{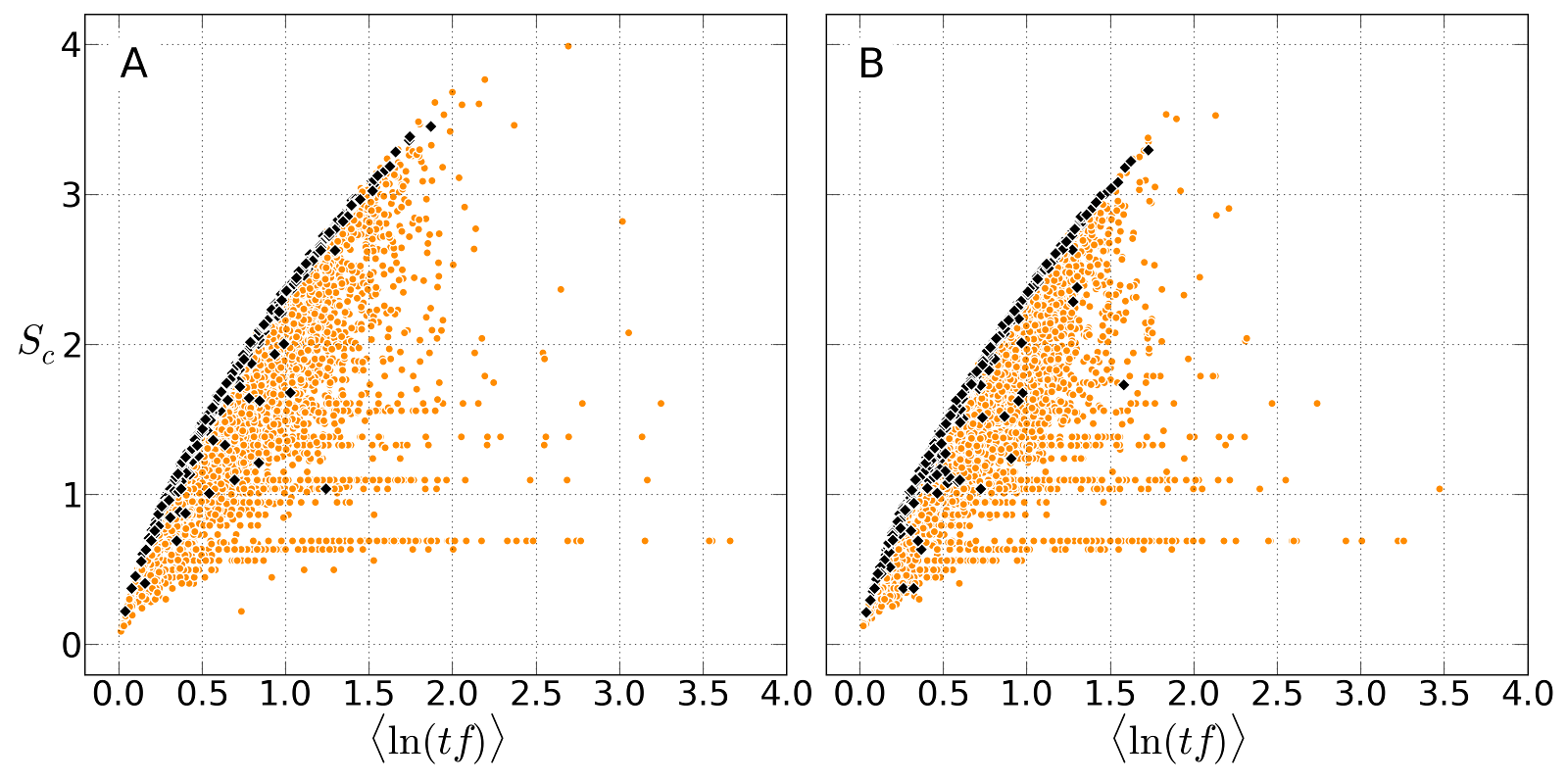}
\caption{Relation between the conditional entropy $S_{c}$ and the average logarithm of the term-frequency, $\avg{ \ln (tf) }$. Black diamonds represent SW common concepts (CC) which are scattered in a very tiny region at the boundary of the plot. Panel A refers to the 2009--2012 collection, while panel B to the 2013 one.}
\label{sfig:ent_cond_vs_log_avg_tf}
\end{figure}

\begin{figure*}[h!]
\centering
\includegraphics[width=\columnwidth]{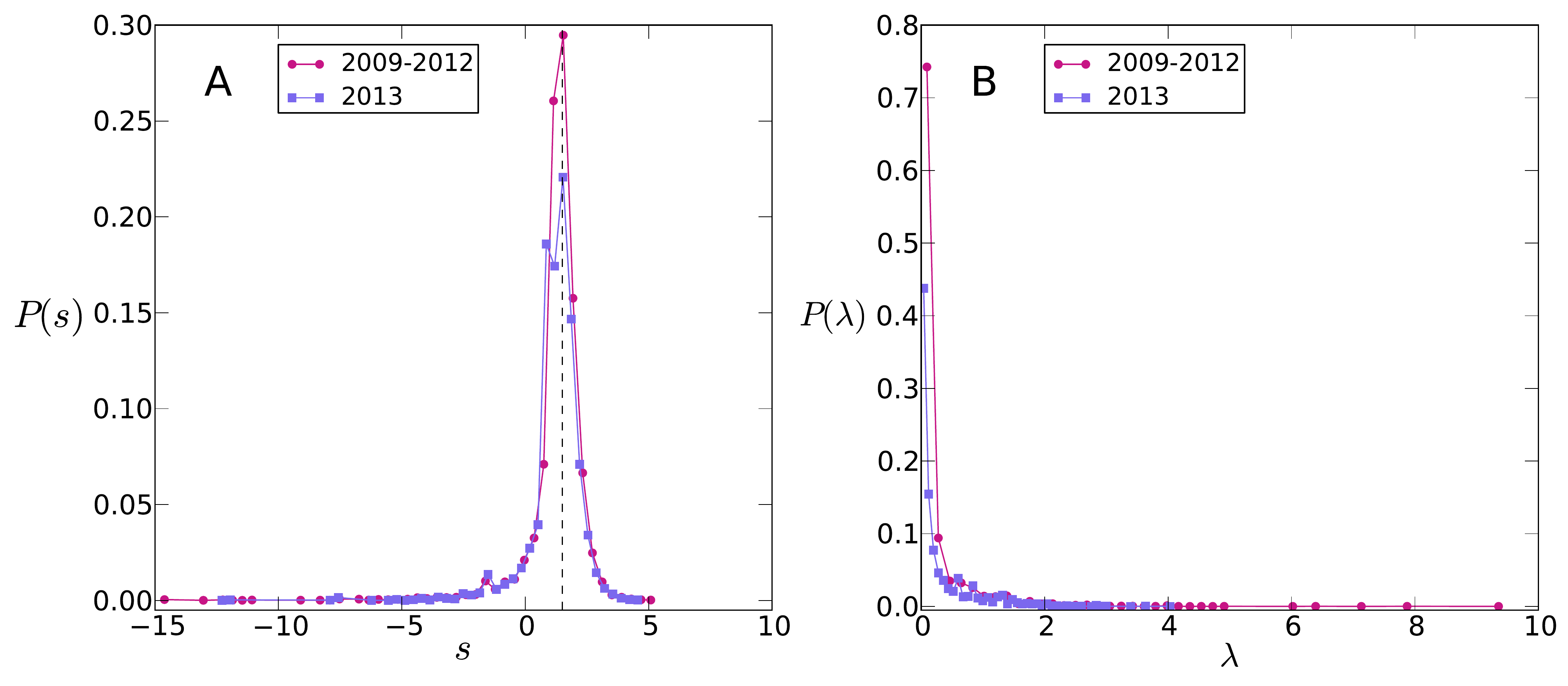}
\caption{Frequency distributions of the maximum entropy parameters $s$ (panel A), and $\lambda$ (panel B) for the discrete $tf$ model of the Physics collections. The distribution of $s$ exhibits a sharp peak around $s=3/2$ (vertical dashed line). The distribution of the exponential cutoff parameter, $\lambda$, presents a concentration around a value of zero.}
\label{sfig:histogram_lambda_phys}
\end{figure*}

The conditional entropy $S_{c}$ used in the entropic filtering is able to capture a peculiar trend of the common concepts (CC) not only when examined against the average term-frequency $\avg{tf}$ (see Fig.~\ref{fig:bidim_entropies_arxiv-conditioned_entropy_vs_avg_tf_and_df}, panel (A)), but also in the case of the average logarithm of the term-frequency, $\avg{ \ln (tf) }$, as displayed in Fig.~\ref{sfig:ent_cond_vs_log_avg_tf}. The computed parameters of the maximum entropy model associated to the discrete $tf$, \ie a powerlaw with cutoff, outline the trend followed by the maximum entropy distribution of the concepts. We recall from what we discussed in the main text that the exponent of the powerlaw part, $s$, is characteristic of the process that drive the observed distribution of the $tf$. The striking maximum shown by its distribution in Fig.~\ref{fig:histo_alpha_inset_fit_selected_concepts} is particularly significant: it is at $s = 3/2$, a value which strongly suggests the presence of a critical branching process \`a la Galton-Watson type behind distribution which gives rise to the observed $tf$ sequence. For the sake of completeness, we display the distribution of the exponential cutoff $\lambda$ in Fig.~\ref{sfig:histogram_lambda_phys}. Note that, in this case, the distribution is peaked around 0.
%
%
%
%
%
%
%
%
%

\pagebreak

\subsubsection{Differences between sets of concepts built using different rankings and across different collections}
\label{s_sssec:phys_diff_ent_idf_2collections}

\begin{figure*}[h!]
\centering
\includegraphics[width=\columnwidth]{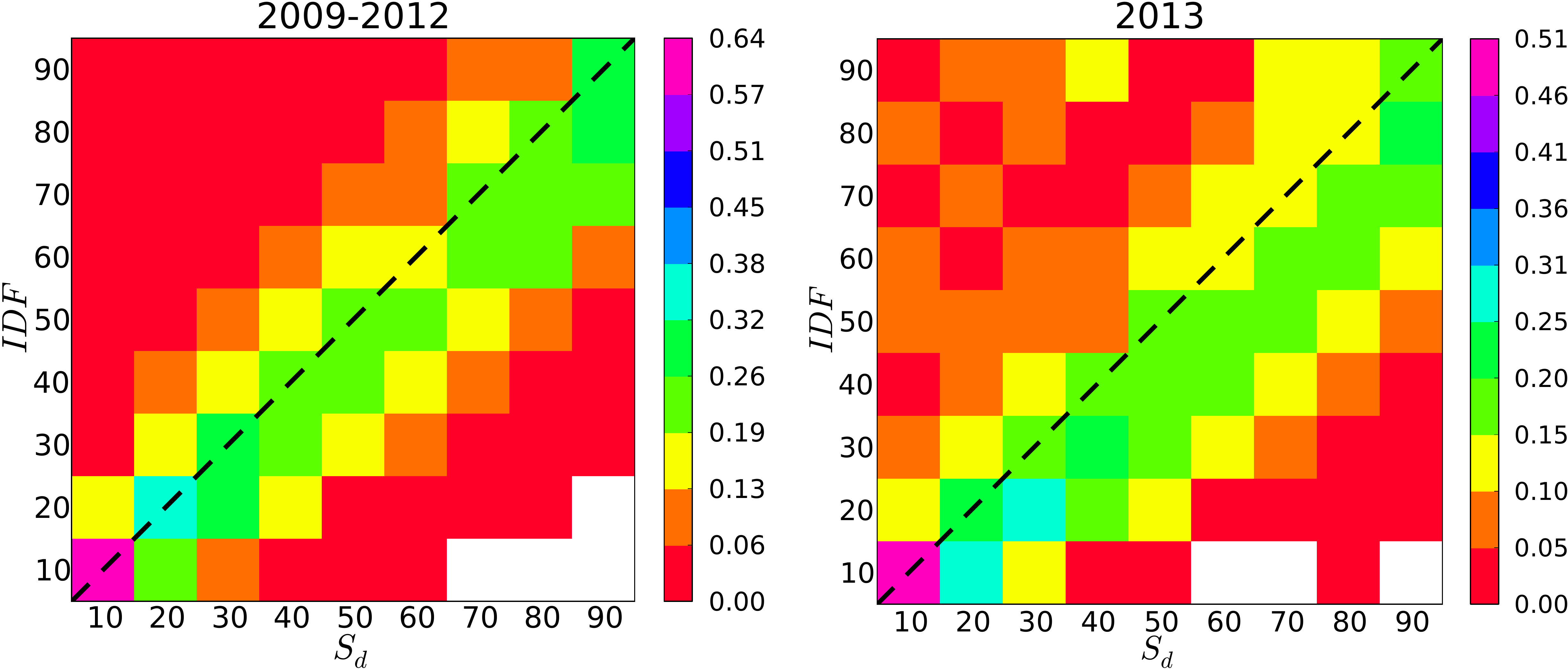}
\caption{Overlap between the lists of concepts ranked according to the residual entropy $S_d$ and $IDF$ for the Physics collection of 2009--2012 (left panel) and 2013 (right panel). The color of the cells denotes the amount of overlap $O$. Matrices are normalized by row, and white entries correspond to absence of overlap. The dashed line indicates the first diagonal.}
\label{sfig:inters_percentiles_idf_vs_maxent_tf}
\end{figure*}

\begin{figure*}[h!]
\centering
\includegraphics[width=0.5\columnwidth]{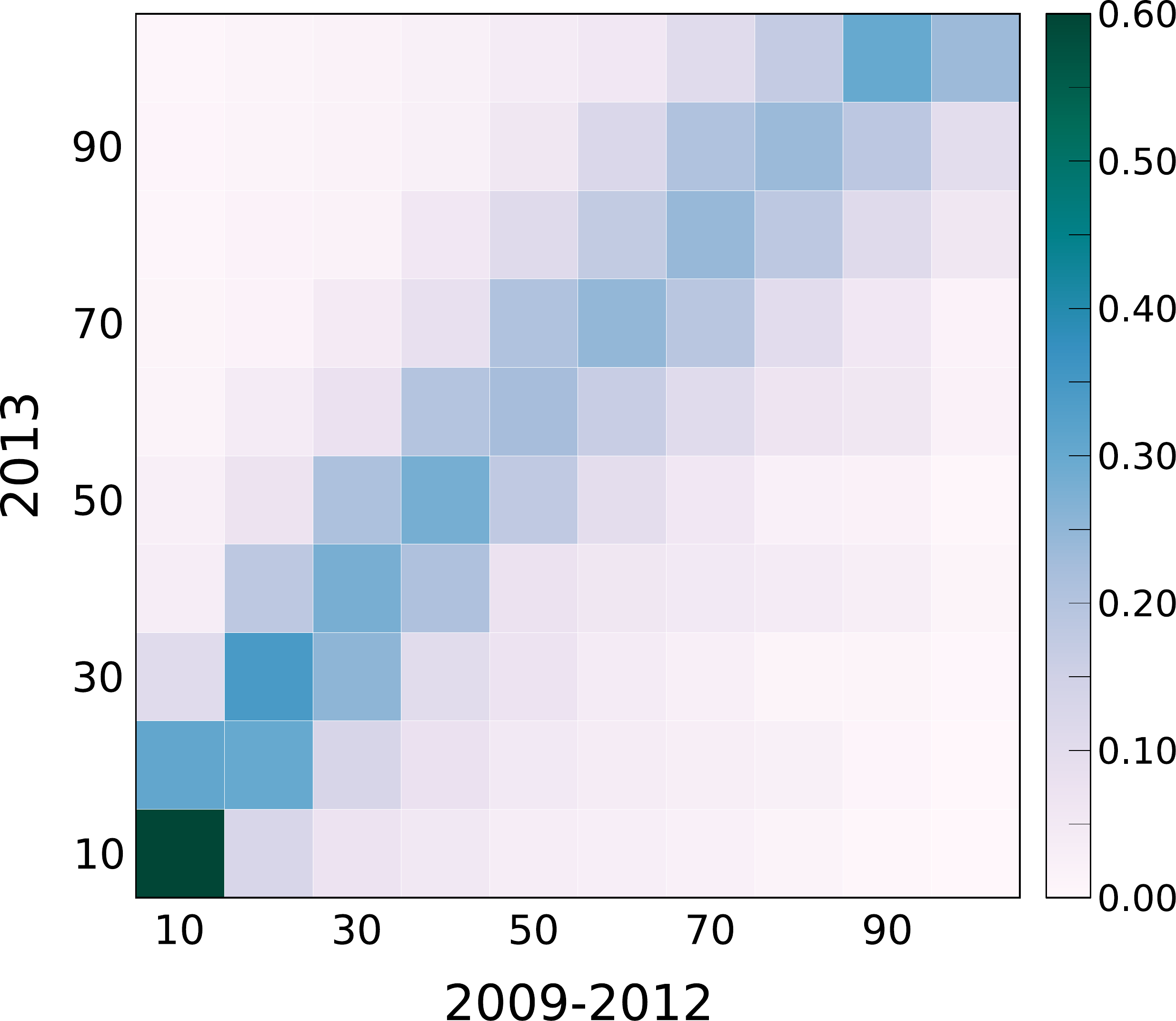}
\caption{Overlap, $O$, between lists of concepts ranked falling in the $n$-th and $m$-th percentile slices of the residual entropy, $S_d$, ranking. The horizontal axis accounts for the Physics collection of 2009--2012, while the vertical axis accounts for the 2013 one. The matrix is normalized by row and white entries correspond to absence of overlap.}
\label{sfig:inters_percentiles_maxent_tf_phys_datas}
\end{figure*}

In this section we quantify the overlap between the lists of concepts belonging to the $n$-th and $m$-th percentile slices ranked using different criterion. Also, we compute the overlap between lists belonging to the two collections of Physics. In Fig.~\ref{sfig:inters_percentiles_idf_vs_maxent_tf} we plot the overlap score $O$, (\eqref{seq:overlap}), of the concepts' lists belonging to two percentile slices of $IDF$ and $S_d$ of a given collection. In the case of $IDF$, concepts are ranked from the most frequent (\ie having the smallest $IDF$) to the least one. In the case of residual entropy, instead, we rank concepts from the closest to its maximum entropy (\ie smallest $S_d$) to the furthest away. According to the definition of $O$, matrices are normalized by row. The analysis of the overlap matrix denotes the presence of a certain degree of similarity in the region near the main diagonal. Within such region, with the sole exception of $O_{10,10}$, the average overlap is around 15$\div$20\% indicating that -- in general -- more frequent concepts tend to fall in higher percentiles of the residual entropy. The $O_{10,10}$ element, instead, has a value around 63\% for Physics 2009--2012 and 50\% for Physics 2013, denoting a remarkable affinity between these sets. This means that \emph{generic} concepts are, to some extent, also those appearing more often within the collection, albeit this is not always the case.
%
%
%
As reported in Tabs.~\ref{tab:stats_topics_arxiv} and \ref{stab:stats_topics_phys2013}, despite the remarkable difference in the number of documents, $N_a$, the number of concepts available in each collection is more or less the same. In the light of that, one question arises: how similar are the two sets of concepts when they are ranked using $S_d$? The analysis of the list similarities across percentile slices is presented in Fig.~\ref{sfig:inters_percentiles_maxent_tf_phys_datas}. The overlap heatmap/matrix displays a certain amount of overlap between contiguous slices, albeit its intensity tends to fade away as we move towards less generic concepts. The lists of very generic concepts, instead, are highly overlapped suggesting that -- if the difference in time is not too large, -- common concepts in Physics tend to remain the same across time, even for non time-overlapping collections.
%
%
%

\newpage

\subsubsection{The 2013 collection similarity network}
\label{s_sssec:phys_2013_similarity_nets}

\begin{figure*}[h!] 
  \centering 
  \includegraphics[width=0.9\textwidth]{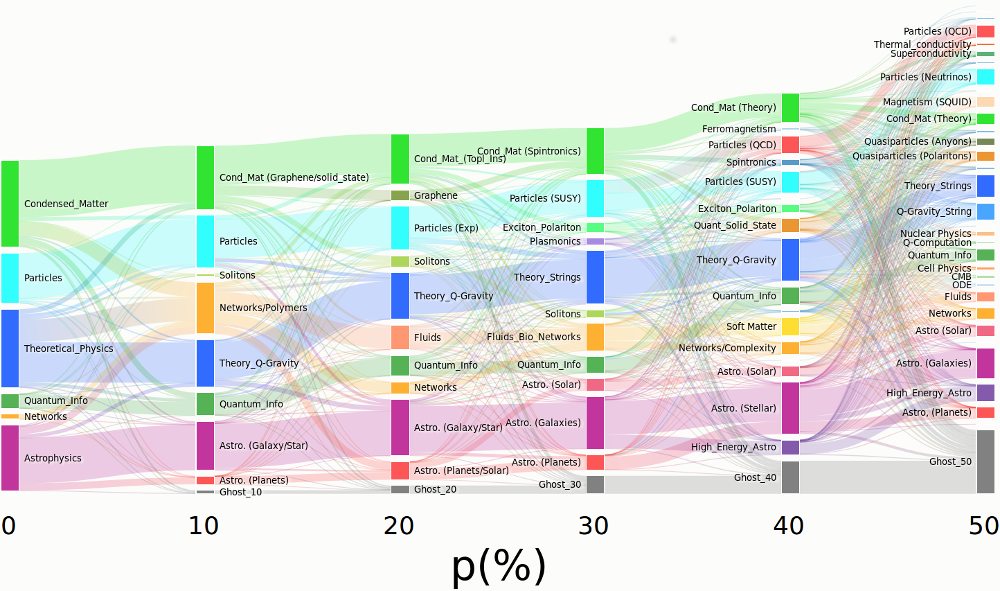} 
  \caption{Static Sankey diagram reporting the topics found by Louvain method in the network of similarity in the Physics 2013 collection. Each community is identified with a colored box whose height corresponds to the number of articles belonging to it. The topic of the community is manually assigned from the ten most representative concepts according to the local document frequency in the communities. The boxes labeled ``ghost'' are composed by articles that do not contain any significant concept at a given percentile $p$, therefore are not part of the network. The thickness of the bands between boxes indicates the number of shared articles. Interactive version available at \cite{sankey-interactive}}
  \label{sfig:sankey_net_phys2013} 
\end{figure*}

The structural properties of the similarity network of the 2013 collection are reported in Tab.~\ref{stab:net_topol_prop_phys2013}. As commented in the main, the first row ($p = 0\%$) corresponds to the original system. The other rows, instead, refer to the networks obtained removing a percentage $p$ of concepts using the entropic filtering. We can notice how, the link density $\rho$ is dramatically affected by the filtering. Indeed, it goes from 36\% to 7\% when $p$ passes from 0\% to 10\%. Consequently, the maximum and average values of the degree -- \ie the number of connections of a node -- $k_{max}$ and $\avg{k}$ drop significantly with $p$, while the average distance between documents $\avg{l}$ increases \cite{latora-book-2017}. The increase of $\avg{l}$ with $p$ -- together with the fragmentation into distinct components $M$ -- is the byproduct of the existence of ``\emph{cultural holes}'' among distinct topics of Physics and, more in general, science itself \cite{vilhena-soc_sci-2014}.
\indent Moreover, at each percentile level, the community structure of the similarity network is retrieved using the Louvain method \cite{blondel-jstat-2008}, a popular and effective approach to discover the communities of a network. The Louvain algorithm has a bias associated with the labelling of the nodes. To mitigate such bias, for each network, we run the algorithm 100 times shuffling the nodes IDs at each realization, and ultimately storing the realization with the highest modularity \cite{newman-pnas-2006}. The community structure of the similarity network is portrayed in the Sankey diagram of Fig.~\ref{sfig:sankey_net_phys2013}.
%
%
%
%
\begin{table*}[hb!]
\setlength{\tabcolsep}{8pt}
\begin{tabular}{@{\extracolsep{\fill}}c|cc d{3} d{3} c d{3} d{3} c}
$p \, (\%)$ & $N_{con}$ & $N_a$ & \multicolumn{1}{c}{$\rho \, (\%)$} & \multicolumn{1}{c}{$\avg{k}$} & $k_{max}$ &  \multicolumn{1}{c}{$\avg{C}$} & \multicolumn{1}{c}{$\avg{l}$} & $M$\\ \hline
{\cellcolor{gray!20}} 0 & {\cellcolor{gray!20}} 11637 & {\cellcolor{gray!20}} 52979 & {\cellcolor{gray!20}} 36.493 & {\cellcolor{gray!20}} 19333.522 & {\cellcolor{gray!20}} 46504 & {\cellcolor{gray!20}} 0.557 & {\cellcolor{gray!20}} 1.635 & {\cellcolor{gray!20}} 1 \\
10 & 9594 & 52337 & 7.340 & 3841.235 & 17532 & 0.327 & 1.935 & 1   \\ 
20 & 8528 & 51522 & 3.752 & 1933.031 & 10399 & 0.319 & 2.008 & 1   \\
30 & 7462 & 49821 & 2.057 & 1024.818 & 8109  & 0.332 & 2.160 & 1   \\
40 & 6396 & 47173 & 1.197 & 564.823  & 5669  & 0.343 & 2.378 & 2   \\
50 & 5330 & 41775 & 0.638 & 266.419  & 2771  & 0.390 & 2.687 & 7   \\
60 & 4264 & 34939 & 0.482 & 168.307  & 1999  & 0.508 & 2.914 & 20  \\
70 & 3197 & 24710 & 0.363 & 89.766   & 1140  & 0.755 & 3.409 & 59  \\
80 & 2132 & 14789 & 0.257 & 37.989   & 495   & 0.783 & 4.242 & 153 \\
90 & 1066 & 5703  & 0.228 & 13.027   & 104   & 0.848 & 7.124 & 342 \\
\end{tabular}
\caption{Topological indicators of the Physics 2013 similarity networks. The first row ($p=0\%$) corresponds to the original network, while the others to the networks obtained using the maximum entropy filter. In the columns we report: the percentage of filtered concepts $p$, the number of concepts $N_{con}$, the number of articles containing at least one concept (nodes) $N_a$. the link density $\rho$, the average and maximum degrees, $\avg{k}$ and $k_{max}$, the average clustering coefficient $\avg{C}$, the average path length $\avg{l}$ and the number of connected components $M$. The minimum edge weight is equal to $w_{min} = 0.01$.}
\label{stab:net_topol_prop_phys2013}
\end{table*}
%




\subsubsection{TopicMapping analysis}
\label{s_sssec:topicmapping_physics}

\begin{figure*}[h!]
\centering
\includegraphics[width=0.87\columnwidth]{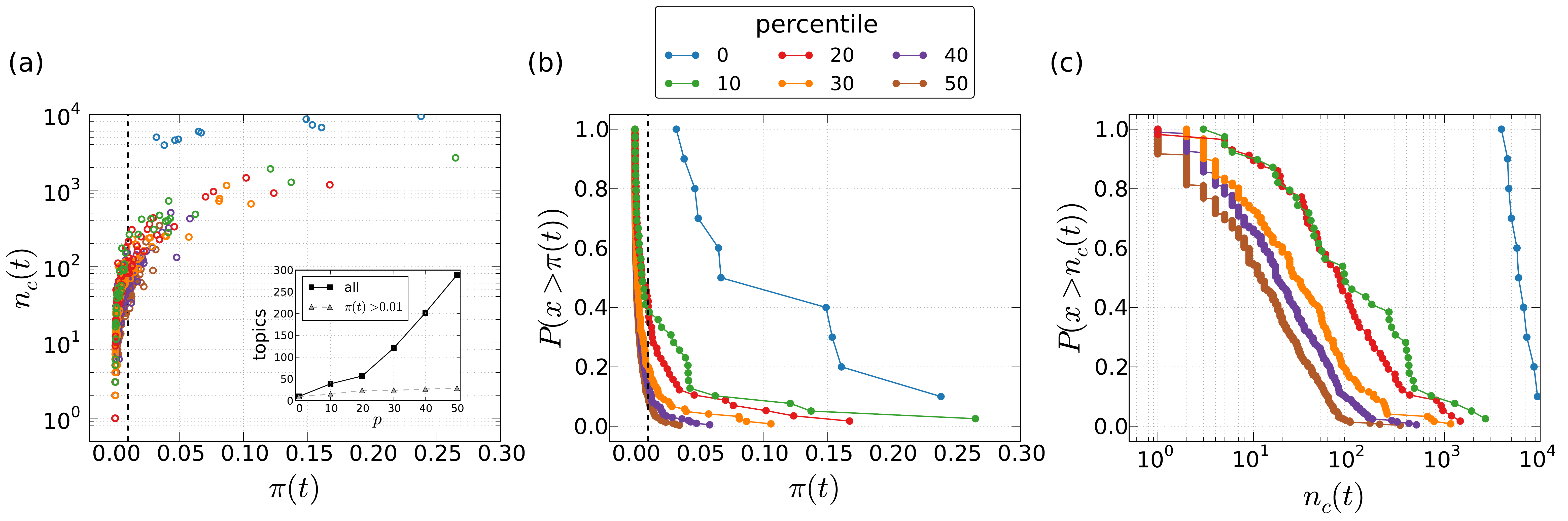}
\caption{Statistics about topics for the Physics 2009--2012 dataset. (a) Relation between the number of concepts $n_c (t)$ and the probability $\pi(t)$ associated to each topic $t$. Every circle represents a topic whose color denotes the filtering percentile $p$. The dashed vertical line corresponds to the topic probability $\pi(t) = 0.01$ below which topics are not considered meaningful. In the inset, the total number of topics for each percentile (squares) is shown along with the number of meaningful topics with probability $\pi(t) > 0.01$ (triangles). The complementary cumulative distribution functions of the topic probability, $\pi(t)$, and the number of concepts per topic, $n_c (t)$, are displayed in panels (b) and (c), respectively. The different colors denote the percentiles, $p$, of filtered concepts.}
\label{sfig:topicmapping_statistics_phys_2009-2012}
\end{figure*}

The statistical properties of the topics found by the TopicMapping (TM) algorithm vary according to the filtering percentile $p$ as reported in Tabs.~\ref{tab:stats_topics_arxiv} and \ref{stab:stats_topics_phys2013}. A topic $t$ is constituted by a given number of concepts, $n_c(t)$, and it has an associated probability, $\pi(t) = \sum_{c} \pcond{\pi}{t}{c}\; \pi(c)$, which indicates its importance. The relation between these quantities is shown in Fig.~\ref{sfig:topicmapping_statistics_phys_2009-2012} (a): each circle represents a topic discovered at a given filtering percentile $p$ as represented by colors. As $p$ increases, we notice a higher density of topics falling close to the significance threshold $\pi(t) = 0.01$.
In the same picture, the number of words per topic, $n_c(t)$ decrease as $p$ increases. Moreover, the significance of a topic $t$ also decreases as displayed in the inset of panel (a), as well as in panel (b). The plot of the cumulative distribution of the number of words per topic, $P(x > n_c(t))$, reveals a progressive fragmentation of topics into smaller and more specific (\ie having less words) ones passing from $n_c \gtrsim 3000$ words per topic at $p=0$ to $n_c \sim 100$ for half of them at $p=10\%$, instead.

%
%
%
%
\begin{figure*}[ht!] 
  \centering 
  \includegraphics[width=0.75\columnwidth]{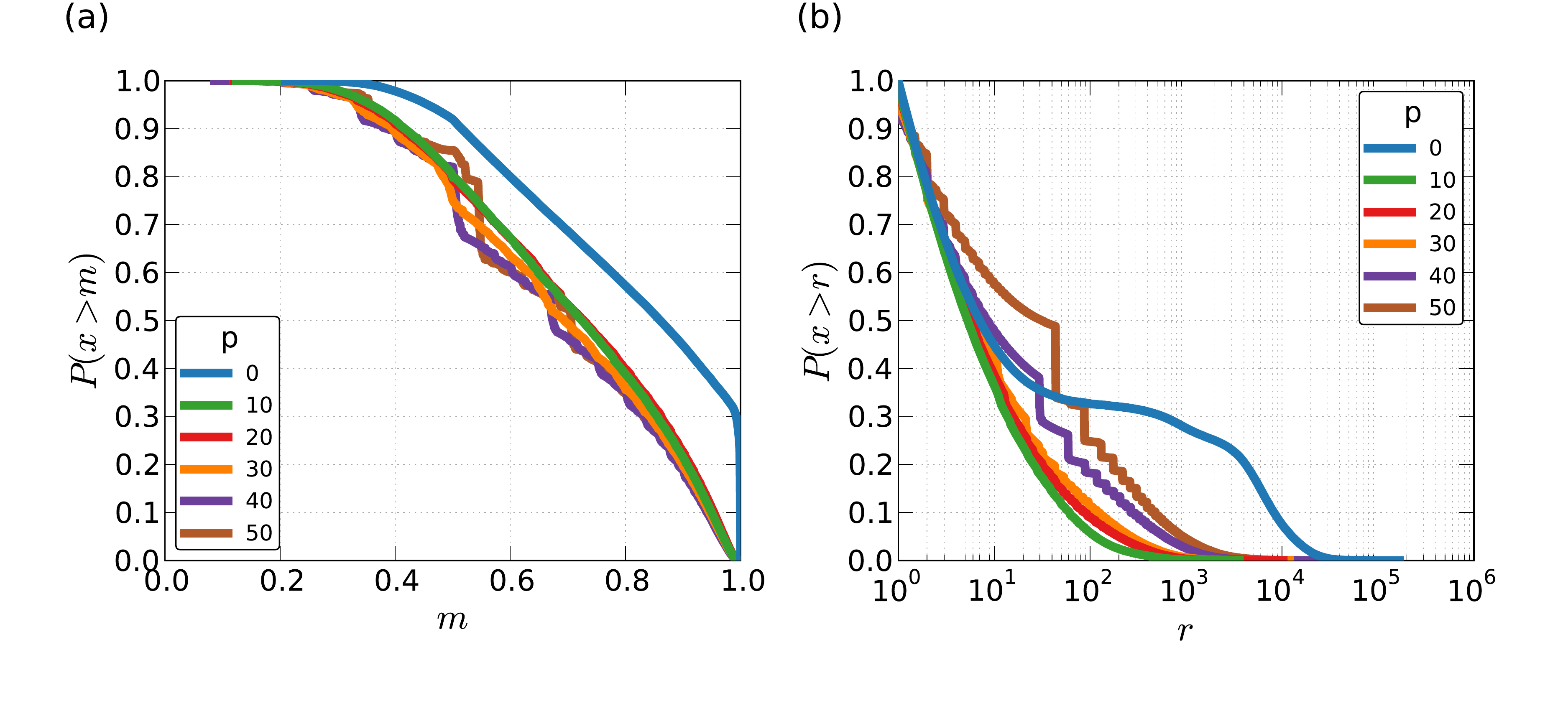} 
  \caption{(panel a) Complementary cumulative distribution functions (ccdfs) of the maximum of the probability that a topic, $t$, belongs to a document $\alpha$, $m = \max_{t \in T}(\pcond{\pi}{t}{\alpha})$. (panel b) ccdfs of the ratio, $r$, between $m$ and the second highest value of the probability. Colors account for different intensities of filtering $p$ on the Physics 2009--2012 collection.}
  \label{sfig:ccdfs_topics_given_docs_phys_2009-2012} 
\end{figure*}

\begin{figure*}[h!] 
  \centering 
  \includegraphics[width=0.9\textwidth]{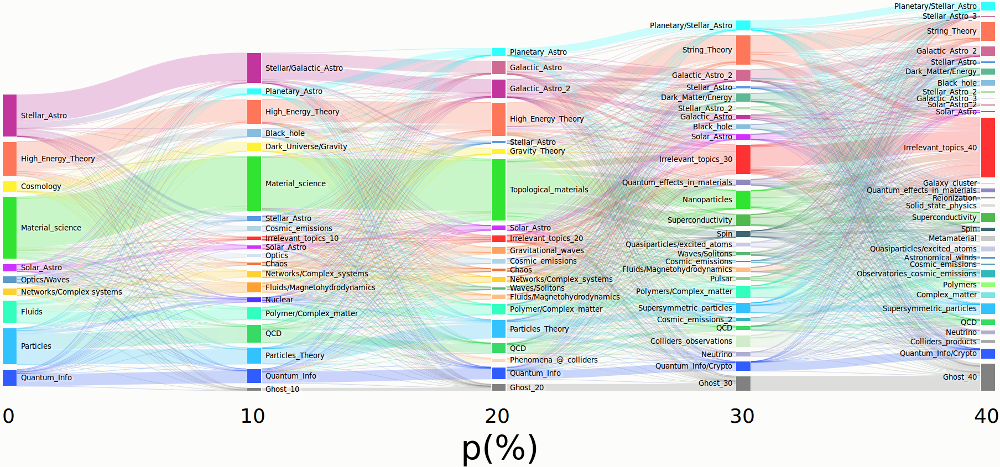} 
  \caption{Static Sankey diagram representing the topics found by TM on the Physics collection of articles from 2013. Each topic is identified with a colored box whose height corresponds to the number of articles associated to it. Each article $\alpha$ is assigned to the topic with maximum probability $\pcond{\pi}{t}{\alpha}$, \ie the topic that describes the highest portion of the article. Topics are manually labeled from the ten most representative concepts according to the probabilities of concepts given the topic, $\pcond{\pi}{w}{\alpha}$. For the ease of visualization, only topics with probability $\pi(t) > 0.01$ are shown, whereas the remaining ones are incorporated together in a single ``super-topic'' denoted as ``Irrelevant\_topics''. The boxes labeled ``ghost'' are composed by articles that do not contain any significant concept at a given percentile $p$, therefore are not part of the dataset used by TM. The thickness of the bands between boxes indicates the number of shared articles. Interactive version available at \cite{sankey-interactive}}
  \label{sfig:sankey_TM_phys2013} 
\end{figure*}

\indent As commented in the main, the TM algorithm associates a topic $t$ to an article $\alpha$ with a probability $\pcond{\pi}{t}{\alpha}$. Thus, each document $\alpha$ is made by a mixture of topics whose composition is described by the shape of $\pcond{\pi}{t}{\alpha}$. Hence, to ensure the assignment of each article to a single topic, one has to check first that $\pcond{\pi}{t}{\alpha}$ has a maximum in $t$, $m$, and then that such maximum is undoubtedly greater than any other value. In Fig.~\ref{sfig:ccdfs_topics_given_docs_phys_2009-2012}, we analyze the distribution of $m = \max_{t \in T}(\pcond{\pi}{t}{\alpha})$ (panel a), and the ratio, $r$, between $m$ and the second highest value of $\pcond{\pi}{t}{\alpha}$ (panel b). 
%
%
%
%
We see how at least $30\%$ of the articles have $m$ greater than $0.5$ (panel a). Moreover, even when $m < 0.5$, the ratio $r$ is bigger than 2 in, at least, $80\%$ of the articles (panel b). These results enable the assignation of each article to its main topic, especially for $p > 10\%$ which is fundamental for the correct interpretation of the results shown in the Sankey diagram of Fig.~\ref{sfig:sankey_TM_phys2013}.
%
%
%
%
\begin{table*}[h!]
\centering
%
\newcolumntype{d}[1]{D{.}{.}{#1} }
\setlength{\tabcolsep}{10pt}
\begin{tabular}{@{\extracolsep{\fill}}c|ccccccd{2}}
$p \, (\%)$ & $N_{con}$ & $N_a$ & \multicolumn{1}{c}{$T$} & \multicolumn{1}{c}{$T^*$} & \multicolumn{1}{c}{$\langle N_a \rangle_{T^{*}}$} & \multicolumn{1}{c}{$\langle N_{con} \rangle_{T^{*}}$} & \multicolumn{1}{c}{$F$} \\ \hline
{\cellcolor{gray!20}} 0 & {\cellcolor{gray!20}} 13173 & {\cellcolor{gray!20}} 52979 & {\cellcolor{gray!20}} 10 & {\cellcolor{gray!20}} 10 & {\cellcolor{gray!20}} 5298  & {\cellcolor{gray!20}} 4212 & {\cellcolor{gray!20}} 1.00\\
10 & 9593 & 52337 & 28 & 18 & 2861 & 1586 & 0.98 \\
20 & 8527 & 51522 & 45 & 19 & 2632 & 430 & 0.97 \\
30 & 7461 & 49821 & 90 & 25 & 1738 & 245 & 0.87 \\
40 & 6396 & 47173 & 141 & 29 & 1185 & 150 & 0.73 \\
50 & 5330 & 41775 & 235 & 25 & 857 & 99 & 0.51 \\
60 & 4264 & 34939 & 300 & 24 & 590 & 61 & 0.40 \\
70 & 3197 & 24710 & 539 & 23 & 329 & 30 & 0.30 \\
80 & 2132 & 14789 & 1178 & 11 & 238 & 25 & 0.18 \\
90 & 1066 & 5703 & 1839 & 7 & 62 & 8 & 0.08 \\
\end{tabular}
\caption{Characteristics of the topic modeling on Physics 2013 collection. The row $p=0\%$ corresponds to the original corpus/dataset, while $p>0\%$ to those filtered using the maximum entropy. In the columns we report: the percentage of filtered concepts $p$, the number of concepts $N_{con}$, of documents having at least one concept $N_a$, of topics found by LDA, $T$, and number of ``meaningful'' ones $T^*$. For the latter, we report also the average number of documents $\langle N_a \rangle_{T^{*}}$, and concepts $\langle N_{con} \rangle_{T^{*}}$ per topic. Finally, we report the fraction of documents assigned to a meaningful topic, $F$.}
\label{stab:stats_topics_phys2013}
\end{table*}
%
%
%
%
%
\begin{figure*}[hb!]
\centering
\includegraphics[width=\columnwidth]{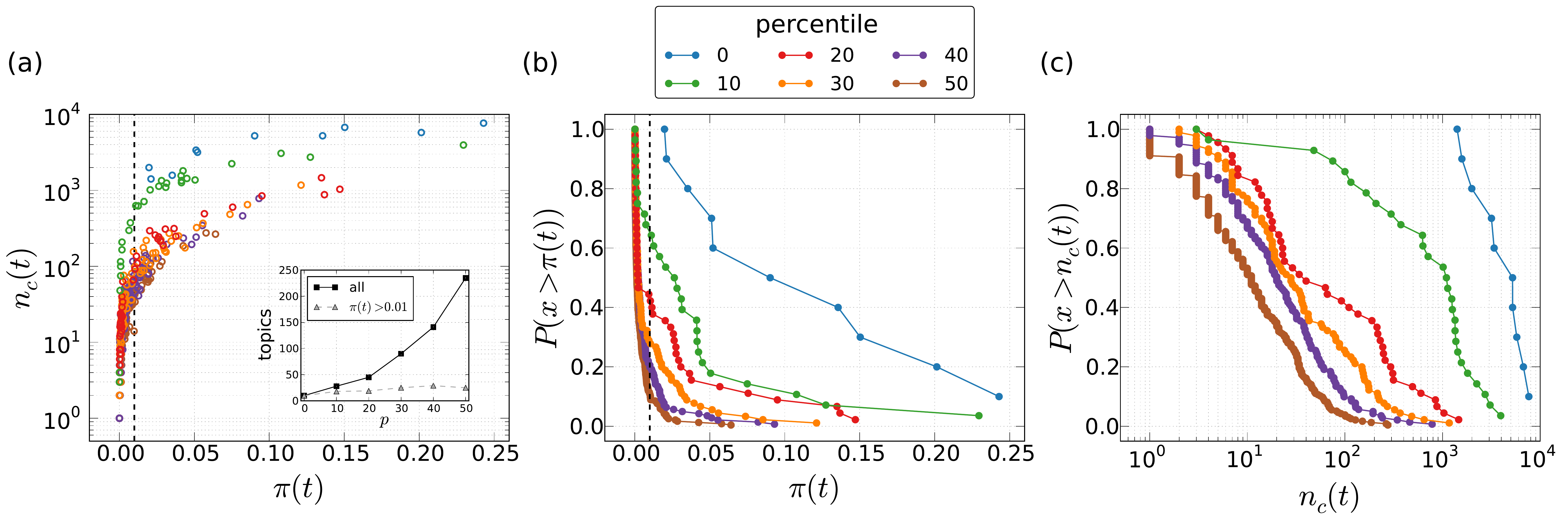}
\caption{Statistics about topics for the Physics articles in 2013. (a) Relation between the number of concepts $n_c (t)$ and the probability $\pi(t)$ associated to each topic $t$. Every circle represents a topic whose color denotes the filtering percentile $p$. The dashed vertical line corresponds to the topic probability $\pi(t) = 0.01$ below which topics are not considered meaningful. In the inset, the total number of topics for each percentile is shown by squares along with the number of important topics with probability $\pi(t) > 0.01$ shown by triangles. The complementary cumulative distribution functions of the topic probability $\pi(t)$ and the number of concepts per topic $n_c (t)$ are represented in (b) and (c), respectively. The different colors denote the percentiles of filtered concepts.}
\label{sfig:topicmapping_statistics_phys2013}
\end{figure*}
%
%
%
%
%
\begin{figure*}[h!] 
  \centering 
  \includegraphics[width=0.75\columnwidth]{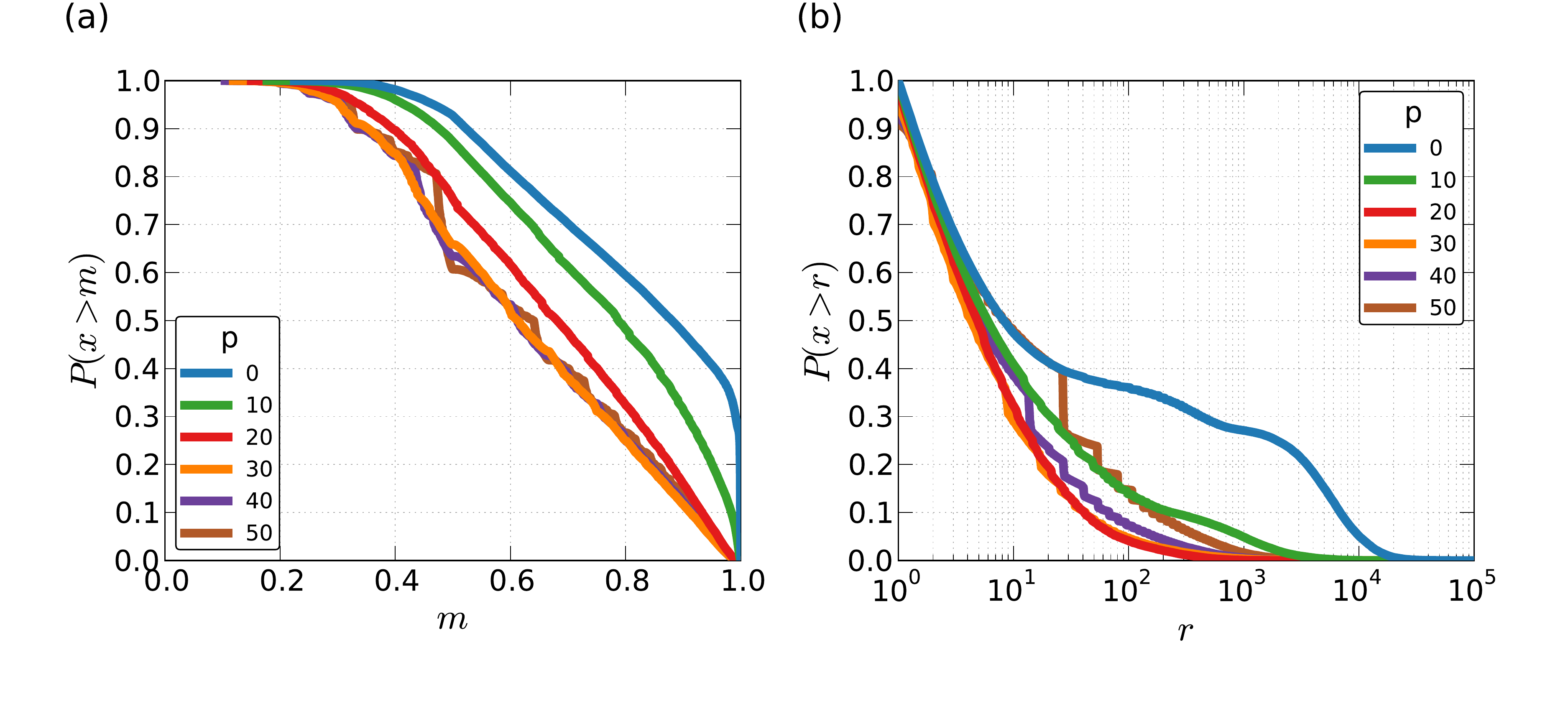} 
  \caption{(panel a) Complementary cumulative distribution functions (ccdfs) of the maximum of the probability that a topic, $t$, belongs to a document $\alpha$, $m = \max_{t \in T}(\pcond{\pi}{t}{\alpha})$. (panel b) ccdfs of the ratio, $r$, between $m$ and the second highest value of the probability. Colors account for different intensities of filtering $p$ on the Physics 2013 collection.}
  \label{sfig:ccdfs_topics_given_docs_phys2013} 
\end{figure*}
Similar trends can be observed also for the 2013 collection, as shown in  Tab.~\ref{stab:stats_topics_phys2013}, and Figs.~\ref{sfig:topicmapping_statistics_phys2013}--\ref{sfig:ccdfs_topics_given_docs_phys2013}. The outcome of the TM on the 2013 collection (Fig.~\ref{sfig:sankey_TM_phys2013}) shows interesting features even when considering the original pool of concepts. At $p=0$ we recognize specialized topics like ``Stellar\_Astro'' and ``Solar\_Astro'' -- two branches of Astrophysics, -- or  ``Optics/Waves'', ``Cosmology,'' and ``Fluids'' to cite others. Beside such specific topics, we can find more broader ones such as ``Material\_science'', ``High\_Energy\_Theory,'' and ``Particles.'' As $p$ increases, instead, topics become more fragmented. For example, ``Particles'' evolves into ``Particles\_Theory'',``QCD,'' and ``Nuclear;'' while ``Stellar\_Astro'' splits into ``Stellar\_Astro,'' ``Cosmic emissions,'' ``Stellar/Galactic\_Astro,'' and ``Planetary\_Astro.'' The fragmentation does not happen in the same way for all the topics. For example, ``Material\_science'' remains unaltered until $p=30\%$, before splitting into smaller topics. On the other hand, ``Quantum\_Info'' never fragments suggesting its isolation from the rest of the topics. Finally, despite the pruning of concepts boosts the emergence of more fine-grained topics, an excessively aggressive filtering is counterproductive since it discards too much information. As a consequence, the number of ``Irrelevant\_topics'' -- and inevitably the number of documents belonging to them -- grows inesorably, suppressing the fraction of papers assigned to meaningful topics, $F$.

\subsubsection{Comparison between Louvain based and TopicMapping based topics}
\label{s_sssec:comparison_louvain_topicmapping_phys2013}

\begin{figure}[p!] 
  \centering 
  \includegraphics[height=0.95\textheight]{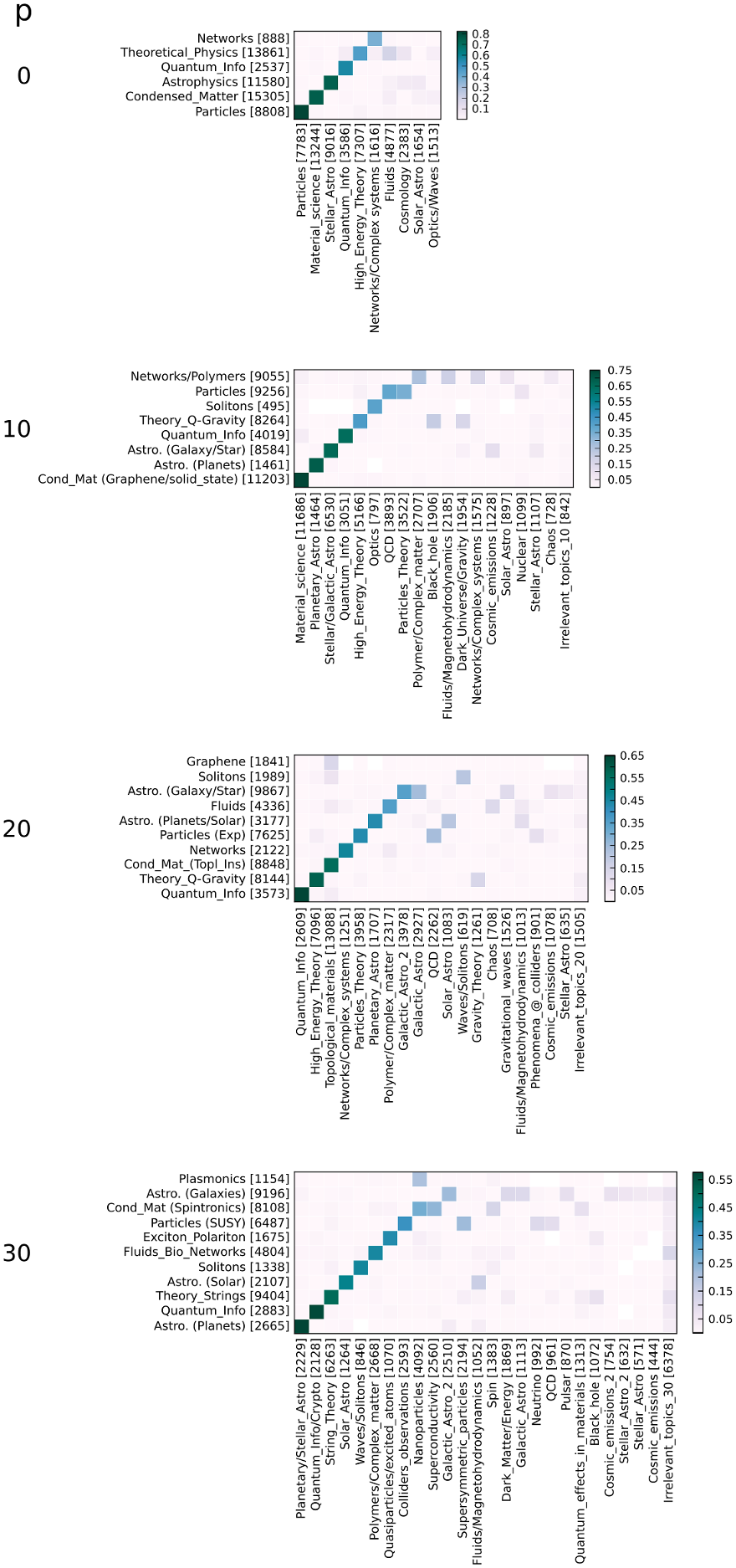} 
  \caption{Jaccard score, $J$, between the sets of articles belonging to a community (row) and those associated to a TM topic (column) for a given filtering percentile $p$ of the Physics 2013 collection. The number of articles in each set is indicated within square brackets. We consider only TM topics for which $\pi(t) > 0.01$.}
  \label{sfig:jaccard_tf_vs_TM_phys2013} 
\end{figure}

Despite the filtering induced fragmentation of topics takes place in both Louvain and TopicMapping cases, this does not ensure that the topics found are the same. To measure the similarity between the topics obtained with both methodologies, we compute the Jaccard score $J$, of \eqref{seq:jaccard}, between the set of documents belonging to a community/topic found with Louvain, $\mathcal{A}$; and the set of documents belonging to a topic found with TM, $\mathcal{B}$. Concering the latter, it is worth remembering that TM does not assign \emph{directly} documents to topics but, instead, assigns topics to each \sout{topic} document $\alpha$ with a certain probability $\pcond{\pi_{t \in T}}{t}{\alpha}$. However, given the results shown in Figs.~\ref{sfig:ccdfs_topics_given_docs_phys_2009-2012} and \ref{sfig:ccdfs_topics_given_docs_phys2013}, it is reasonable to assume that the vast majority of documents can be associated to a single, prevalent topic. Therefore, we assign each document $\alpha$ to the topic $\tilde{t}$ maximizing $\pcond{\pi_{t \in T}}{t}{\alpha}$. The heatmaps of $J$ for various filter intensities, $p$, for the Physics 2013 collection are displayed in Fig.~\ref{sfig:jaccard_tf_vs_TM_phys2013}.

By looking at Fig.~\ref{sfig:jaccard_tf_vs_TM_phys2013} there is one thing that we notice at first glance. There is always a number of topics that are retrieved by both methods corresponding to the dark entries. Another phenomenon that we observe is that as $p$ increases, more and more topics found with Louvain are mapped into multiple TM topics. For example, at $p = 20\%$, ``Astro. (Galaxy/Star)'' gets mapped into ``Galactic\_Astro,'' ``Galactic\_Astro\_2,'' ``Gravitational\_waves,'' ``Cosmic\_emissions,'' and ``Stellar\_Astro'' with a maximum score of 0.35. In some other cases, instead, the topics found by Louvain do not have any equivalent counterpart in TM as for ``Graphene'' at $p=20\%$, and ``Plasmonics'' at $p=30\%$ which, at most, have values of $J$ equal to 0.13 and 0.20, respectively.

\begin{figure}[p!] 
  \centering 
  \includegraphics[height=0.87\textheight]{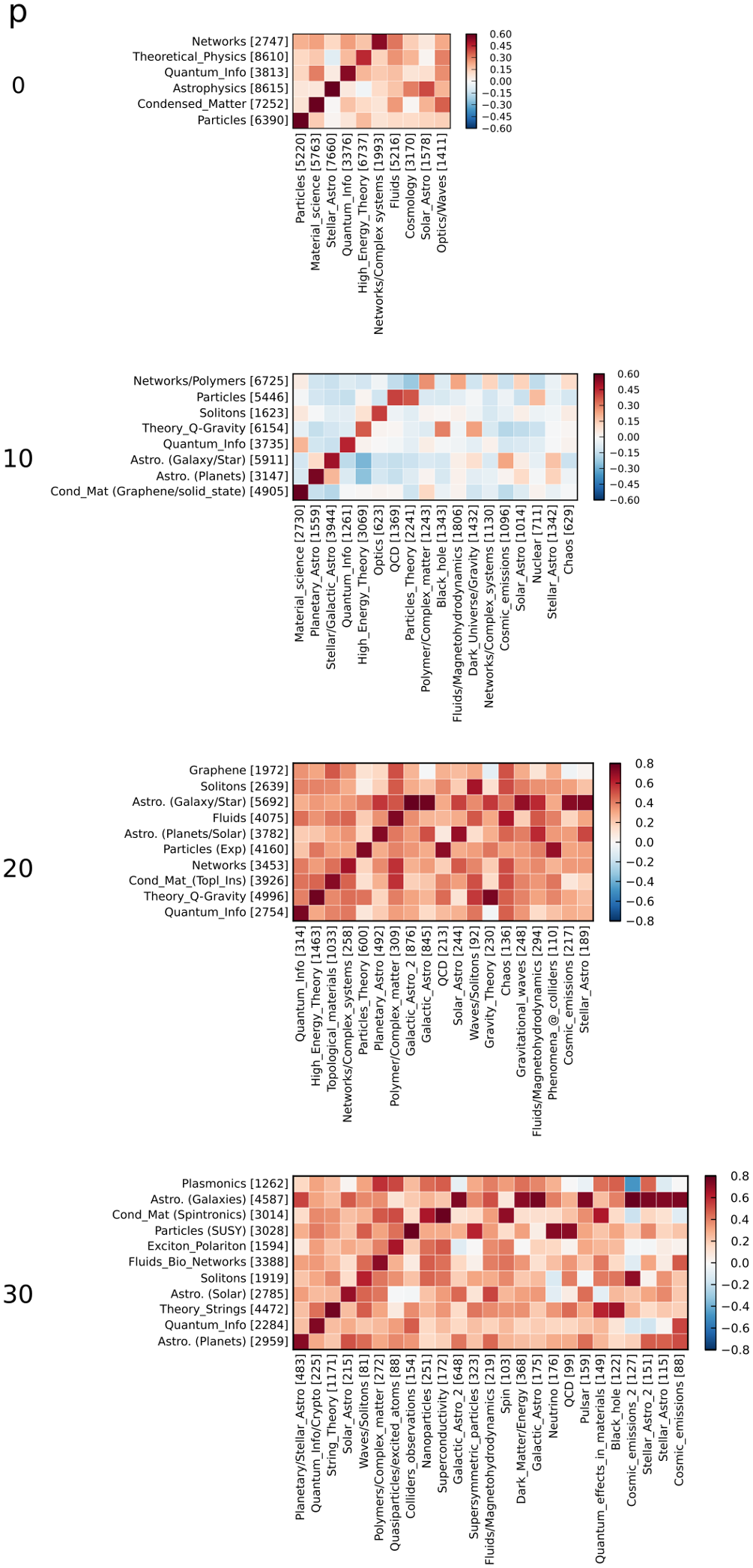} 
  \caption{Kendall correlation coefficient $\tau_b$ for sets of ranked concepts belonging to topics identified by Louvain method (rows) and TM algorithm (columns) of the Physics 2013 collection. Each map refers to a given filtering percentile $p$. The number of concepts in every topic is indicated within square brackets. The $\tau_b$ is computed on the rankings of concepts appearing both in the Louvain and the TM topic. Only those topics for which $\pi(t) > 0.01$ are included in the heatmaps.}
  \label{sfig:tau_tf_vs_TM_phys2013} 
\end{figure}

The Jaccard score measures the similarity between the sets of documents assigned to topics. However, the intimate nature of a topic is determined by the concepts belonging to it. To gain more insight on the correspondence between topics found with Louvain and TM we have decided to compute also the Kendall coefficient, $\tau_b$, -- introduced in \eqref{seq:kendall_tau_b} of Sec.~\ref{s_ssec:comparison_rankings_kendall} -- among the ranked lists of concepts belonging to topics found with Louvain and those found by TM. The resulting heatmaps at different filtering intensities $p$ are diplayed in Fig.~\ref{sfig:tau_tf_vs_TM_phys2013}. A phenomenology similar to that observed for $J$ persists also for the rankings especially at $p\leq 10\%$. At higher levels of filtering, we notice the presence of almost exclusively positive correlations together with the existence of several dark entries which denote highly correlated rankings. Such phenomenon could be explained by the fact that the size of the sets used to compare ranks falls drastically.

Summing up, although the similarity between topics tends to fade as filtering becomes more aggressive, we notice the presence of a core of topics which are found by both methodologies. The behavior of the Kendall coefficient $\tau_b$, instead, seems to point out the presence of correlations between the rankings of concepts used to define topics retrieved by both methods.

\subsubsection{Comparison between communities of networks obtained with different filtering criteria}
\label{s_sssec:comparison_comm_physics}

\begin{figure*}[ht!]
\centering
\includegraphics[width=\columnwidth]{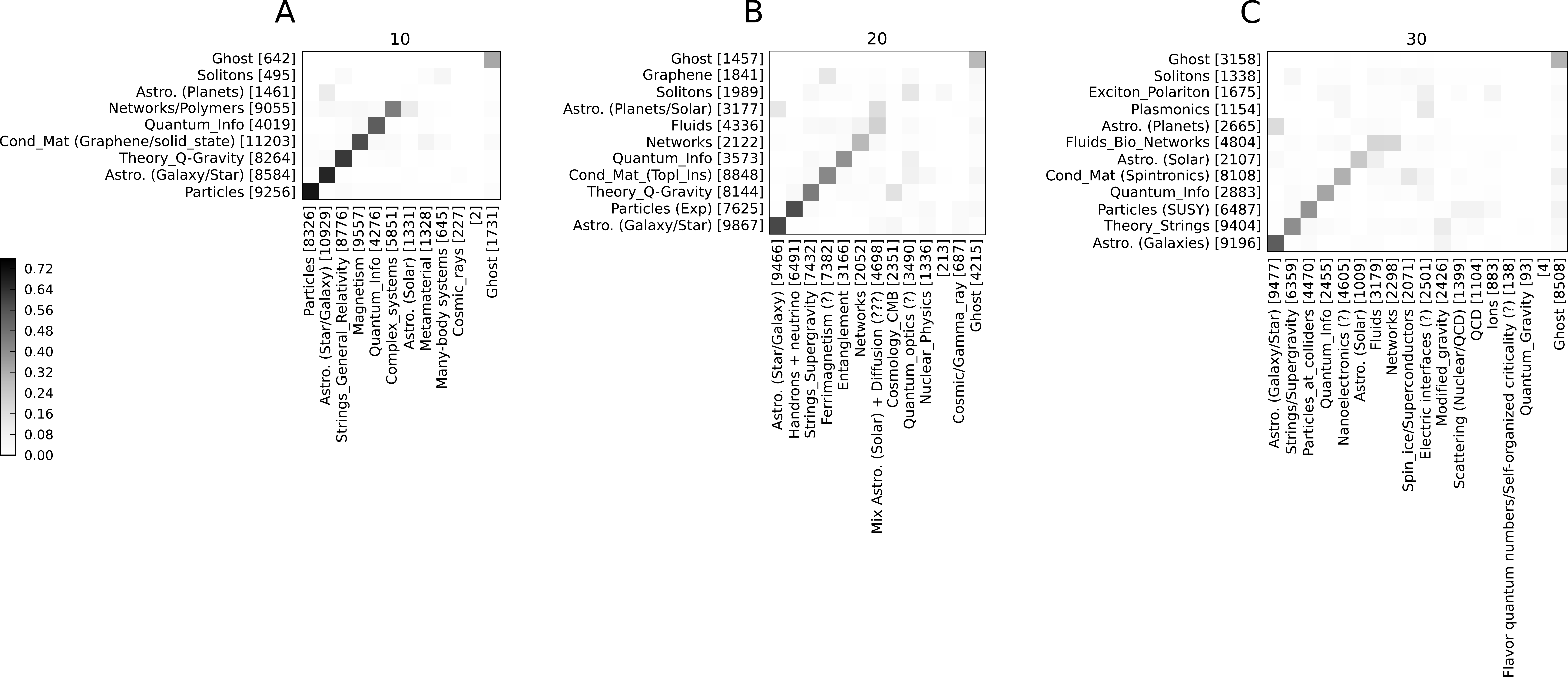}
\caption{Overlap among communities of the similarity network obtained filtering concepts either using entropy (y-axis) or $IDF$ (x-axis). The color of the cells denotes the Jaccard score $J$. Each label accounts for the main topic and the size of a give community. Each matrix refers to removing, respectively, the 10\% (A), 20\% (B) and 30\% (C) of the concepts. The values of the overlap in panel (A) range from 0.70 to 0.45 on the main diagonal, while off-diagonal elements are below 0.11, except for the overlap between ``ghost'' communities which is 0.34. Panel (B) features overlaps between 0.59 and 0.22 on the main diagonal, while the other values are below 0.17 and the overlap between ``ghost'' communities is 0.29. Finally, panel (C) displays values ranging from 0.54 to 0.19 on the main diagonal and below 0.18 outside, apart from the ``ghost'' communities overlap which is 0.30.}
\label{sfig:comparison_comm_physics}
\end{figure*}

In Fig.~\ref{sfig:comparison_comm_physics} we report the heatmaps of the Jaccard score $J$ (see Sec.~\ref{s_ssec:comparison_communities}) computed between the communities of the similarity networks obtained by pruning out a given fraction $p$ of concepts according either to their $IDF$ ranking (horizontal axis) or to their $S_d$ one (vertical axis). The results are obtained for the 2013 collection. Each column corresponds to a different amount of removed concepts spanning from 10\% to 30\%. Overall, the Jaccard heatmaps tell us that there is always a certain degree of similarity among the communities found after filtering according to $IDF$ and $S_d$. However, the overlap fades away as the system begins to display a richer topic/community organization in response to the increasing amount of concepts removed.
%
%
%
%
%
More specifically, as $p$ increases, we notice the coexistence of communities present in both cases, and others typical of a given filtering criterion. Such coexistence is yet another proof (as already shown in Secs.~\ref{s_ssec:rel-full_ent-cond_ent}, \ref{s_ssec:comparison_sc_idf} and \ref{s_sssec:phys_diff_ent_idf_2collections}) that using residual entropy to filter the network is not equivalent to the filtering based on the other previously existing methods. In the case of TM, the final result does not change significantly as we notice from Fig.~\ref{sfig:jaccard_tf_vs_TM_phys2013}. For computational reasons we do not perform the comparison with Louvain for the 2009--2012 collection, although our guess suggests that similar trends should be found.

\subsubsection{Ranking of concepts within papers}
\label{s_sssec:conc_ranking_papers}

In the following, we want to understand if the entropic selection of concepts could be used also to rank concepts and, in turns, use those rankings to classify documents. To this aim, we select ten highly cited documents from the collection of articles submitted during 2013 on \texttt{arXiv} (see Tab.~\ref{stab:selected_papers}), and for each of them we consider rankings based on \textsc{TF}, $IDF$, \textsc{TF-IDF} and $S_d$ respectively. Next, we order concepts from the most generic to the least one according to the four rankings. This translates into sorting either in descending order (\textsc{TF}, \textsc{TF-IDF}) or in ascending one ($IDF$, $S_d$). The results are reported in Tab.~\ref{stab:concepts_ranked_in_papers_various_measures}.

%
%
%
\begin{table*}[h!]
\centering
\resizebox{\textwidth}{!}{
\begin{tabular*}{\hsize}{@{\extracolsep{\fill}} c|llll}
\texttt{arXiv} ID & $N^\#$ & Princ. category & Other categories & Venue \\ \hline
\texttt{1306.5856} & 1692 & cond-mat.mtrl-sci & -- & {\em Nat Nanotech} {\bf 8}, 235--246 (2013) \\ 
\texttt{1301.0842} & 406 & astro-ph.EP & -- & {\em Astroph Jour} {\bf 766}, 81 (2013) \\ 
\texttt{1308.0321} & 503 & cond-mat.quant-gas & cond-mat.str-el, quant-ph & {\em Phys Rev Lett} {\bf 111}, 185301 (2013) \\ 
\texttt{1301.1340} & 278 & hep-ph & -- &  {\em Rep Prog Phys} {\bf 76}, 056201 (2013) \\ 
\texttt{1301.0319} & 457 & astro-ph.SR & astro-ph.IM &  {\em Astrophys J Suppl Ser} {\bf 208}, 4 (2013) \\ 
\texttt{1304.6875} & 724 & astro-ph.HE & astro-ph.SR, cond-mat.quant-gas, gr-qc & {\em Science} {\bf 340}, Issue 6131 (2013) \\  
\texttt{1306.2314} & 273 & astro-ph.CO & -- & {\em Phys Rev D} {\bf 88}, 043502 (2013) \\ 
\texttt{1311.6806} & 242 & astro-ph.EP & -- & {\em PNAS} {\bf 110}, 19175 (2013) \\ 
\texttt{1302.5433} & 199 & cond-mat.supr-con & cond-mat.mes-hall & {\em J Phys Conden. Matter} {\bf 25}, 233201 (2013) \\ 
\texttt{1303.3572} & 27 & cond-mat.str-el & hep-ph, quanth-ph & {\em Phys Rev B} {\bf 89}, 045127 (2014) \\ 
\end{tabular*}
} 
\caption{Main attributes of the manuscripts selected to study the rankings of concepts within documents. For each document we report its \texttt{arXiv} ID, the number of citations, $N^\#$, its principal category, and eventual secondary ones. Finally, we provide the publication venue.}
\label{stab:selected_papers}
\end{table*}

Qualitatively speaking, the ranked lists of concepts presented in Tab.~\ref{stab:concepts_ranked_in_papers_various_measures} seems to confirm, on one hand, the inability of $S_d$ and $IDF$ to fully grasp the subject of each article. On the other hand, instead, \textsc{TF} and \textsc{TF-IDF} perform remarkably better in the same task. However, these conclusions are not surprising: both $S_d$ and $IDF$ are global measures, defined on the whole collection in order to quantify the importance of a concept for the entire collection. Conversely, both \textsc{TF} and \textsc{TF-IDF} are local measures that capture the significance of a concept within papers.

%
%
%
\begin{center}
{\tiny
\newcolumntype{b}{>{\columncolor{purple!40!white}}p{3.4cm}}
\begin{longtable}{|p{1.9cm}|p{4.2cm}|p{2.5cm}|p{3.6cm}|b|}
\caption{List of the ten most generic concepts for the papers listed in Tab.~\ref{stab:selected_papers}. We rank concepts using: residual entropy $S_d$, Inverse Document Frequency {\small \textsc{IDF}}, term frequency {\small \textsc{TF}} and {\small \textsc{TF-IDF}}. Concepts indicated as common by ScienceWISE, are marked by an asterisk. The column corresponding to the best ranking is highlighted.}
\label{stab:concepts_ranked_in_papers_various_measures}\\
\hline 
\multicolumn{1}{|p{1.9cm}}{\texttt{arXiv} \textsc{ID}} & \multicolumn{1}{|p{4.2cm}}{$S_d$} & \multicolumn{1}{|p{2.5cm}}{{\small \textsc{IDF}}} & \multicolumn{1}{|p{3.6cm}}{{\small \textsc{TF}}} & \multicolumn{1}{|b|}{{\small \textsc{TF-IDF}}} \\

\endfirsthead

\hline
\multicolumn{5}{|c|}{{\bfseries continued from previous page}} \\ \hline 
\multicolumn{1}{|p{1.9cm}}{\texttt{arXiv} \textsc{ID}} & \multicolumn{1}{|p{4.2cm}}{$S_d$} & \multicolumn{1}{|p{2.5cm}}{{\small \textsc{IDF}}} & \multicolumn{1}{|p{3.6cm}}{{\small \textsc{TF}}} & \multicolumn{1}{|b|}{{\small \textsc{TF-IDF}}} \\ \hline
\endhead

\hline \multicolumn{5}{|c|}{{\bfseries Continued on next page}} \\ \hline
\endfoot

\hline \hline
\endlastfoot

\hline
%
%
%
 \texttt{1306.5856} & 
\multicolumn{4}{p{12cm}|}{\texttt{Raman spectroscopy as a versatile tool for studying the properties of graphene}} \\ 
 \cline{2-5} 
 & Experimental data * & Energy * & Phonon & Phonon \\ 
 & Regularization & Measurement * & Graphene & Graphene \\ 
 & Intensity & Field * & Electron * & Graphite \\ 
 & Temperature * & Potential * & Energy * & Raman spectroscopy \\ 
 & Field * & Mass * & Graphite & Raman scattering \\ 
 & Optics * & Particles * & Resonance * & Electron * \\ 
 & Electromagnet * & Temperature * & Frequency * & Carbonate * \\ 
 & Energy * & Probability * & Measurement * & Resonance * \\ 
 & Mass * & Units * & Scattering * & Wave vector \\ 
 & Wavelength * & Vector * & Intensity & Selection rule \\ 
 \hline 
 \texttt{1301.0842} & 
\multicolumn{4}{l|}{\texttt{The false positive rate of Kepler and the occurrence of planets}} \\ 
 \cline{2-5} 
 & Order of magnitude * & Measurement * & Planet & Planet \\ 
 & Numerical simulation & Field * & Star & Kepler Objects of Interest \\ 
 & Space telescopes & Potential * & Kepler Objects of Interest & False positive rate \\ 
 & Temperature * & Mass * & Periodate * & Star \\ 
 & Statistical error & Temperature * & Frequency * & Eclipsing binary \\ 
 & Field * & Probability * & Signal to noise ratio & Eclipses \\ 
 & Optics * & Units * & False positive rate & Signal to noise ratio \\ 
 & Mass * & Frequency * & Eclipsing binary & Neptune \\ 
 & Frequency * & Periodate * & Eclipses & Earth-like planet \\ 
 & Fluctuation * & Velocity * & Orbit * & Stellar classification \\ 
 \hline 
 \texttt{1308.0321} & 
\multicolumn{4}{p{12cm}|}{\texttt{Realization of the Hofstadter Hamiltonian with ultracold atoms in optical lattices}} \\ 
 \cline{2-5} 
 & Experimental data * & Energy * & Atom * & Atom * \\ 
 & Intensity & Measurement * & Magnetic field * & Magnetic field * \\ 
 & Strong interactions & Field * & Potential * & Optical lattice \\ 
 & Field * & Potential * & Measurement * & Cyclotron \\ 
 & Optics * & Mass * & Optical lattice & Ultracold atom \\ 
 & Energy * & Particles * & Hamiltonian & Spin Quantum Hall Effect \\ 
 & Mass * & Units * & Spin * & Band mapping \\ 
 & Wavelength * & Electron * & Orbit * & Time-reversal symmetry \\ 
 & Frequency * & Frequency * & Cyclotron & Hamiltonian \\ 
 & Factorisation & Periodate * & Energy * & Superlattice \\ 
 \hline 
 \texttt{1301.1340} & 
\multicolumn{4}{l|}{\texttt{Neutrino Mass and Mixing with Discrete Symmetry}} \\ 
 \cline{2-5} 
 & Order of magnitude * & Energy * & Symmetry * & Neutrino mass \\ 
 & Experimental data * & Measurement * & Neutrino mass & Neutrino \\ 
 & Weak interaction & Field * & Mass * & Charged lepton \\ 
 & Vacuum * & Potential * & Neutrino & Symmetry * \\ 
 & Field * & Mass * & Charged lepton & Sterile neutrino \\ 
 & Electromagnet * & Particles * & Field * & See-saw \\ 
 & Energy * & Probability * & Leptons * & Leptons * \\ 
 & Mass * & Units * & Sterile neutrino & Mixing patterns \\ 
 & Equation of motion * & Vector * & Vacuum expectation value & Grand unification theory \\ 
 & Momentum * & Electron * & See-saw & Vacuum expectation value \\ 
 \hline 
 \texttt{1301.0319} & 
\multicolumn{4}{p{12cm}|}{\texttt{Modules for Experiments in Stellar Astrophysics (MESA):\newline Giant Planets, Oscillations, Rotation, and Massive Stars }} \\ 
 \cline{2-5} 
 & Order of magnitude * & Energy * & Mass * & Star \\ 
 & Stellar physics & Measurement * & Star & White dwarf \\ 
 & Right Hand Side\newline of the exression * & Field * & Frequency * & Massive stars \\ 
 & Regularization & Potential * & White dwarf & Stellar evolution \\ 
 & Intensity & Mass * & Angular momentum * & Angular momentum * \\ 
 & Temperature * & Particles * & Massive stars & Mass * \\ 
 & Field * & Temperature * & Temperature * & Planet \\ 
 & Optics * & Probability * & Pressure * & Red supergiant \\ 
 & Energy * & Units * & Luminosity & Asteroseismology \\ 
 & Mass * & Electron * & Stellar evolution & Zero-age main\newline sequence stars \\ 
 \hline 
 \texttt{1304.6875} & 
\multicolumn{4}{l|}{\texttt{A Massive Pulsar in a Compact Relativistic Binary}} \\ 
 \cline{2-5} 
 & Order of magnitude * & Energy * & Mass * & White dwarf \\ 
 & Stellar physics & Measurement * & White dwarf & Neutron star \\ 
 & Solar mass & Field * & Orbit * & Pulsar \\ 
 & Temperature * & Potential * & Neutron star & Orbit * \\ 
 & Statistical error & Mass * & Pulsar & Gravitational wave \\ 
 & Degree of freedom & Particles * & General relativity & General relativity \\ 
 & Field * & Temperature * & Gravitational wave & Companion \\ 
 & Optics * & Probability * & Star & Mass * \\ 
 & Energy * & Units * & Gravitation * & Low-mass X-ray binary \\ 
 & Mass * & Vector * & Companion & Binary star \\ 
 \hline 
 \texttt{1306.2314} & 
\multicolumn{4}{p{12cm}|}{\texttt{Warm Dark Matter as a solution to the small scale crisis:\newline new constraints from high redshift Lyman-alpha forest data}} \\ 
 \cline{2-5} 
 & Astrophysics and cosmology * & Measurement * & Simulations * & WDM particles \\ 
 & Numerical simulation & Mass * & Resolution * & Simulations * \\ 
 & Regularization & Particles * & Cold dark matter & Cold dark matter \\ 
 & Intensity & Temperature * & Temperature * & Intergalactic medium \\ 
 & Temperature * & Probability * & Intergalactic medium & Mean transmitted flux \\ 
 & Statistical error & Universe * & Quasar & Ultraviolet background \\ 
 & Degree of freedom & Velocity * & WDM particles & Quasar \\ 
 & Optics * & Objective * & Wavenumber * & Free streaming \\ 
 & Mass * & Formate * & Free streaming & Redshift bins \\ 
 & Fluctuation * & Optics * & Matter power spectrum & Matter power spectrum \\ 
 \hline 
 \texttt{1311.6806} & 
\multicolumn{4}{l|}{\texttt{Prevalence of Earth-size planets orbiting Sun-like stars}} \\ 
 \cline{2-5} 
 & Stefan-Boltzmann constant & Energy * & Signal to noise ratio & Kepler Objects of Interest \\ 
 & Solar mass & Measurement * & Kepler Objects of Interest & Signal to noise ratio \\ 
 & Intensity & Potential * & Light curve & Light curve \\ 
 & Temperature * & Mass * & Photometry & Habitable zone \\ 
 & Statistical error & Temperature * & Eclipses & Photometry \\ 
 & Energy * & Probability * & Stellar radii & Stellar radii \\ 
 & Mass * & Periodate * & Extrasolar planet & Eclipses \\ 
 & Wavelength * & Universe * & Habitable zone & Eclipsing binary \\ 
 & Fluctuation * & Objective * & Eclipsing binary & Extrasolar planet \\ 
 & Uniform distribution & Statistics & Event * & High resolution\newline \'echelle spectrometer \\ 
 \hline 
 \texttt{1302.5433} & 
\multicolumn{4}{p{12cm}|}{\texttt{Majorana Fermions in Semiconductor Nanowires:\newline Fundamentals, Modeling, and Experiment}} \\ 
 \cline{2-5} 
 & Order of magnitude * & Energy * & Majorana fermion & Majorana fermion \\ 
 & Bohr magneton & Measurement * & Energy * & Nanowire \\ 
 & Experimental data * & Field * & Nanowire & Majorana bound state \\ 
 & Right Hand Side\newline of the exression * & Potential * & Superconductor & Superconductor \\ 
 & Critical value & Mass * & Topology * & Semiconductor \\ 
 & Regularization & Particles * & Field * & Josephson effect \\ 
 & Temperature * & Temperature * & Semiconductor & Topology * \\ 
 & Expectation Value & Probability * & Hamiltonian & Superconductivity \\ 
 & Degree of freedom & Units * & Superconductivity & Topological\newline superconductor \\ 
 & Field * & Vector * & Measurement * & Heterostructure \\ 
 \hline 
 \texttt{1303.3572} & 
\multicolumn{4}{p{12cm}|}{\texttt{3-dimensional bosonic topological insulators and its exotic electromagnetic response}} \\ 
 \cline{2-5} 
 & Right Hand Side\newline of the exression * & Energy * & Bosonization & Dyon \\ 
 & Regularization & Field * & Dyon & Electromagnetism \\ 
 & Strong interactions & Potential * & Charge * & U(1) * \\ 
 & Degree of freedom & Mass * & Condensation & Witten effect \\ 
 & Vacuum * & Particles * & Fermion * & Bosonization \\ 
 & Field * & Units * & Statistics & Condensation \\ 
 & Electromagnet * & Vector * & U(1) * & Projective construction \\ 
 & Energy * & Periodate * & Electromagnetism & Time-reversal symmetry \\ 
 & Mass * & Symmetry * & Symmetry * & Fermion * \\ 
 & Fluctuation * & Statistics & Time-reversal symmetry & Mean field \\ 
 \hline 
\end{longtable}} 
\end{center}
Tab.~\ref{stab:concepts_ranked_in_papers_various_measures} shows how the generality of a concept depends on the criterion used to rank it. It is also worth to see how the selective removal of concepts reverberates on the rankings. To this aim, we report in Tab.~\ref{stab:concepts_ranked_in_papers_residual_entropy_percentiles} the ten most generic concepts as a function of the filtering intensity $p$ going from the original set ($p = 0\%$) to the optimal level ($p_{opt} = 30\%$) computed using the fragmentation criterion presented in Sec.~\ref{s_ssec:optimal_filtering}. At first glance, we observe how increasing the aggressiveness of the filter produces an immediate decrease of the number of concepts marked as common by SW. More importantly, we clearly see how the pruning removes also concepts classifiable as generic that have not been marked as such by SW.


%
%
%
\begin{center}
{\tiny
\newcolumntype{a}{>{\columncolor{orange!40!white}}p{4.1cm}}
%
\begin{longtable}{|p{1.9cm}|p{4.0cm}|p{3.4cm}|p{4.0cm}|a|}
\caption{List of the ten most generic concepts per paper as a function of the entropic filtering intensity $p$. $p=0$ denotes the original set, while $p_{opt}$ corresponds to the optimal level of filtering. Concepts indicated as common by ScienceWISE are marked by an asterisk.}
\label{stab:concepts_ranked_in_papers_residual_entropy_percentiles}\\
\hline 
\multicolumn{1}{|p{1.9cm}}{\texttt{arXiv} \textsc{ID}} & \multicolumn{1}{|p{4.0cm}}{$ p = 0 $} & \multicolumn{1}{|p{3.4cm}}{$ p = 10 $} & \multicolumn{1}{|p{4.0cm}}{$ p = 20 $} & \multicolumn{1}{|a|}{$ p_{opt} = 30 $} \\

\endfirsthead

\hline
\multicolumn{5}{|c|}{{\bfseries continued from previous page}} \\ \hline 
\multicolumn{1}{|p{1.9cm}}{\texttt{arXiv} \textsc{ID}} & \multicolumn{1}{|p{4.0cm}}{$ p = 0 $} & \multicolumn{1}{|p{3.4cm}}{$ p = 10 $} & \multicolumn{1}{|p{4.0cm}}{$ p = 20 $} & \multicolumn{1}{|a|}{$ p_{opt} = 30 $} \\ \hline
\endhead

\hline \multicolumn{5}{|c|}{{\bfseries Continued on next page}} \\ \hline
\endfoot

\hline \hline
\endlastfoot

\hline
%
%
%
 \texttt{1306.5856} & 
\multicolumn{4}{l|}{\texttt{Raman spectroscopy as a versatile tool for studying the properties of graphene}} \\ 
 \cline{2-5} 
 & Experimental data * & Electronic transition & Electron hole pair & Monochromator \\ 
 & Regularization & Irradiance & Topological insulator & Surface plasmon resonance \\ 
 & Intensity & Group velocity * & Thermal Expansion & Bilayer graphene \\ 
 & Temperature * & Reciprocal lattice & Transistors & Graphene layer \\ 
 & Field * & Diffraction * & Backscattering & Superlattice \\ 
 & Optics * & Nanostructure & Scanning tunneling\newline microscope & Van Hove singularity \\ 
 & Electromagnet * & Hydrostatics & Graphite & Surface plasmon \\ 
 & Energy * & Electron scattering & Nitriding & Exciton \\ 
 & Mass * & Circular polarization * & Dirac point & Nanomaterials \\ 
 & Wavelength * & Space-time singularity & Normal mode & Intervalley scattering \\ 
 \hline 
 \texttt{1301.0842} & 
\multicolumn{4}{l|}{\texttt{The false positive rate of Kepler and the occurrence of planets }} \\ 
 \cline{2-5} 
 & Order of magnitude * & Planet formation & Stellar classification & Luminosity class \\ 
 & Numerical simulation & Near-infrared & Early-type star & Eclipsing binary \\ 
 & Space telescopes & Error function & Probability density function * & Matched filter \\ 
 & Temperature * & Companion & Companion stars & Asteroseismology \\ 
 & Statistical error & Spectrographs & Star counts & High accuracy radial velocity\newline planetary search \\ 
 & Field * & Angular distance & Earth-like planet & Hot Jupiter \\ 
 & Optics * & Stellar magnitude & Orbital elements & Triple system \\ 
 & Mass * & Extinction & Eccentricity & Neptune \\ 
 & Frequency * & Kolmogorov-Smirnov test & Primary stars & Periastron \\ 
 & Fluctuation * & Solar neighborhood & Giant planet & Planet \\ 
 \hline 
 \texttt{1308.0321} & 
\multicolumn{4}{l|}{\texttt{Realization of the Hofstadter Hamiltonian with ultracold atoms in optical lattices}} \\ 
 \cline{2-5} 
 & Experimental data * & Atomic number * & Coriolis force * & Landau-Zener transition \\ 
 & Intensity & Helicity & Topological insulator & Chern number \\ 
 & Strong interactions & Quantum Hall Effect & Mott insulator & Superlattice \\ 
 & Field * & SU(2) * & Cyclotron & Magnetic trap \\ 
 & Optics * & Freezing & Edge excitations & Spin Hall effect \\ 
 & Energy * & Lorentz force * & Topological order & Spin Quantum Hall Effect \\ 
 & Mass * & Spontaneous emission & Berry phase & Quadrupole magnet \\ 
 & Wavelength * & Optical lattice & Fractal & Band mapping \\ 
 & Frequency * & Quadrupole & Landau-Zener transition & Lowest Landau Level \\ 
 & Factorisation & Bose-Einstein condensate & Chern number & Hofstadter's butterfly \\ 
 \hline 

 \texttt{1301.1340} & 
\multicolumn{4}{l|}{\texttt{Neutrino Mass and Mixing with Discrete Symmetry}} \\ 
 \cline{2-5} 
 & Order of magnitude * & Neutron * & Zenith & Flavour physics \\ 
 & Experimental data * & Antisymmetrizer & Supersymmetry breaking & Atmospheric neutrino \\ 
 & Weak interaction & Mass spectrum & Upper atmosphere & Infinite group \\ 
 & Vacuum * & Supersymmetry & Weak neutral current \newline interaction & Clebsch-Gordan\newline coefficients \\ 
 & Field * & Baryon number & Renormalisation group \newline equations & Neutrino telescope \\ 
 & Electromagnet * & Subgroup & CP violation & CP violating phase \\ 
 & Energy * & Permutation & Euler angles & Proton decay \\ 
 & Mass * & Quark mass & Rotation group * & Neutrino mixing angle \\ 
 & Equation of motion * & Irreducible representation & Superpotential & Complex conjugate\newline representation \\ 
 & Momentum * & Embedding & Reactor neutrino experiments & Neutralino \\ 
 \hline 
 \texttt{1301.0319} & 
\multicolumn{4}{p{12cm}|}{\texttt{Modules for Experiments in Stellar Astrophysics (MESA):\newline Giant Planets, Oscillations, Rotation, and Massive Stars}} \\ 
 \cline{2-5} 
 & Order of magnitude * & Planet formation & Diffusion equation & Complete mixing \\ 
 & Stellar physics & Accretion & Gravitational energy & Kelvin-Helmholtz timescale \\ 
 & Right Hand Side \newline of the exression * & Low-mass stars & Circumstellar envelope & Radiative Diffusion \\ 
 & Regularization & Massive stars & Early-type star & Optical bursts \\ 
 & Intensity & Diffusion coefficient & Helium shell flashes & Asteroseismology \\ 
 & Temperature * & Accretion disk & Modified gravity & Stellar oscillations \\ 
 & Field * & Irradiance & Evolved stars & Zero-age main\newline sequence stars \\ 
 & Optics * & Sloan Digital Sky Survey & Neutron star & Classical nova \\ 
 & Energy * & Viscosity & Hertzsprung-Russell diagram & Large Synoptic\newline Survey Telescope \\ 
 & Mass * & Hydrostatics & Supernova & Giant branches \\ 
 \hline 
 \texttt{1304.6875} & 
\multicolumn{4}{l|}{\texttt{A Massive Pulsar in a Compact Relativistic Binary}} \\ 
 \cline{2-5} 
 & Order of magnitude * & Accretion & Radio telescope & Lunar Laser Ranging\newline experiment \\ 
 & Stellar physics & Massive stars & Moment of inertia * & Mass discrepancy \\ 
 & Solar mass & Black hole & Mass function & Matched filter \\ 
 & Temperature * & Irradiance & Comparison stars & Zero-age main\newline sequence stars \\ 
 & Statistical error & Sloan Digital Sky Survey & Circumstellar envelope & Grism \\ 
 & Degree of freedom & Cooling & Probability density function * & Radio pulsar \\ 
 & Field * & Companion & Companion stars & Laser Interferometer \newline Gravitational-Wave\newline Observatory \\ 
 & Optics * & Spectrographs & Roche Lobe & Radiation damping \\ 
 & Energy * & Space-time singularity & Mass accretion rate & Barycenter \\ 
 & Mass * & Stellar surfaces & Peculiar velocity & Space velocity \\ 
 \hline 
 \texttt{1306.2314} & 
\multicolumn{4}{p{12cm}|}{\texttt{Warm Dark Matter as a solution to the small scale crisis:\newline new constraints from high redshift Lyman-alpha forest data}} \\ 
 \cline{2-5} 
 & Astrophysics and cosmology * & Simulations * & Dark matter particle & Nuisance parameter \\ 
 & Numerical simulation & Cutoff scale & Mass function & Satellite galaxy \\ 
 & Regularization & Mean transmitted flux & Matter power spectrum & Free streaming \\ 
 & Intensity & Sloan Digital Sky Survey & Luminosity function & Quasar \\ 
 & Temperature * & Cooling & A dwarfs & Active Galactic Nuclei \\ 
 & Statistical error & Spectrographs & Supernova & Planck data \\ 
 & Degree of freedom & Dark matter & Tellurate & Halo finding algorithms \\ 
 & Optics * & Wavenumber * & Monte Carlo Markov chain & Baryon acoustic oscillations \\ 
 & Mass * & Absorption feature & Cold dark matter & Void \\ 
 & Fluctuation * & Flavour & Stellar feedback & Strong gravitational lensing \\ 
 \hline 
 \texttt{1311.6806} & 
\multicolumn{4}{l|}{\texttt{Prevalence of Earth-size planets orbiting Sun-like stars}} \\ 
 \cline{2-5} 
 & Stefan-Boltzmann constant & Simulations * & Hertzsprung-Russell diagram & Eclipsing binary \\ 
 & Solar mass & Companion & Earth-like planet & Asteroseismology \\ 
 & Intensity & Stellar surfaces & Monte Carlo Markov chain & Limb darkening \\ 
 & Temperature * & Stellar magnitude & Hydrogen 21 cm line & Planet \\ 
 & Statistical error & Host star & Keck Array & High resolution \newline échelle spectrometer \\ 
 & Energy * & Orbit Eccentricity & Parallax & Eclipses \\ 
 & Mass * & Droplet * & Eclipsing binary & Habitable zone \\ 
 & Wavelength * & Angular separation & Asteroseismology & Gaussian process \\ 
 & Fluctuation * & Teams * & Limb darkening & Mars \\ 
 & Uniform distribution & Galactic structure & Planet & Orange dwarf \\ 
 \hline 
 \texttt{1302.5433} & 
\multicolumn{4}{p{12cm}|}{\texttt{Majorana Fermions in Semiconductor Nanowires:\newline Fundamentals, Modeling, and Experiment}} \\ 
 \cline{2-5} 
 & Order of magnitude * & Tight-binding model & Second quantization & P-wave \\ 
 & Bohr magneton & Quantum dots & Feshbach resonance & Quantum decoherence \\ 
 & Experimental data * & Neutron * & Zero mode & Nanowire \\ 
 & Right Hand Side \newline of the exression * & Rest mass * & Topological insulator & Chern number \\ 
 & Critical value & Nanostructure & Proximity effect & Local density of states \\ 
 & Regularization & Winding number & Networks * & Topological\newline superconductor \\ 
 & Temperature * & Helicity & Critical current & Andreev reflection \\ 
 & Expectation Value & Quantum Hall Effect & Scaling limit & Josephson effect \\ 
 & Degree of freedom & Coarse graining & Pair potential & Weak antilocalization \\ 
 & Field * & Chiral symmetry & Quantum critical point & Non-Abelian statistics \\ 
 \hline 
 \texttt{1303.3572} & 
\multicolumn{4}{p{12cm}|}{\texttt{3-dimensional bosonic topological insulators and its exotic electromagnetic response}} \\ 
 \cline{2-5} 
 & Right Hand Side \newline of the exression * & Dirac fermion & Band insulator & Hall conductance \\ 
 & Regularization & Quantum Hall Effect & Topological insulator & Electric magnetic \\ 
 & Strong interactions & SU(2) * & Mott insulator & Long-range entanglement \\ 
 & Degree of freedom & Effective field theory & Charge conservation & Topological field theory \\ 
 & Vacuum * & Parton & Magnetic monopole & Axion \\ 
 & Field * & Deconfinement & Edge excitations & Exciton \\ 
 & Electromagnet * & Screening effect & Topological order & Short-range entanglement \\ 
 & Energy * & Effective Lagrangian & Berry phase & Symmetry protected \newline topological order \\ 
 & Mass * & Directional derivative & Fractional charge & Group cohomology \\ 
 & Fluctuation * & Global symmetry & Electromagnetism & Charge quantization \\ 
 \hline 
\end{longtable}} 
\end{center}

The information presented in Tab.~\ref{stab:concepts_ranked_in_papers_residual_entropy_percentiles} confirms the power of our filtering methodology. In analogy to what we have done in Tab.~\ref{stab:concepts_ranked_in_papers_various_measures}, we check if $S_d$ still outperforms other rankings also in the filtered networks. For this reason, in Tab.~\ref{stab:concepts_ranked_in_papers_various_measures_optimal_filtering} we report the rankings of the concepts at the optimal level of filtering ($p_{opt} = 30\%$). A quick glance at its columns tells us that, albeit being more specific, concepts ranked using $S_d$ are still capable of describing the content of the document, outperforming measures like \textsc{TF} and \textsc{TF-IDF}.

%
%
%
\begin{center}
{\tiny
\newcolumntype{a}{>{\columncolor{orange!40!white}}p{4.0cm}}
\newcolumntype{b}{>{\columncolor{cyan!40!white}}p{3.2cm}}
\begin{longtable}{|p{1.9cm}|a|p{3.8cm}|p{3.9cm}|b|}
\caption{Ten most generic concepts ranked upon different indices among the set of concepts available at the $S_d$ optimal level of filtering ($p_{opt} = 30\%$). Columns are the same as Tab.~\ref{stab:concepts_ranked_in_papers_various_measures}. We highlight the columns of $S_d$ (best ranking) and \textsc{TF-IDF} (standard ranking).}

\label{stab:concepts_ranked_in_papers_various_measures_optimal_filtering}\\
\hline
\multicolumn{1}{|p{1.9cm}}{\texttt{arXiv} \textsc{ID}} & \multicolumn{1}{|a|}{$ S_d (p_{opt} = 30) $} & \multicolumn{1}{|p{3.8cm}}{{\small \textsc{IDF}} $ (p_{opt} = 30) $} & \multicolumn{1}{|p{3.9cm}}{{\small \textsc{TF}} $ (p_{opt} = 30) $} & \multicolumn{1}{|b|}{{\small \textsc{TF-IDF}} $ (p_{opt} = 30) $} \\

\endfirsthead

\hline
\multicolumn{5}{|c|}{{\bfseries continued from previous page}} \\ \hline 
\multicolumn{1}{|p{1.9cm}}{\texttt{arXiv} \textsc{ID}} & \multicolumn{1}{|a|}{$ S_d (p_{opt} = 30) $} & \multicolumn{1}{|p{3.8cm}}{{\small \textsc{IDF}} $ (p_{opt} = 30) $} & \multicolumn{1}{|p{3.9cm}}{{\small \textsc{TF}} $ (p_{opt} = 30) $} & \multicolumn{1}{|b|}{{\small \textsc{TF-IDF}} $ (p_{opt} = 30) $} \\ \hline
\endhead

\hline \multicolumn{5}{|c|}{{\bfseries Continued on next page}} \\ \hline
\endfoot

\hline \hline
\endlastfoot

\hline

 \texttt{1306.5856} & 
\multicolumn{4}{p{12cm}|}{\texttt{Raman spectroscopy as a versatile tool for studying the properties of graphene}} \\ 
 \cline{2-5} 
 & Monochromator & Exciton & Surface enhanced \newline Raman spectroscopy & Surface enhanced \newline Raman spectroscopy \\ 
 & Surface plasmon resonance & Superlattice & Van Hove singularity & Kohn anomaly \\ 
 & Bilayer graphene & Graphene layer & Grüneisen parameter & Grüneisen parameter \\ 
 & Graphene layer & Bilayer graphene & Kohn anomaly & Van Hove singularity \\ 
 & Superlattice & Surface plasmon & Hexagonal boron nitride & Hexagonal boron nitride \\ 
 & Van Hove singularity & Monochromator & Nanocrystalline & Nanocrystalline \\ 
 & Surface plasmon & Van Hove singularity & Graphene layer & Graphene layer \\ 
 & Exciton & Nanocrystal & Exciton & Fullerene \\ 
 & Nanomaterials & S-process & Nanocrystal & Nanocrystal \\ 
 & Intervalley scattering & Fullerene & Fullerene & Depolarization ratio \\ 
 \hline 
 \texttt{1301.0842} & 
\multicolumn{4}{p{12cm}|}{\texttt{The false positive rate of Kepler and the occurrence of planets}} \\ 
 \cline{2-5} 
 & Luminosity class & Planet & Planet & Planet \\ 
 & Eclipsing binary & White dwarf & Kepler Objects of Interest & Kepler Objects of Interest \\ 
 & Matched filter & Eclipses & False positive rate & False positive rate \\ 
 & Asteroseismology & M dwarfs & Eclipses & Eclipsing binary \\ 
 & High accuracy radial velocity\newline planetary search & Periastron & Eclipsing binary & Eclipses \\ 
 & Hot Jupiter & Eclipsing binary & Neptune & Neptune \\ 
 & Triple system & Hot Jupiter & Super-earth & Super-earth \\ 
 & Neptune & Neptune & Triple system & Triple system \\ 
 & Periastron & Asteroseismology & White dwarf & Logarithmic distribution \\ 
 & Planet & Super-earth & Logarithmic distribution & White dwarf \\ 
 \hline 
 \texttt{1308.0321} & 
\multicolumn{4}{p{12cm}|}{\texttt{Realization of the Hofstadter Hamiltonian with ultracold atoms in optical lattices}} \\ 
 \cline{2-5} 
 & Landau-Zener transition & Superlattice & Spin Quantum Hall Effect & Spin Quantum Hall Effect \\ 
 & Chern number & Chern number & Superlattice & Band mapping \\ 
 & Superlattice & Spin Hall effect & Band mapping & Superlattice \\ 
 & Magnetic trap & Lowest Landau Level & Landau-Zener transition & Landau-Zener transition \\ 
 & Spin Hall effect & Spin Quantum Hall Effect & Spin Hall effect & Spin Hall effect \\ 
 & Spin Quantum Hall Effect & Magnetic trap & Magnetic trap & Quadrupole magnet \\ 
 & Quadrupole magnet & Landau-Zener transition & Hofstadter's butterfly & Hofstadter's butterfly \\ 
 & Band mapping & Hofstadter's butterfly & Chern number & Magnetic trap \\ 
 & Lowest Landau Level & Quadrupole magnet & Lowest Landau Level & Lowest Landau Level \\ 
 & Hofstadter's butterfly & Band mapping & Quadrupole magnet & Chern number \\ 
 \hline 
 \texttt{1301.1340} & 
\multicolumn{4}{l|}{\texttt{Neutrino Mass and Mixing with Discrete Symmetry}} \\ 
 \cline{2-5} 
 & Flavour physics & Gamma ray burst & Mixing patterns & Mixing patterns \\ 
 & Atmospheric neutrino & Superfield & Solar neutrino & Tri Bimaximal mixing \\ 
 & Infinite group & Neutralino & Reactor Experiment \newline for Neutrino Oscillation & Solar neutrino \\ 
 & Clebsch-Gordan\newline coefficients & Mantle & Tri Bimaximal mixing & Reactor Experiment \newline for Neutrino Oscillation \\ 
 & Neutrino telescope & Two Higgs Doublet Model & Clebsch-Gordan\newline coefficients & Trimaximal mixing \\ 
 & CP violating phase & Supermultiplet & Super-Kamiokande & SNO+ \\ 
 & Proton decay & CP violating phase & SNO+ & Super-Kamiokande \\ 
 & Neutrino mixing angle & Atmospheric neutrino & Atmospheric neutrino & Clebsch-Gordan\newline coefficients \\ 
 & Complex conjugate\newline representation & Proton decay & Trimaximal mixing & Mikheev-Smirnov-Wolfenstein effect \\ 
 & Neutralino & Massive neutrino & Type I seesaw & Cabibbo Angle \\ 
 \hline 
 \texttt{1301.0319} & 
\multicolumn{4}{p{12cm}|}{\texttt{Modules for Experiments in Stellar Astrophysics (MESA):\newline Giant Planets, Oscillations, Rotation, and Massive Stars}} \\ 
 \cline{2-5} 
 & Complete mixing & Planet & White dwarf & White dwarf \\ 
 & Kelvin-Helmholtz timescale & White dwarf & Planet & Planet \\ 
 & Radiative Diffusion & Gamma ray burst & Zero-age main\newline sequence stars & Red supergiant \\ 
 & Optical bursts & Optical bursts & Red supergiant & Asteroseismology \\ 
 & Asteroseismology & Pre-main-sequence star & Asteroseismology & Zero-age main\newline sequence stars \\ 
 & Stellar oscillations & Asymptotic giant branch & Pre-main-sequence star & Pre-main-sequence star \\ 
 & Zero-age main\newline sequence stars & Zero-age main\newline sequence stars & Asymptotic giant branch & Stellar oscillations \\ 
 & Classical nova & Large Synoptic\newline Survey Telescope & Gamma ray burst & Asymptotic giant branch \\ 
 & Large Synoptic\newline Survey Telescope & Wolf-Rayet star & Stellar oscillations & Gamma ray burst \\ 
 & Giant branches & Asteroseismology & Optical bursts & Classical nova \\ 
 \hline 
 \texttt{1304.6875} & 
\multicolumn{4}{l|}{\texttt{A Massive Pulsar in a Compact Relativistic Binary}} \\ 
 \cline{2-5} 
 & Lunar Laser Ranging\newline experiment & Planet & White dwarf & White dwarf \\ 
 & Mass discrepancy & Pulsar & Pulsar & Pulsar \\ 
 & Matched filter & White dwarf & Low-mass X-ray binary & Low-mass X-ray binary \\ 
 & Zero-age main\newline sequence stars & Albedo & Binary pulsar & Binary pulsar \\ 
 & Grism & VLT telescope & Orbital angular \newline momentum of light & Green Bank Telescope \\ 
 & Radio pulsar & Low-mass X-ray binary & Green Bank Telescope & Orbital angular \newline momentum of light \\ 
 & Laser Interferometer \newline Gravitational-Wave\newline Observatory & Zero-age main\newline sequence stars & Zero-age main\newline sequence stars & Zero-age main\newline sequence stars \\ 
 & Radiation damping & Grism & Millisecond pulsar & Solar system barycenter \\ 
 & Barycenter & Laser Interferometer \newline Gravitational-Wave\newline Observatory & Solar system barycenter & Radio pulsar \\ 
 & Space velocity & Millisecond pulsar & VLT telescope & Dispersion measure \\ 
 \hline 
 \texttt{1306.2314} & 
\multicolumn{4}{p{12cm}|}{\texttt{Warm Dark Matter as a solution to the small scale crisis:\newline new constraints from high redshift Lyman-alpha forest data}} \\ 
 \cline{2-5} 
 & Nuisance parameter & Active Galactic Nuclei & Quasar & WDM particles \\ 
 & Satellite galaxy & Quasar & WDM particles & Ultraviolet background \\ 
 & Free streaming & Gamma ray burst & Free streaming & Quasar \\ 
 & Quasar & Void & Ultraviolet background & Free streaming \\ 
 & Active Galactic Nuclei & Baryon acoustic oscillations & Redshift bins & Redshift bins \\ 
 & Planck data & Reionization & Temperature-density relation & Temperature-density relation \\ 
 & Halo finding algorithms & Satellite galaxy & Reionization & Effective optical depth \\ 
 & Baryon acoustic oscillations & Nuisance parameter & Warm dark matter & Warm dark matter \\ 
 & Void & Population III & Effective optical depth & Reionization \\ 
 & Strong gravitational lensing & Free streaming & Nuisance parameter & WDM particle mass \\ 
 \hline 
 \texttt{1311.6806} & 
\multicolumn{4}{p{12cm}|}{\texttt{Prevalence of Earth-size planets orbiting Sun-like stars}} \\ 
 \cline{2-5} 
 & Eclipsing binary & Planet & Kepler Objects of Interest & Kepler Objects of Interest \\ 
 & Asteroseismology & Eclipses & Eclipses & Habitable zone \\ 
 & Limb darkening & Eclipsing binary & Habitable zone & Eclipses \\ 
 & Planet & Limb darkening & Eclipsing binary & Eclipsing binary \\ 
 & High resolution \newline échelle spectrometer & Mars & Planet & High resolution \newline échelle spectrometer \\ 
 & Eclipses & Asteroseismology & Mars & Mars \\ 
 & Habitable zone & Habitable zone & High resolution \newline échelle spectrometer & Horizon Run simulation \\ 
 & Gaussian process & Gaussian process & Limb darkening & Planet \\ 
 & Mars & Ephemerides & False positive rate & False positive rate \\ 
 & Orange dwarf & High resolution \newline échelle spectrometer & Ephemerides & Orange dwarf \\ 
 \hline 
 \texttt{1302.5433} & 
\multicolumn{4}{p{12cm}|}{\texttt{Majorana Fermions in Semiconductor Nanowires:\newline Fundamentals, Modeling, and Experiment}} \\ 
 \cline{2-5} 
 & P-wave & Nanowire & Nanowire & Nanowire \\ 
 & Quantum decoherence & Carbon nanotubes & Majorana bound state & Majorana bound state \\ 
 & Nanowire & P-wave & Josephson effect & Josephson effect \\ 
 & Chern number & Local density of states & Topological\newline superconductor & Topological\newline superconductor \\ 
 & Local density of states & Chern number & P-wave & P-wave \\ 
 & Topological\newline superconductor & Topological\newline superconductor & Local density of states & Local density of states \\ 
 & Andreev reflection & Andreev reflection & Fermion doubling & Fermion doubling \\ 
 & Josephson effect & Josephson effect & Andreev reflection & Moore-Read \newline Pfaffian wavefunction \\ 
 & Weak antilocalization & Weyl fermion & Non-Abelian statistics & Majorana zero mode \\ 
 & Non-Abelian statistics & Non-Abelian statistics & Moore-Read \newline Pfaffian wavefunction & Non-Abelian statistics \\ 
 \hline 
 \texttt{1303.3572} & 
\multicolumn{4}{p{12cm}|}{\texttt{3-dimensional bosonic topological insulators and its exotic electromagnetic response}} \\ 
 \cline{2-5} 
 & Hall conductance & Exciton & Dyon & Dyon \\ 
 & Electric magnetic & Axion & Witten effect & Witten effect \\ 
 & Long-range entanglement & Hall conductance & Projective construction & Projective construction \\ 
 & Topological field theory & Topological field theory & Group cohomology & Group cohomology \\ 
 & Axion & Electric magnetic & Topological field theory & Topological field theory \\ 
 & Exciton & Long-range entanglement & Exciton & Response theory \\ 
 & Short-range entanglement & Symmetry protected \newline topological order & Response theory & Charge quantization \\ 
 & Symmetry protected \newline topological order & Dyon & Charge quantization & Short-range entanglement \\ 
 & Group cohomology & Short-range entanglement & Axion & Symmetry protected \newline topological order \\ 
 & Charge quantization & Charge quantization & Hall conductance & Long-range entanglement \\ 
 \hline 
\end{longtable}} 
\end{center}
%




%
%
%
%

\subsection{Climate change web documents}
\label{s_ssec:climate}

In this section, we present the results obtained for another dataset originated from a collection of web documents on \emph{climate change}. It is worth stressing that, on average, a web document lacks the same structural organization of a scientific manuscript. This is partially due to the fact that such kind of documents convey information in a perspective different than those of a scientific document on the same topic. Web documents have almost no limitations in length, they can simply report facts (news releases) or provide an opinion on a given subject. Moreover, contrary to the case of Physics, the ScienceWISE platform does not have an ontology on climate change. Therefore, we do not have access to the validated \emph{concepts} but, instead, directly to the \emph{keyword} extracted by ScienceWISE machinery.

\subsubsection{Source}
\label{s_sssec:climate_source}

Our collection of web documents has been extracted from the pool of documents available within the ScienceWISE database. More specifically, SW has a collection of tweets on climate change, posted between January and June 2015, that was compiled using Twitter API \cite{twitter-api} through several harvesting campaigns. From the original pool of tweets, we kept only the \emph{original} ones, thus culling mentions, re-tweets and other similar ``non original'' posts. From the original tweets, we kept only those written in English containing at least one URL. Such procedure returns a list of distinct URLs pointing to web documents (approx. 50 millions) ranked according to the number of tweets pointing to a each URL. We then select the top 100,000 most ``tweeted'' URLs. To ensure tematic consistency with the ``climate change'' topic, only documents with at least one of 165 \emph{specific} concepts in their URL are retained.

The above procedure returns a set of 30705 documents. To discard documents excessively short, we decided to consider only those with, at least, $L_{\min} = 500$ words. After thresholding, the collection shrinks to 18770 documents. To extract the keywords from these documents (and the tweets associated with them) we used the KPEX algorithm \cite{constantin-thesis-2014}, since it is natively implemented in the SW platform. The KPEX extraction returns 822601 unique keywords which were stemmed first and then lemmatized, returning a final set of 152871 keywords. Since SW platform does not have a cured ontology on climate change, n-grams extracted from KPEX are simple \textbf{keywords} and thus cannot be called \textbf{concepts} in a strict sense, albeit we will continue to do so in the rest of the manuscript. The main features of the similarity network obtained from this collection are reported in the first row of Tab.~\ref{stab:net_topol_prop_climate}, while the properties of the topics extracted with TM are displayed in Tab.~\ref{stab:stats_topics_climate}.

\subsubsection{Entropic filtering}
\label{s_sssec:entropic_filter_climate}

\begin{figure}[h!]
\centering
\includegraphics[width=0.50\columnwidth]{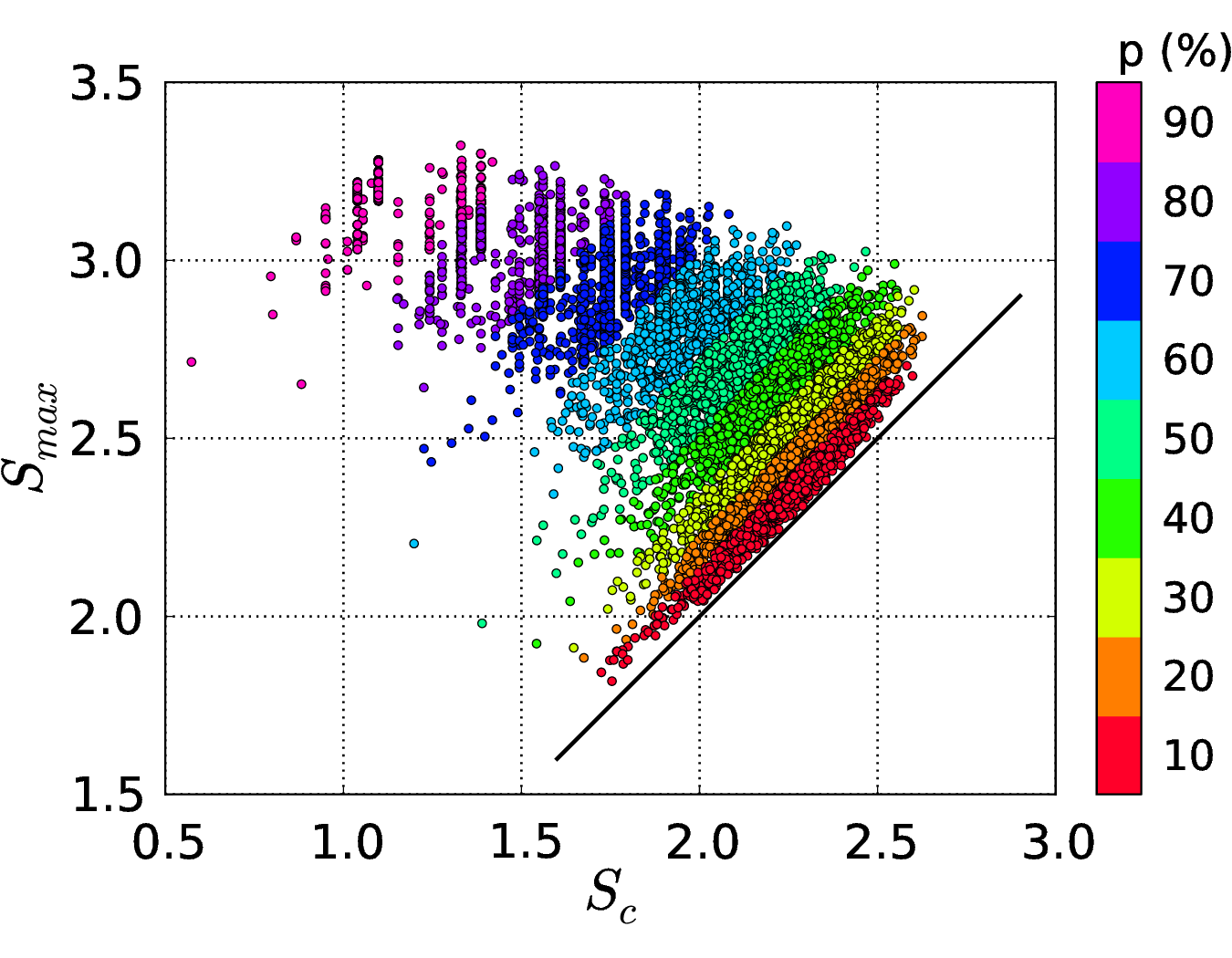}
\caption{Relation between the empirical entropy, $S_c$, and the maximum one, $S_{max}$ for the climate change collection. The colors of the points encode the various percentiles of the residual entropy $S_d$ to which concepts belong to.}
\label{sfig:entropies_rescaled_tf_climate}
\end{figure}

Given the high heterogeneity of the collection in terms of document length $L$, already pointed out in Fig.~\ref{sfig:distro_num_words_both_data}, we decided to use the $tf$ density, $rtf$, instead of its raw value ending in a lognormal maximum entropy distribution (see Sec.~\ref{s_sssec:maxent_rescaled_tf} for details). Moreover, keywords for which $\max(rtf) - \min(rtf) < 0.005$ are ignored, since it is better to discard terms that have similar values of $rtf$. Thus, the number of keywords gets shrunk to 9222. The location of points on the $S_c$, $S_{\max}$ plane is reported in Fig.~\ref{sfig:entropies_rescaled_tf_climate}. As for Physics, we clearly observe a stratification of the residual entropy $S_d$ on the plane confirming the validity of our filtering criterion. Among the generic concepts in the percentile slice $p = 10$ we find ``people,'' ``climate change,'' ``water,'' ``home,'' and ``company.'' On the other hand, among concepts at the percentile slice $p = 50$ we find ``palm,'' ``whale,'' ``Boulder,'' ``metal,'' and ``shop.'' The effects of concepts' selective removal based on $S_d$ on the topological properties of the similarity network and on the topic strucure are reported in Tabs.~\ref{stab:net_topol_prop_climate} and \ref{stab:stats_topics_climate}, respectively.
%
%
%

\subsubsection{Differences between $S_d$ and $IDF$ rankings}
\label{s_sssec:climate_diff_ent_idf}

\begin{figure*}[h!]
\centering
\includegraphics[width=0.5\columnwidth]{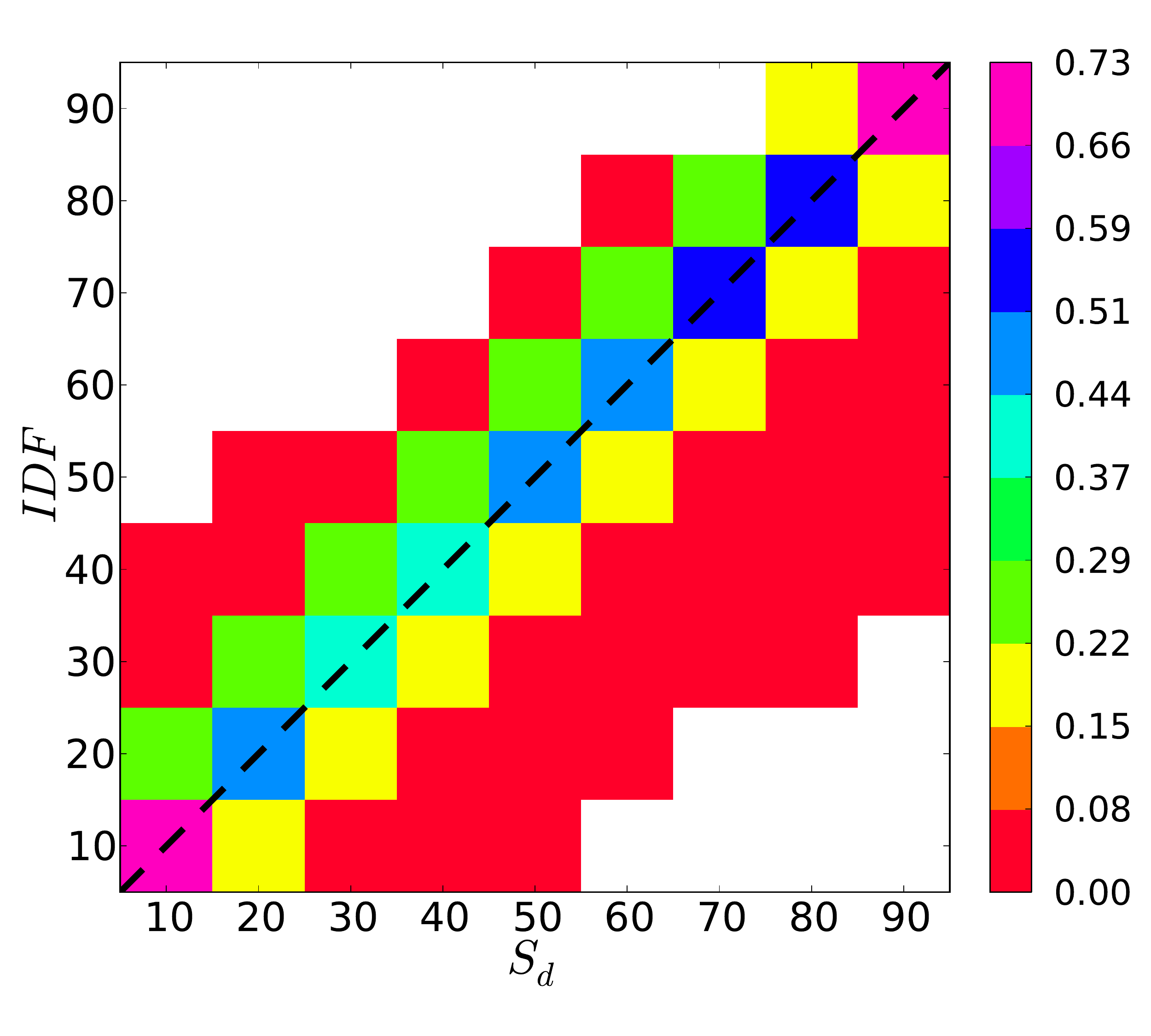}
\caption{Overlap, $O$, between lists of concepts ranked according to residual entropy $S_d$ and inverse document frequency $IDF$ for the climate change collection. The matrix is normalized by row and white entries correspond to absence of overlap. The dashed line indicates the main diagonal.}
\label{sfig:climate_comparison_conc_idf_vs_maxent}
\end{figure*}

Using the same formalism of Sec.~\ref{s_sssec:phys_diff_ent_idf_2collections}, the overlap, $O$, between the list of concepts ranked alternatively using $S_d$ or $IDF$ is shown in Fig.~\ref{sfig:climate_comparison_conc_idf_vs_maxent}. The heatmap presents a narrow peak of $O$ located along the main diagonal. Compared with the Physics collection, the overlap is much higher denoting a much stronger relation between $IDF$ and $S_d$.

\subsubsection{Networks of similarity between documents}
\label{s_sssec:climate_network_similarity}

In Tab.~\ref{stab:net_topol_prop_climate} we report the topological characteristics of the similarity network between climate change documents. As for Physics 2013 collection, the Sankey diagram shown in Fig.~\ref{sfig:sankey_climate_rtf} is useful to understand the progressive specialization of the organization of topics in response to the selective removal of concepts. However, one significative difference captures our attention: the presence of a remarkable condensation phenomenon taking place around the \emph{extreme weather}/\emph{energy storage} communities going from $p=5\%$ to $p = 20\%$. To gain insight on such phenomenon, for each community, $s$, we study the \emph{coverage} $\Gamma_s(\widetilde{\cal{C}})$ of the set, $\widetilde{\cal{C}}$, of top twenty most locally used concepts. The coverage of a set of concepts $\Gamma_s(\widetilde{\cal{C}}) \in [0,1]$ is defined as the union of the sets of documents where those concepts appear divided by the size of number of documents in the community $N_a^s$. Hence, $\Gamma_s(\widetilde{\cal{C}}) = \tfrac{1}{N_a^s} \cup_{c \in \widetilde{\cal{C}}} N_a^s(c)$. The coverage of the community named ``Mixed\_themes'' is $\Gamma = 0.016$ which is pretty small compared to $\Gamma = 0.64$ of ``extreme weather'' community or $\Gamma = 0.87$ of ``ice melting.'' The poor coverage of concepts in community ``Mixed\_themes'' suggests that documents condense into a single community as a result of similarities associated to small groups of concepts weakly linked together. The intimate origin of such condensation, however, is the presence of keywords whose distribution does not resembles a lognormal, as shown in Fig.~\ref{sfig:lognormal_fit_top_10_concepts_uncertain_community_p_20_sankey_climate} for the 10 most frequent ones. Hence, the dissimilarity between the sampled distributions and the lognormal fits is at origin of the feeble ability to describe the content of the webdocs at the community level. Once we get rid of these keywords, the interactions holding together this huge condensed community vanish -- or become weaker, at least -- and the system breaks apart showing communities that address more specific topics.\\
\indent It is worth mentioning that such condensation phenomenon is observed exclusively for topics corresponding to the communities of the similarity networks. The condensation does not take place in the case of topics extracted using TM, where the accretion of the ``Irrelevant topics'' -- shown in Sec.~\ref{s_sssec:topicmapping_climate} -- is consequence of the significance of the topics.

%
%
%
\begin{table*}[h!]
\centering
\setlength{\tabcolsep}{8pt}
\newcolumntype{d}[1]{D{.}{.}{#1}}
\begin{tabular}{@{\extracolsep{\fill}}c|cc d{3} d{3} c d{3} d{3} c}
$p \, (\%)$ & $N_{con}$ & $N_a$ & \multicolumn{1}{c}{$\rho \, (\%)$} & \multicolumn{1}{c}{$\avg{k}$} & $k_{max}$ &  \multicolumn{1}{c}{$\avg{C}$} & \multicolumn{1}{c}{$\avg{l}$} & $M$\\ \hline
%
{\cellcolor{gray!20}} 0 & {\cellcolor{gray!20}} 152871 & {\cellcolor{gray!20}} 18770 & {\cellcolor{gray!20}} 10.111 & {\cellcolor{gray!20}} 1938.624 & {\cellcolor{gray!20}} 11047 & {\cellcolor{gray!20}} 0.399 & {\cellcolor{gray!20}} 1.904 & {\cellcolor{gray!20}} 1  \\
5 & 8760 & 18770 & 9.960 & 1869.425 & 10199 & 0.400 & 1.902 & 1 \\
10 & 8299 & 18762 & 7.610 & 1427.629 & 8677 & 0.480 & 1.936 & 1 \\ 
15 & 7838 & 18743 & 5.351 & 1002.891 & 6789 & 0.569 & 2.003 & 1 \\
20 & 7377 & 18622 & 2.478 & 461.369 & 3863 & 0.658 & 2.221 & 1 \\
25 & 6916 & 18308 & 0.763 & 139.691 & 1362 & 0.308 & 2.565 & 1 \\
30 & 6455 & 17888 & 0.512 & 91.521 & 1160 & 0.302 & 2.771 & 1  \\
40 & 5533 & 16117 & 0.268 & 43.179 & 911 & 0.330 & 3.235 & 14 \\
50 & 4611 & 13527 & 0.157 & 21.206 & 713 & 0.274 & 3.938 & 43 \\ 
60 & 3689 & 10493 & 0.105 & 10.979 & 349 & 0.360 & 5.242 & 147 \\
70 & 2767 & 7318 & 0.088 & 6.415 & 189 & 0.481 & 8.132 & 443 \\ 
80 & 1845 & 4337 & 0.074 & 3.217 & 46 & 0.803 & 10.146 & 925 \\ 
90 & 923 & 1876 & 0.102 & 1.919 & 29 & 0.954 & 1.207 & 744 \\
\end{tabular}
\caption{Topological indicators of the similarity networks between climate change webdocs. The first row ($p=0\%$) corresponds to the original network, while the others to the networks obtained using the maximum entropy filter. In the columns we report: the percentage of filtered concepts $p$, the number of concepts $N_{con}$, the number of web documents containing at least one concept (nodes) $N_a$. the link density $\rho$, the average $\avg{k}$, and maximum degrees $k_{max}$, the average clustering coefficient $\avg{C}$, the average path length $\avg{l}$ and the number of connected components $M$. The minimum edge weight is equal to $w_{min} = 0.01$.}
\label{stab:net_topol_prop_climate}
\end{table*}
%
%
%
%
%
%
\begin{figure*}[h!]
\centering
\includegraphics[width=\columnwidth]{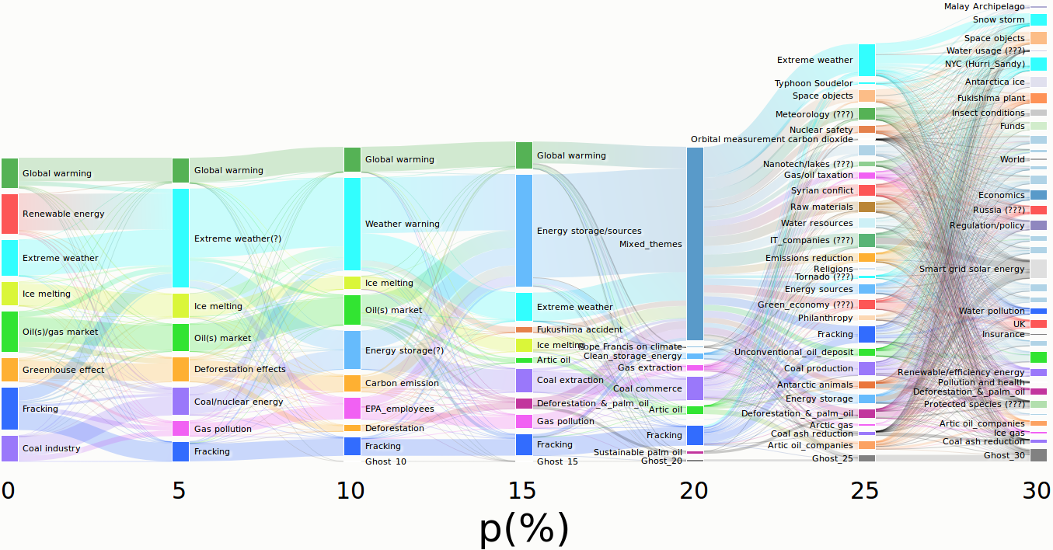}
\caption{Static Sankey diagram of the climate change collection. Each community is represented as a colored box whose height is proportional to the number of web documents it contains. A topic is assigned to each box according to the 10 most used keywords, \ie those appearing in more papers. The thickness of the bands between boxes indicates the number of shared webdocs. Each column denotes a different intensity of filtering $p$. Concepts are pruned according to their residual entropy $S_d$ computed from the $tf$ density, $rtf$. The minimum fluctuation of $rtf$ is equal to 0.005. Interactive version available at \cite{sankey-interactive}.}
\label{sfig:sankey_climate_rtf}
\end{figure*}
%
%
%
%
%
\begin{figure}[h!]
\centering
\includegraphics[width=\columnwidth]{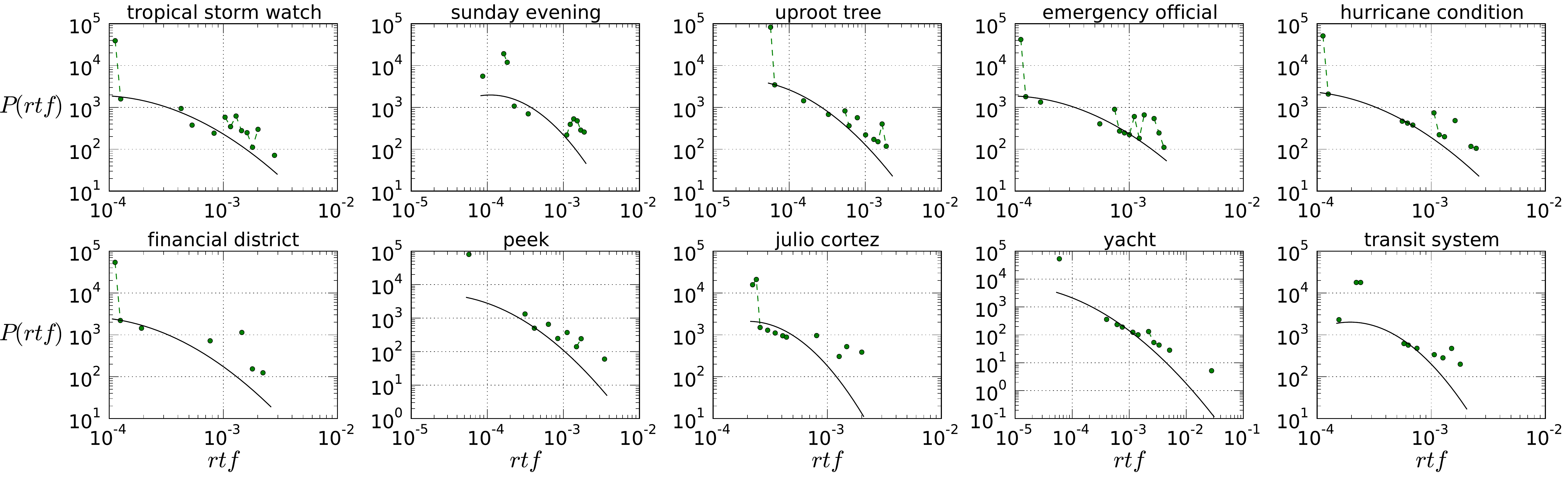}
\caption{Distribution of the top 10 most frequent concepts within the community of uncertain label ``Mixed\_themes'' found at $p=20\%$ in Fig.~\ref{sfig:sankey_climate_rtf}. The lognormal fit of each distribution (see Sec.~\ref{s_sssec:maxent_rescaled_tf}) is plotted with a black line. All the distributions deviate considerably from their lognormal fit, corresponding to high KL distances between the lognormal distribution and the observed one.}
\label{sfig:lognormal_fit_top_10_concepts_uncertain_community_p_20_sankey_climate}
\end{figure}

\pagebreak

\subsubsection{TopicMapping analysis}
\label{s_sssec:topicmapping_climate}

\begin{figure*}[h!]
\centering
\includegraphics[width=\columnwidth]{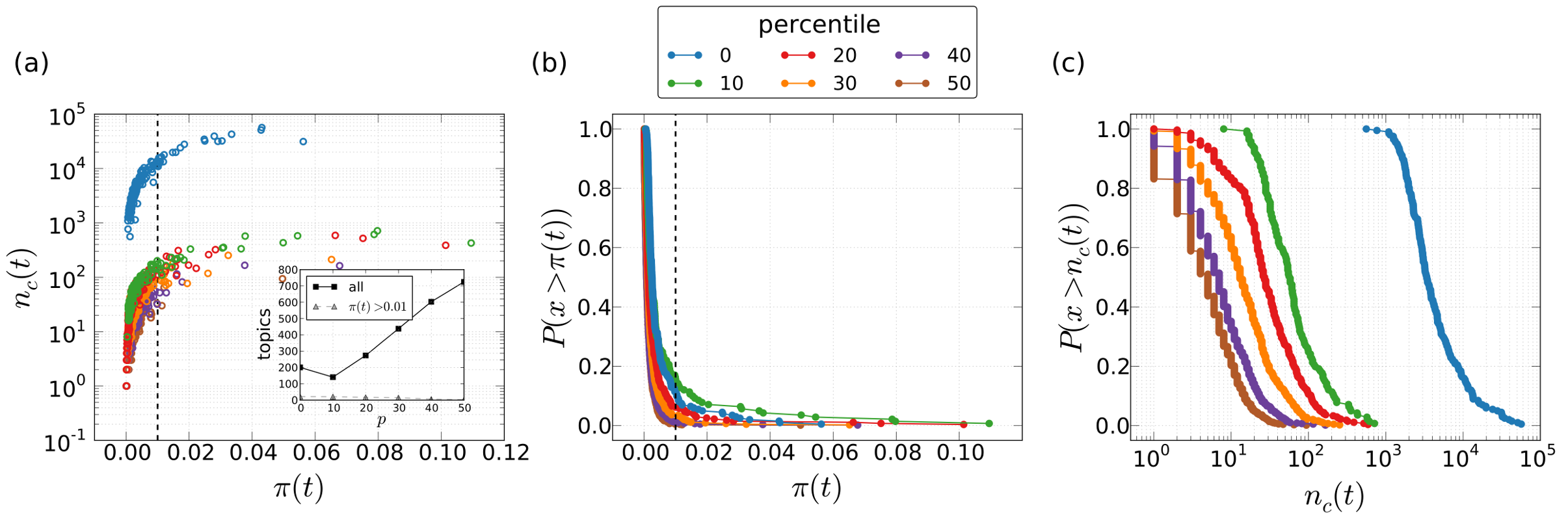}
\caption{Statistics about topics for the climate change dataset. (a) Relation between the number of concepts $n_{w} (t)$ and the probability $\pi(t)$ associated to each topic $t$. Every circle represents a topic whose color denotes the filtering percentile $p$. The dashed vertical line corresponds to the topic probability $\pi(t) = 0.01$ below which topics are not considered meaningful. In the inset, the total number of topics for each percentile (squares) is shown along with the number of meaningful topics with probability $\pi(t) > 0.01$ (triangles). The complementary cumulative distribution functions of the topic probability, $\pi(t)$, and the number of concepts per topic, $n_{w} (t)$, are displayed in panels (b) and (c), respectively. The different colors denote the percentiles, $p$, of filtered concepts.}
\label{sfig:topicmapping_statistics_climate}
\end{figure*}

\begin{figure*}[h!] 
  \centering 
  \includegraphics[width=0.75\columnwidth]{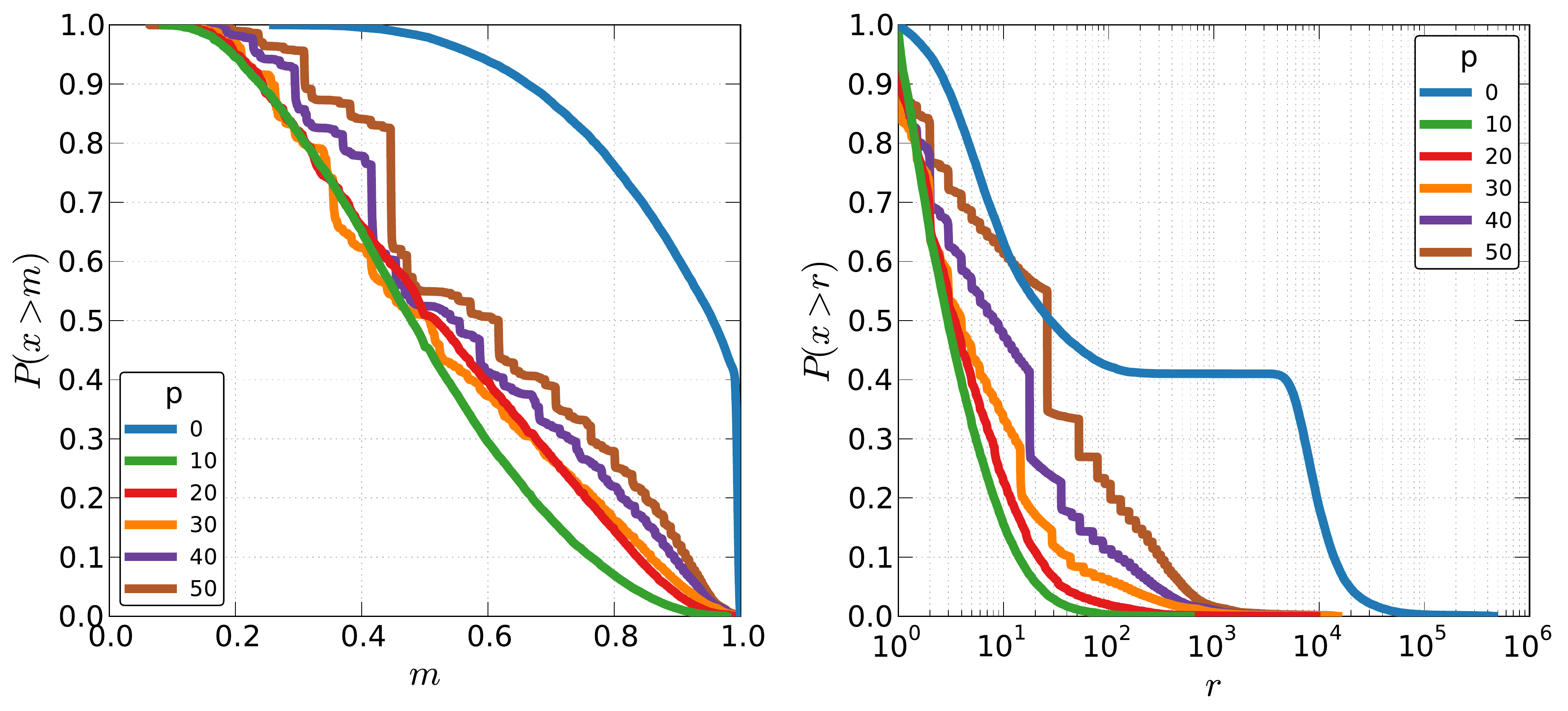} 
  \caption{(Panel a) Complementary cumulative distribution functions (ccdfs) of the maximum of the probability that a topic, $t$, belongs to a document $\alpha$, $m = \max_{t \in T}(\pcond{\pi}{t}{\alpha})$. (Panel b) ccdfs of the ratio, $r$, between $m$ and the second highest value of the probability. Colors account for different intensities of filtering $p$ on the climate change collection.}
  \label{sfig:ccdfs_topics_given_docs_climate} 
\end{figure*}

\begin{figure*}[h!] 
  \centering 
  \includegraphics[width=0.9\textwidth]{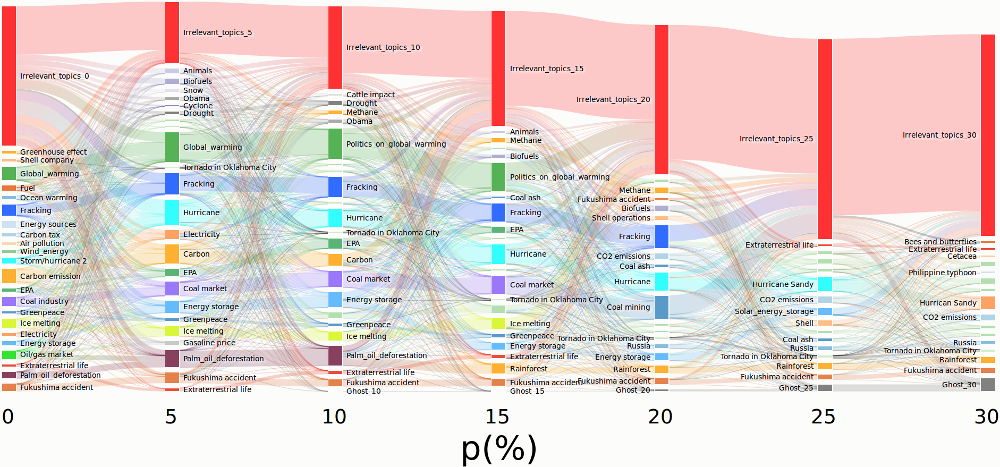} 
  \caption{Static Sankey diagram representing the topics found by TM on the climate change collection of webdocs. Each topic is identified with a colored box whose height corresponds to the number of articles associated to it. Each article $\alpha$ is assigned to the topic with maximum probability $\pcond{\pi}{t}{\alpha}$, \ie the topic that describes the highest portion of the article. Topics are manually labeled from the ten most representative concepts according to the probabilities of concepts given the topic, $\pcond{\pi}{w}{\alpha}$. For the ease of visualization, only topics with probability $\pi(t) > 0.01$ are shown, whereas the remaining ones are incorporated together in a single ``super-topic'' denoted as ``Irrelevant\_topics''. The boxes labeled ``ghost'' are composed by articles that do not contain any significant concept at a given percentile $p$, therefore are not part of the dataset used by TM. The thickness of the bands between boxes indicates the number of shared articles. Interactive version available at \cite{sankey-interactive}.}
  \label{sfig:sankey_TM_climate} 
\end{figure*}

As stated previously, the overall properties of the topics extracted with TM for several filtering intensities, are summarized in Tab.~\ref{stab:stats_topics_climate}. In Fig.~\ref{sfig:topicmapping_statistics_climate}, instead, we present the relation between the number of words per topic, $n_{w} (t)$, and its probability, $\pi(t)$. We dig more in the assignation of documents to topics with the cumulative distribution of the probability $\pcond{\pi}{t}{\alpha}$, and the ratio between its two highest values, $r$, reported in Fig.~\ref{sfig:ccdfs_topics_given_docs_climate}. Finally, we provide a comprehensive view of the organization of the collection into topics as we increase the filtering intensity in the Sankey diagram of Fig.~\ref{sfig:sankey_TM_climate} \cite{sankey-interactive}. The comparison of the Sankeys obtained using Louvain (Fig.~\ref{sfig:sankey_climate_rtf}) and TM (Fig.~\ref{sfig:sankey_TM_climate}) reveals two striking differences. One is the absence of the topic named ``Mixed\_themes'' found by Louvain for $p=20\%$. The other is the assignment of a remarkable fraction of documents to the ``Irrelevant topics'' already in absence of filtering. The absence of the huge ``Mixed\_themes'' topic, suggests that the progressive condensation phenomena taking place up to $p=25\%$ could be due to the inability of modularity to disentangle correctly the documents falling into that community. On the other hand, the presence of a non negligible amount of documents belonging to irrelevant topics -- regardless of the filtering intensity, -- is the symptom of the absence of an ontology. This proves the importance of validating the keywords extracted by KPEX to ensure the maximization of the information used to assign documents to topics encoded into the quantity $F$ reported in Tab.~\ref{stab:stats_topics_climate}.

%
%
\begin{table*}[h!]
\centering
%
\newcolumntype{d}[1]{D{.}{.}{#1} }
\begin{tabular*}{0.5\hsize}{@{\extracolsep{\fill}}c|ccccccd{2}}
$p \, (\%)$ & $N_{con}$ & $N_a$ & \multicolumn{1}{c}{$T$} & \multicolumn{1}{c}{$T^*$} & \multicolumn{1}{c}{$\langle N_a \rangle_{T^{*}}$} & \multicolumn{1}{c}{$\langle N_{con} \rangle_{T^{*}}$} & \multicolumn{1}{c}{$F$} \\ \hline
{\cellcolor{gray!20}} 0 & {\cellcolor{gray!20}} 822545 & {\cellcolor{gray!20}} 18770 & {\cellcolor{gray!20}} 201 & {\cellcolor{gray!20}} 22 & {\cellcolor{gray!20}} 432 & {\cellcolor{gray!20}} 26004 & {\cellcolor{gray!20}} 0.51 \\
5 & 8760 & 18770 & 112 & 23 & 637 & 359 & 0.78 \\
10 & 8299 & 18762 & 141 & 21 & 632 & 321 & 0.71 \\
15 & 7838 & 18743 & 193 & 20 & 554 & 248 & 0.59 \\
20 & 7377 & 18622 & 274 & 17 & 511 & 245 & 0.46 \\
25 & 6916 & 18308 & 365 & 14 & 358 & 136 & 0.27 \\
30 & 6455 & 17888 & 438 & 15 & 298 & 129 & 0.24 \\
35 & 5994 & 17126 & 515 & 10 & 282 & 102 & 0.15 \\
40 & 5533 & 16117 & 603 & 6 & 363 & 105 & 0.19 \\
50 & 4611 & 13527 & 725 & 3 & 280 & 63 & 0.04 \\
60 & 3689 & 10493 & 1035 & 2 & 201 & 44 & 0.02 \\
70 & 2767 & 7319 & 1745 & 2 & 107 & 25 & 0.01 \\
80 & 1845 & 4340 & 2275 & 0 & 0 & 0 & 0.00 \\
90 & 923 & 1879 & 1520 & 0 & 0 & 0 & 0.00 \\
\bottomrule
\end{tabular*}
\caption{Characteristics of the topic modeling on climate change dataset. The row $p=0\%$ corresponds to the original corpus/dataset, while $p>0\%$ to those filtered using the maximum entropy. In the columns we report: the percentage of filtered concepts $p$, the number of concepts $N_{con}$, of documents having at least one concept $N_a$, of topics found by LDA, $T$, and number of ``meaningful'' ones $T^*$. For the latter, we report also the average number of documents $\langle N_a \rangle_{T^{*}}$, and concepts $\langle N_{con} \rangle_{T^{*}}$ per topic. Finally, we report the fraction of documents assigned to a meaningful topic, $F$.}
\label{stab:stats_topics_climate}
\end{table*}
%

%
%
%
%
%
%
%
%
%
%
%
%
%
%
%

\newpage

\subsubsection{Comparison between Louvain based and TopicMapping based topics}
\label{s_sssec:comparison_louvain_topicmapping_climate}

\begin{figure}[hp!] 
  \centering 
  \includegraphics[height=0.95\textheight]{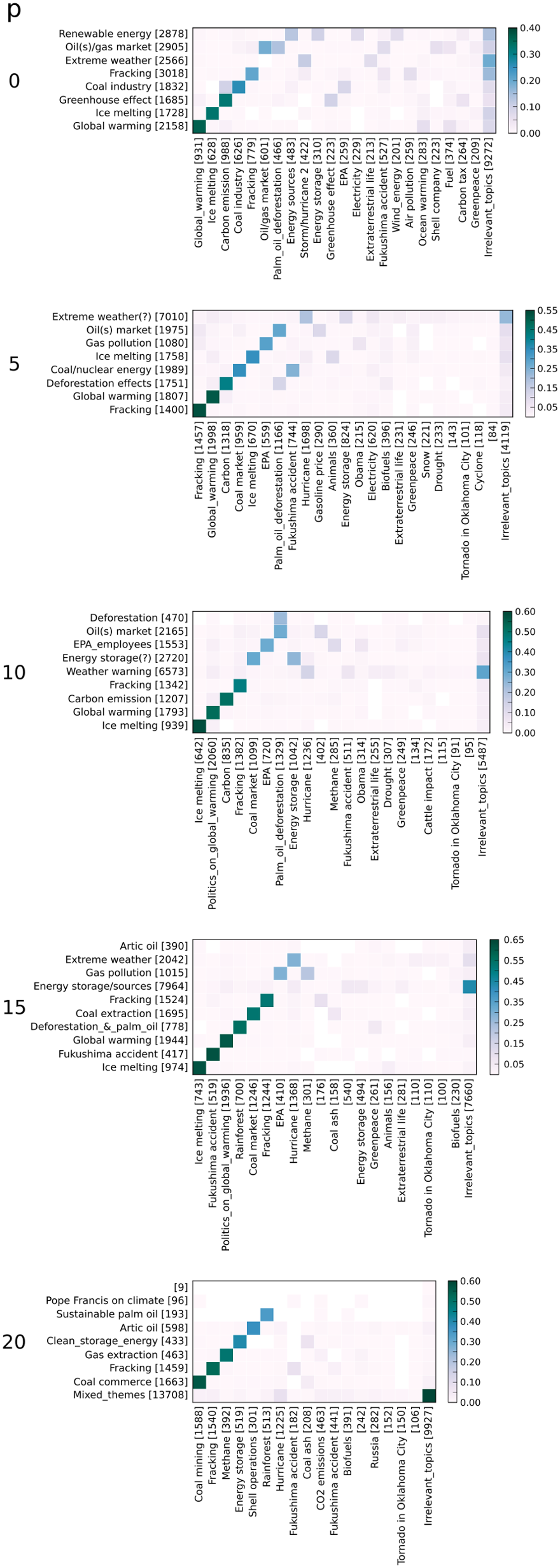} 
  \caption{Jaccard score, $J$, between the sets of articles belonging to a community (row) and those associated to a TM topic (column) for a given filtering percentile $p$ in the climate change collection. The number of articles in each set is indicated within square brackets. We consider only TM topics for which $\pi(t) > 0.01$.}
  \label{sfig:jaccard_rtf_vs_TM_climate} 
\end{figure}

\begin{figure}[hp!] 
  \centering 
  \includegraphics[height=0.87\textheight]{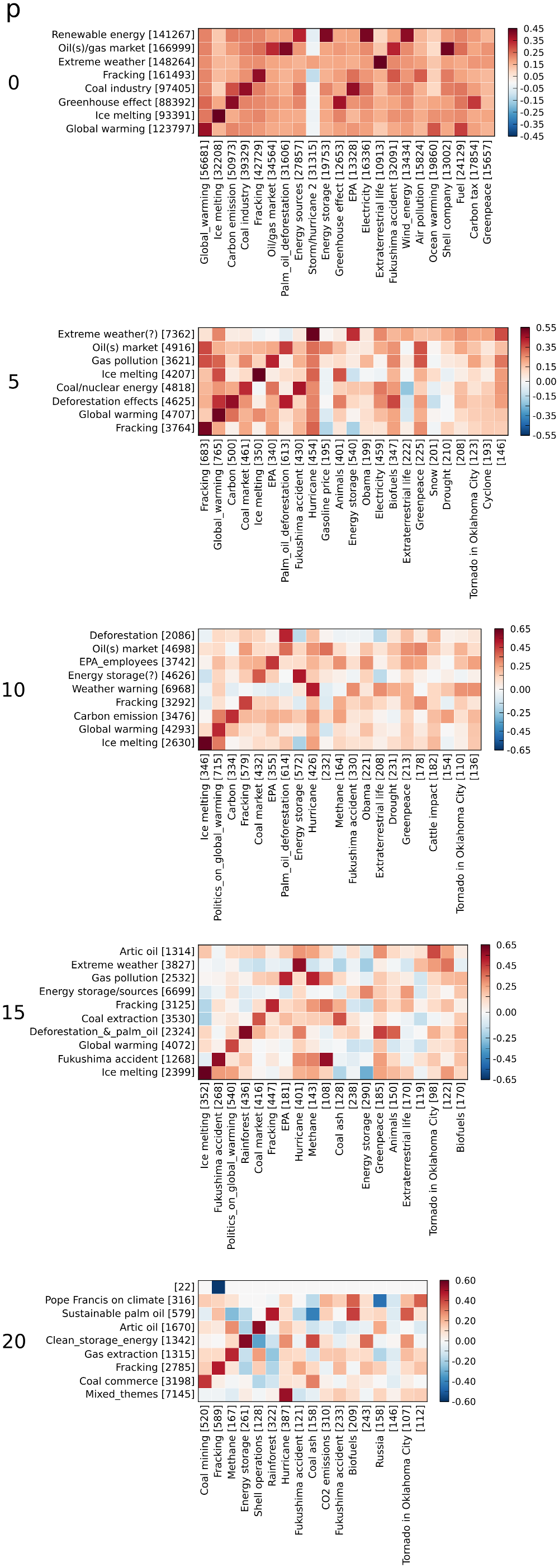} 
  \caption{Kendall correlation coefficient, $\tau_b$, for sets of ranked concepts belonging to topics identified by Louvain method (rows) and TM algorithm (columns) of the climate change collection. Each map refers to a given filtering percentile $p$. The number of concepts in every topic is indicated within square brackets. The $\tau_b$ is computed on the rankings of concepts appearing both in the Louvain and the TM topic. Only those topics for which $\pi(t) > 0.01$ are included in the heatmaps.}
  \label{sfig:tau_tf_vs_TM_climate} 
\end{figure}

In analogy with the analysis done in Sec.~\ref{s_sssec:comparison_louvain_topicmapping_phys2013} for the Physics 2013 collection, in Figs.~\ref{sfig:jaccard_rtf_vs_TM_climate} and \ref{sfig:tau_tf_vs_TM_climate} we compare the topics obtained with Louvain and TM for the climate change collection. As seen for Physics, both the Jaccard score, $J$, and the Kendall coefficient confirm that there is a certain level of overlap between the topics found by the two methodologies. However, the presence of a remarkable fraction of documents assigned to the ``Irrelevant\_topics'' topic done by TM, and the ``Mixed\_themes'' community found by Louvain at $p=20\%$ are exclusive hallmarks of this collection. Concerning the first, from the heatmap of the Jaccard score (Fig.~\ref{sfig:jaccard_rtf_vs_TM_climate}) we notice that except for $p=0\%$, the ``Irrelevant\_topics'' are mainly mapped into one of the Louvain topics. At $p=20\%$, instead, the ``Irrelevant\_topics'' and the ``Mixed\_themes'' are highly overlapping.


\pagebreak

%
%
%
%


\section{Filtering}
\label{s_sec:filtering}

In this section we provide some details concerning the choice of the optimal level of filtering (Sec.~\ref{s_ssec:optimal_filtering}) and how to implement the entropic filtering (Sec.~\ref{s_ssec:filtering_implementation}).

\subsection{Optimal filtering}
\label{s_ssec:optimal_filtering}

\begin{figure}[ht!]
\centering
\includegraphics[width=0.7\columnwidth]{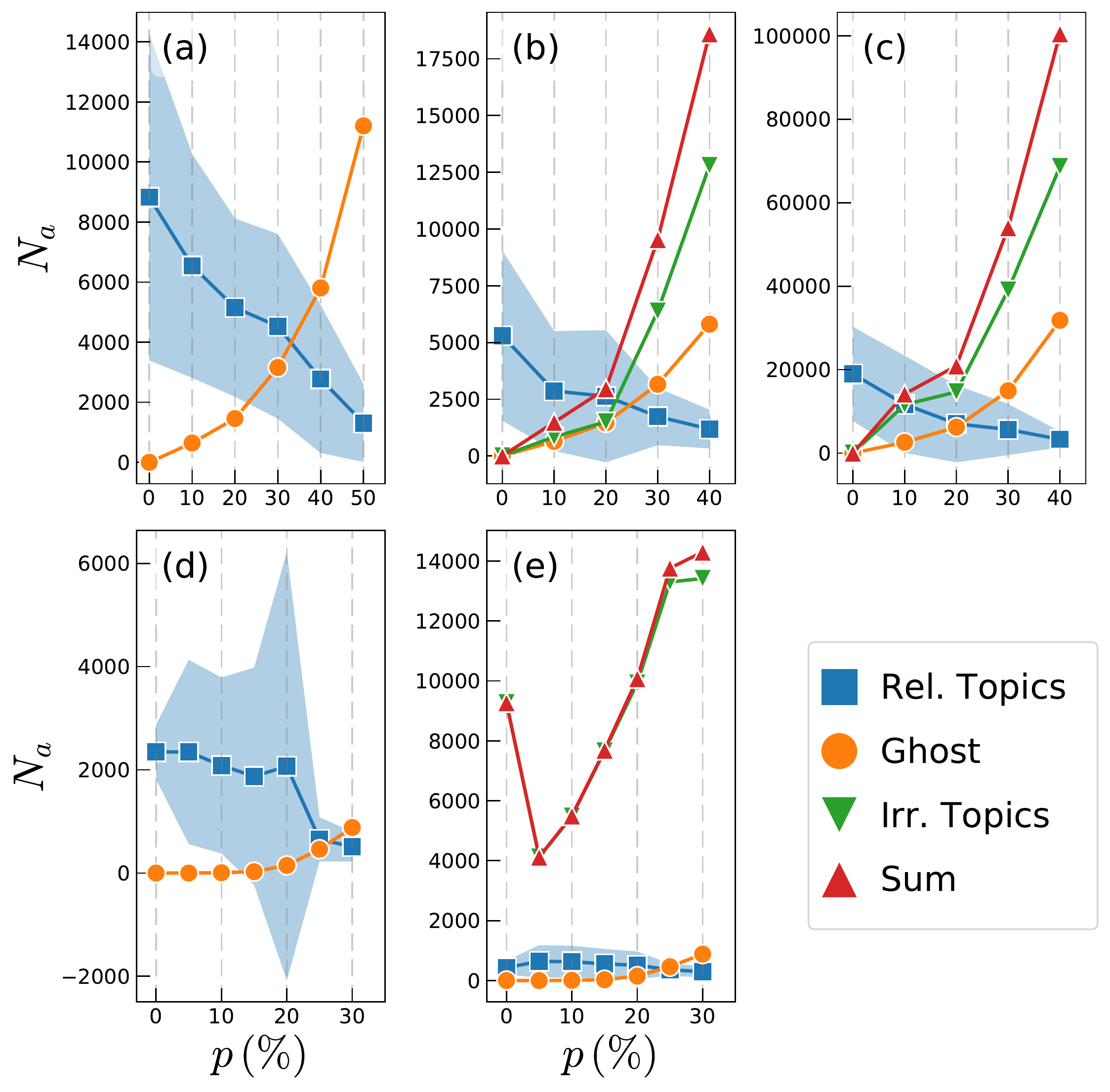}
\caption{Number of document per class of topics, $N_a$, as a function of filter aggressiveness, $p$, for all the collections considered. The squares indicate the average of the number of document per Relevant Topic, $\avg{N_a}$, while the shaded area denotes its standard deviation. Red triangles refers to the sum of the number of documents falling in the Ghost and Irrelevant Topics categories (if available.) Each panel refers to a distinct collection and topic extraction method: (a) Physics 2013 with Louvain, (b) Physics 2013 with TM, (c) Physics 2009-2012 with TM, (d) Climate with Louvain, and (e) Climate with TM, respectively.}
\label{sfig:optimal_filtering_bothdata_sizes}
\end{figure}

\begin{figure}[ht!]
\centering
\includegraphics[width=0.5\columnwidth]{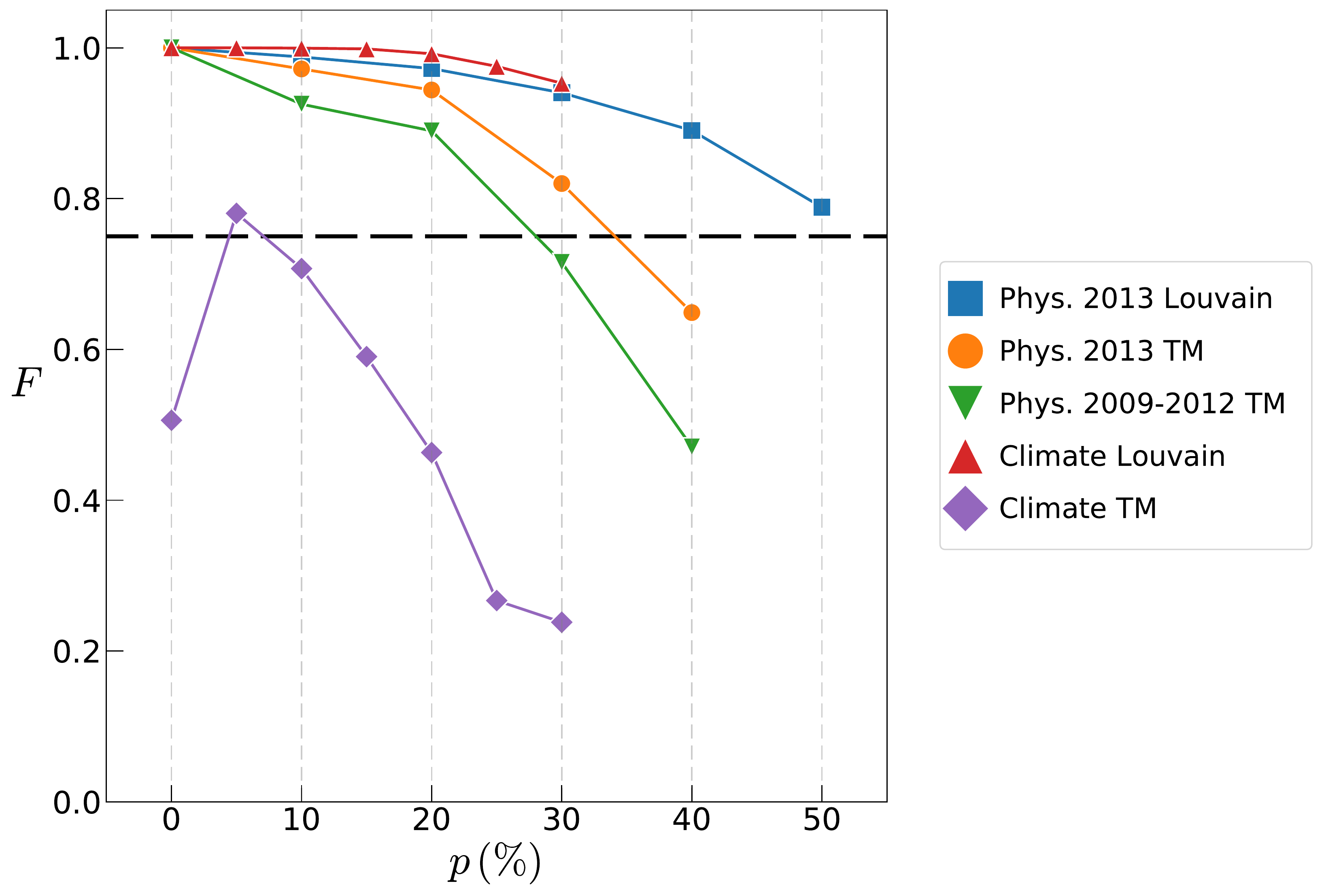}
\caption{Fraction of documents assigned to meaningful topics, $F$, as a function of the filter intensity $p$ for all the collections and methods used to extract topics. The vertical dashed line $F = 0.75$ denotes the threshold value below which too much information is lost.}
\label{sfig:optimal_filtering_bothdata_infos}
\end{figure}

The results shown so far depend on the amount of filtering applied. Hence, it is resonable to ask what is the optimal intensity of filtering to use (if it exists.) The different nature of the methods used to extract the topic structures, makes the formulation of a unique criterion not straightforward. For this reason, we considered distinct criteria for topics found using either TM or Louvain, and we summarize their outcome in Figs.~\ref{sfig:optimal_filtering_bothdata_sizes} and \ref{sfig:optimal_filtering_bothdata_infos}. Notably, in some cases different criteria return similar results (Physics) while in other (Climate) they do not.\\
\indent The first criterion stems from the idea of finding a trade off between the fragmentation of topics/communities into more specific (thus smaller) groups, and the size of the topics that are not relevant. In the latter category we find those documents belonging to the Ghost and/or to the Irrelevant Topic (in the case of TM extraction.) In Fig.~\ref{sfig:optimal_filtering_bothdata_sizes} we consider all the collection analyzed in combination with the two methods used to topic extraction. If we assume that a plausible filtering, $p_{opt}$, corresponds to the case in which the average size of a relevant topic is about the same of the irrelevant ones, then for Physics collection we get that -- depending on the case -- $20\% \leq p_{opt} \leq 30\%$. For Climate, instead, the crossing between the blue (squares) and orange (cicles) lines points to a value of $p_{opt} \simeq 25\%$ for Louvain, while no optimum seems to exist in the case of TM where the red/green lines (triangles) never intersect with the blue (squares) one.\\
\indent The second criterion, instead, consist in computing the fraction of documents assignable to a meaningful topic, $F \in [0,1]$. For a given intensity of filtering $p$, such fraction reads:
\begin{equation}
\label{seq:frac_mean_topics}
F(p) = \dfrac{N_{good}(p)}{N_{tot}} = %
\begin{cases}
&\dfrac{N_{tot} - N_a^G(p)}{N_{tot}} = 1 - \dfrac{N_a^G(p)}{N_{tot}} \qquad \text{for Louvain,}\\[3ex]
&\dfrac{N_{tot} - \Bigl(N_a^G(p) + N_a^I(p)\Bigr)}{N_{tot}} = 1 - \dfrac{N_a^G(p) + N_a^I(p)}{N_{tot}}\qquad \text{for TopicMapping.}
\end{cases}
\end{equation}
Where, $N_{good}$ is the number of documents assigned to meaningful topics, $N_a^G$ is the number of documents without any concept (\ie belonging to the Ghost,) and $N_a^I$ is the number of documents assigned to Irrelevant Topics. A value of $F=1$ means that all the documents in the collection are assigned to a meaningful topic while, conversely, $F=0$ corresponds to the case where none of the documents can be assigned to a topic. In the light of that, we can think of $F$ as a measure of the effective amount of classifiable information in the corpus under scrutiny. The value of $F$ as a function of $p$ is shown in Fig.~\ref{sfig:optimal_filtering_bothdata_infos}. If we tollerate the loss of -- at most -- 25\% of the information (\ie $F = 0.75$), then the optimal level of filtering $p_{opt}$ corresponds to the point where the data cross with the horizontal black dashed line.\\
\indent By looking at the picture, we immediately notice two things. One is that not all the collections/methods cross the $F=0.75$ threshold. The other is that the $F$ of the topic structure of Climate change collection obtained using TM displays a peak for $p=5\%$. Topic structures that do not display any crossing are those obtained using Louvain method, suggesting that the fragmentation approach is more suited than $F$ to estimate $p_{opt}$. On the other hand, $F$ outperforms fragmentation for Climate collection and TM method because the latter fails to grasp the huge disproportion between the amount of documents belonging to the meaninful topics, and the same quantity for the irrelevant ones. For those collections/methods where $F$ displays a crossing, the correspoinding values of $p_{opt}$ are higher than those obtained with the fragmentation approach, especially for the TM based topic structure. In a nutshell, it seems that the fragmentation approach is more suited for topics obtained using the Louvain method, while the $F$ performs better in the case of TopicMapping.
%
%
%
%
%
%

\newpage

\subsection{Numerical implementation with pseudocode}
\label{s_ssec:filtering_implementation}

In this section we present a step-by-step description of the algorithm used to implement the entropic filtering of concepts. We comment the cases of null model based on power-law distribution with a cutoff first (Sec.~\ref{s_sssec:filter_implementation_tf}) and lognormal then (Sec.~\ref{s_sssec:filter_implementation_rtf}). Our pseudocode is written using the Python programming language \cite{python} and we make use of several functions available in the \texttt{scipy} ecosystem \cite{scipy_python}.

The core of the method is the comparison between two entropies: the actual/experimental one $S_c$ and the expected/theoretical one $S_{\max}$ drawn from the the distribution obeying the maximum entropy principle. Given a collection of documents $\mathcal{D}$, for each concept $c$ appearing inside a document $d \in \mathcal{D}$, ScienceWISE provides its \textit{boosted term-frequency} $tf_{d}(c)$. Such quantity encodes the relevance of the concept taking into account where it appears. If we consider each document formed by three parts: \emph{title}, \emph{abstract} and \emph{body}, the boosted $tf$ is given by the sum of these contributions:
\begin{itemize}
 \item the number of times $c$ appears in the body;
 \item the number of times $c$ appears in the abstract multiplied by a factor of three;
 \item the number of times $c$ appears in the title, multiplied by a factor of five.
\end{itemize}
It is worth mentioning that, in the case of $tf$ density, this quantity is divided for the length of document $\alpha$, $L(\alpha)$, hence:
$$rtf_c(\alpha) = \dfrac{ tf_{c} (\alpha) } { L(\alpha)} \,.$$

\subsubsection{Discrete $tf$}
\label{s_sssec:filter_implementation_tf}

Given a sequence of $M$ values $\mathcal{X} = \lbrace x_1, x_2, \ldots, x_M \rbrace$, the corresponding probability mass function is given by:
$$P(\mathcal{X} = x) = P(x) = \tfrac{N(x)}{M}\,,$$
where $N(x)$ is the number of times the variable $\mathcal{X}$ has value $x$, while $M$ is the total number of values of $\mathcal{X}$. In our case, $\mathcal{X}$ is the $tf$ sequence of a concept $c$ and $P(\mathcal{X} = x)$ is the probability that $tf_c = x$, \ie the ratio between the number of documents $N(x)$ where a concept appears $x$ times and the total number of documents where $c$ appears, $M$. Given such definition, we denote with $\avg{\mathcal{X}}$, $\sigma_\mathcal{X}$ and $\avg{\ln \left( \mathcal{X} \right) }$, respectively: the average, standard deviation and average of the logarithm of $\mathcal{X}$. The algorithm is made by the following steps:
\begin{enumerate}
 \item \ulbf{Collection of the $tf$:}\newline\newline
    For each concept $c$, we collect the values of its $tf$ into a list, $l_{tf}$. After that, we compute the standard deviation of the set of values in such list, $\sigma_{l_{tf}}$. If the standard deviation is equal to zero, then it means that either the concept has appeared in only one paper or that it has appeared always the same number of times within the papers. Hence, we discard such concepts since their entropy is zero. For the remaining concepts, we count how many times $tf_{c} = k \in l_{tf} \; \forall \, k$ and store such number into another list named $tf_c^{exp}$. Thus, the probability that concept $c$ has a given $tf$ is:
    \begin{equation}
    \label{seq:prob_tf}
    p_c^{exp}(k) = \dfrac{tf_c^{exp}(k)}{\sum_{k^\prime} tf_c^{exp}(k^\prime)} \,.
    \end{equation}
 \item \ulbf{Extraction of distribution parameters:}\newline\newline
    In order to get a power law with a cutoff using the Lagrange multiplier method, we need to impose two constraints: the expected value of $\avg{tf}_{th}$ and of $\avg{\ln \left( tf \right)}_{th}$ have to match the same quantities computed on the data, as discussed in Sec.~\ref{s_sssec:maxent_discrete_tf}. The analytical form of the maximum entropy distribution becomes:
    \begin{equation}
    \label{seq:pmf}
    p(tf_c = k) \equiv p_c^{th}(k) = \frac{1}{Z} \, \frac{e^{-\lambda k}}{k^n} \,.
    \end{equation}
    where $Z$ is the normalization constant corresponding to the polylogarithm function $\polilog{n}$ of order $n$ and argument $\text{e}^{-\lambda}$, defined as:
    \begin{equation}
    \label{seq:polylog}
    Z \equiv \polilog{n} = \sum_{k=1}^\infty \dfrac{\text{e}^{-\lambda k}}{k^n} \,, \\
    \end{equation}
    The theoretical distribution $p_c^{th}(k)$ depends on two parameters: $n$ and $\lambda$. There are two ways to compute their values:
    \begin{enumerate}
      \item Exploit the fact that the theoretical maximum entropy distribution must reproduce the expectation values $\avg{l_{tf}}$ and $\avg{\ln \left( l_{tf} \right)}$. Therefore, we can find $n$ and $\lambda$ by solving numerically the following system:
	    \begin{equation}
	    \label{seq:maxent_system}
	      \begin{cases}
	      \avg{l_{tf}} =&  \dfrac{\polilog{n-1}}{\polilog{n}}\,,  \\
	      &\quad\\
	      \avg{\ln \left( l_{tf} \right)} =&  \dfrac{- \partial_n \polilog{n}}{\polilog{n}} =  \dfrac{\sum\limits_{k=1}^\infty \dfrac{\text{e}^{-\lambda k}}{k^n} \ln(k)}{\polilog{n}} \,.
	      \end{cases}
	    \end{equation}
      Since the polylogarithm function appears in the above system, we need to use the Python package named \texttt{mpmath}, which implements functions and methods with arbitrary precision float arithmetics. Thus, we define the two equations that have to the be solved simultaneously as:
\begin{lstlisting}
from mpmath import polylog, diff, findroot, fdiv
from math import log as mln
from math import exp as mexp

def eqs(n,z):
  eqA = fdiv(polylog(n-1,z),polylog(n,z)) - avg_tf
  eqB = fdiv(- diff(lambda v: polylog(v,z), n),polylog(n,z)) - avg_ln_tf
  
return (eqA, eqB)	
\end{lstlisting}
      where \texttt{fdiv} performs the division in \texttt{mpmath}, while \texttt{diff} is used to calculate numerically the derivative of the function \texttt{polylog} with respect to $n$. Then, we use the \texttt{findroot} function of \texttt{mpmath} to numerically solve the system of equations with:
\begin{lstlisting}
sol=findroot(eqs, ci, solver="secant")	
\end{lstlisting}
      Typically, the initial values of the parameters are \texttt{ci = (0.5,mexp(-0.1))}. The solution of \eqref{seq:maxent_system} is stored in \texttt{sol}, having $n$ and $\text{e}^{-\lambda}$ as its first and second element. The two parameters, together with the empirical values of $\avg{ l_{tf} }$ and $\avg{\ln \left( l_{tf} \right)}$, are then passed to the \texttt{max\_ent} function defined below to compute the maximum entropy.
      \item Use the \emph{maximum likelihood estimators} which employs the full data sequence to determine the parameters directly in $p_c^{th}$, without relying only on two constraints to do so. In this case, following the technique presented in \cite{powerlaw_newman,powerlaw_python} we use the Python \texttt{powerlaw} package to compute the parameters.
      \end{enumerate}

  \item \ulbf{Computation of Entropies:}\newline\newline
    Given the parameters $n$ and $\lambda$, we can compute the maximum entropy of a concept $c$ as:
    \begin{equation}
    S_{max} = \ln \left[ \polilog{n} \right] + \lambda \, \avg{tf_c^{exp}} + n \, \avg{\ln \left(tf_c^{exp}\right)} \,.
    \end{equation}
    which, implemented in Python, reads as follows:
\begin{lstlisting}
def max_ent(n,z,avg_tf,avg_ln_tf): 
  return mln( fp.polylog(n,z) ) - mln(z)*avg_tf + n*avg_ln_tf
\end{lstlisting}
    The empirical entropy, $S_{c}$, is instead computed using Shannon formula (\eqref{eq:entropy}) and distribution $p_c^{exp}(k)$.
\end{enumerate}

\subsubsection{Density of $tf$}
\label{s_sssec:filter_implementation_rtf}

As commented previously (see Sec.~\ref{s_sssec:maxent_rescaled_tf}), the maximum entropy distribution associated to the case of a rescaled term-frequency sequence, $rtf$, is a lognormal, defined as:
\begin{equation}
\label{seq:lognorm}
p(x;\mu,\sigma) = \frac{1}{ \sqrt{2 \pi} \, \sigma \, x} \exp \left[-\frac {(\ln x - \mu)^{2}} {2 \, \sigma^{2}} \right] \qquad \ \text{with} \; x>0 \,.
\end{equation}
Given a sequence of $M$ continuous values $\mathcal{X} = \lbrace x_1, x_2, \ldots, x_M \rbrace$, we define the probability to observe a value between $x$ and $x + \Delta x$ as $P(x,x + \Delta x)$. To compute such quantity, we have to consider the probability density function $p(x)$ and integrate it across the interval, such that:
\begin{equation}
\label{seq:prob_continuous}
P(x,x + \Delta x) = \int_{x}^{x + \Delta x} {p(y) \, dy} \,.
\end{equation}
Under this assumption, the algorithm is made by the following steps:
\begin{enumerate}
 \item \ulbf{Collection of $rtf$:}\newline\newline
    For each concept $c$, collect its $rtf$ values into a list, $l_{rtf}$. In analogy with the case of discrete $tf$, we discard those concepts having $\max(rtf)-\min(rtf) \leq 0.005$. Then, we create a binning $\lbrace \Delta k \rbrace$ of the interval $[\min(rtf), \max(rtf)]$ and compute the empirical probability, $P$, that the $rtf$ assumes a value between $k$ and $k + \Delta k$ , using \eqref{seq:prob_continuous}.
 \item \ulbf{Extraction of fit parameters:}\newline\newline
    Since the form of the lognormal distribution, \eqref{seq:lognorm}, the parameters $\mu$ and $\sigma$ that determine it are directly calculated from the empirical $rtf$ list, $l_{rtf}$, as $\mu \equiv \avg{ \ln( l_{rtf} ) } $ and $\sigma \equiv \sigma{ \left( \ln( l_{rtf} ) \right) } $, where the last term denotes of the standard deviation of the logarithm of the term-frequency density $l_{rtf}$.
 \item \ulbf{Computation of the residual entropy:}\newline\newline
  After obtaining parameters $\mu$ and $\sigma$, we compute the residual entropy, $S_d$, using a discrete version of the Kullback-Leibler divergence given by:
  \begin{equation}
  S_d = \sum P(k,k + \Delta k) \ln{ \frac{P(k,k + \Delta k)} {Q(k,k + \Delta k)} } \Delta k \,,
  \end{equation}
  where the sum is performed over the set of intervals used for the binning $\lbrace \Delta k \rbrace$. It is worth stressing that such binning is the same for both $P$ and $Q$. Such operation is achieved by the following code:
\begin{lstlisting}
def discrete_KL(data_distro, th_distro, bin_widths):
  return np.sum(data_distro*np.log(np.true_divide(data_distro, th_distro))*bin_widths)
	  
num_bins_fixed_kl = 15

binning = np.logspace(np.log10(min(rescaled_tfs)*0.999),\
                      np.log10(max(rescaled_tfs)*1.001),\
                      num_bins_fixed_kl+1)

vs_r_tfs, bs_r_tfs = np.histogram(r_tfs, bins = binning, density=True)
	
centers_bins = (binning[1:]+binning[:-1])/2.

bin_ranges = binning[1:] - binning[:-1]

# Removal of bins with no data for the experimental distro
indx_nnz_vs_r_tfs = np.nonzero(vs_r_tfs)

vs_r_tfs_nnz = vs_r_tfs[indx_nnz_vs_r_tfs]

centers_bins_nnz = centers_bins[indx_nnz_vs_r_tfs]

bin_ranges_nnz = bin_ranges[indx_nnz_vs_r_tfs]

# Only calculated for the middle point of the bins for nonzero integral
# values of the data histogram

th_pdf = lognorm.pdf( centers_bins_nnz, loc=0, s=sigma, scale=scale )

dKL = discrete_KL( vs_r_tfs_nnz, th_pdf, bin_ranges_nnz )
\end{lstlisting}
\texttt{data\_distro} and \texttt{th\_distro} contain the values of the probability distribution functions evaluated at the center of the intervals $\lbrace \Delta k \rbrace$ for the observed sequence $l_{rtf}$ and the theoretically expected one.
\end{enumerate}

\subsubsection{Generation of similarity networks}
\label{s_sssec:filter_build_similarity}

After computing the $S_{max}$, for each concept $c$, we compute its \textit{residual entropy}, $S_d$, and store its value on a list $l_{sdiff}$. We then compute the percentiles of $l_{sdiff}$ using the \texttt{numpy} function named \texttt{percentile}. We compute the percentiles $\tilde{P}$ from $\tilde{P}_{\min}$ to $\tilde{P}_{\max}$ using a step of $\tilde{P}_{\text{step}}$, and use those values to create separate files containing the lists of concepts having $S_d \geq \tilde{P}$. Finally, we build the similarity network between documents using exclusively those concepts contained in the file corresponding to the $i$-th percentile $\tilde{P}$.
%

%
%
%
%


\end{document}